\documentclass[11pt]{article}
\usepackage{geometry}             
\usepackage{graphicx}
\usepackage{amsmath}
\usepackage[colorlinks]{hyperref}
\usepackage{array}
\usepackage{alltt}
\usepackage{amssymb}
\usepackage{epstopdf}
\newcommand{\p}[1]{${\tt #1}$}

\usepackage{bm} 
\newcommand{\bmv}[1]{\bm{ #1}} 

\title{Poincar\'{e} Analyticity and the Complete Variational Equations}

\author{D. Kaltchev$^{a,}$\footnote{Corresponding author.}\;
and\; 
A.~J.~Dragt$^{b,}$\thanks{Work supported in part by U.S. Department of Energy Grant DE-FG02-96ER40949.}
 \\ \\
{\it $^a$\,TRIUMF, 4004 Wesbrook Mall, Vancouver, B.C., Canada V6T 2A3 }\\
\texttt{kaltchev@triumf.ca}\\ \\
{\it $^b$\,Physics Department, University of Maryland, College Park, Maryland 20742, USA }\\ 
\texttt{dragt@umd.edu}
}

\begin{document}
\maketitle
\section*{Abstract}
According to a theorem of Poincar\'{e}, the solutions to differential equations are analytic functions of (and therefore have Taylor expansions in) the initial  conditions and various parameters providing the right sides of the differential equations are analytic in the variables, the time, and the parameters.   We describe how these Taylor expansions may be obtained, to any desired order, by integration of what we call the {\em complete} variational equations.  As illustrated in a Duffing equation stroboscopic map example, these Taylor expansions, truncated at an appropriate order thereby providing polynomial approximations, can well reproduce the behavior (including infinite period doubling cascades and strange attractors) of the solutions of the underlying differential equations. 

\section{Introduction}
\setcounter{equation}{0}

In his prize-winning essay[1,2] and subsequent monumental work on celestial mechanics[3], Poincar\'{e} established and exploited the fact that solutions to differential equations are frequently analytic functions of the initial conditions and various parameters (should any occur).  This result is often referred to as {\em Poincar\'{e} analyticity} or {\em Poincar\'{e}'s theorem} on analyticity.

Specifically, consider any set of $m$ first-order differential equations of the form 
\begin{equation}
  \dot{z}_a = f_a(z_1,\cdots,z_m;t;\lambda_1,\cdots,\lambda_n),\quad a= 1,\cdots,m.
\end{equation}
Here $t$ is the independent variable, the $z_a$ are dependent variables, and the $\lambda_b$ are possible parameters.  Let the quantities $z^0_a$ be initial conditions specified at some initial time $t=t^0$,
\begin{equation}
z_a(t^0)=z^0_a.
\end{equation}
Then, under mild conditions imposed on the functions $f_a$ that appear on the right side of (1.1) and thereby define the set of differential equations, there exists a \emph{unique} solution 
\begin{equation}
z_a(t) = g_a(z^0_1, \cdots, z^0_m; t^0, t;\lambda_1,\cdots,\lambda_n), \ \ a = 1,m
\end{equation}
of (1.1) with the property
\begin{equation}
z_a(t^0) = g_a(z^0_1, \cdots, z^0_m; t^0, t^0;\lambda_1,\cdots,\lambda_n) = z^0_a, \ \ a = 1,m.
\end{equation}

Now assume that the functions $f_a$ are analytic (within some domain) in the quantities $z_a$, the time $t$, and the parameters $\lambda_b$.  Then, according to Poincar\'{e}'s Theorem,  the solution given by (1.3) will be analytic (again within some domain) in the initial conditions $z^0_a$, the times $t^0$ and $t$, and the parameters $\lambda_b$.  

Poincar\'{e} established this result on a case-by-case basis as needed using  {\em Cauchy's} method of {\em majorants}.  It is now more commonly established in general using {\em Picard iteration}, and appears as background material in many standard texts on ordinary differential equations [4]. 

If the solution $z_a(t)$ is analytic in the initial conditions $z^0_a$ and the parameters $\lambda_b$, then it is possible to expand it in the form of a Taylor series, with time-dependent coefficients, in the variables $z^0_a$ and $\lambda_b$.  The aim of this paper is to describe how these Taylor coefficients can be found as solutions to what we will call the {\em complete variational} equations.

To aid further discussion, it is useful to also rephrase our goal in the context of maps.  Suppose we rewrite the set of first-order differential equations (1.1) in the more compact vector form
\begin{equation}
\dot{\mbox{\boldmath $z$}} = \mbox{\boldmath $f$} (\mbox{\boldmath $z$}; t;\mbox{\boldmath $\lambda$}).
\end{equation}

Then, again using vector notation, their solution can be written in the form
\begin{equation}
\mbox{\boldmath $z$}(t) = \mbox{\boldmath $g$} (\mbox{\boldmath $z$}^0; t^0,t;\mbox{\boldmath $\lambda$}).
\end{equation}
That is, the quantities $\mbox{\boldmath $z$}(t)$ at any time $t$ are uniquely  specified by the initial quantities $\mbox{\boldmath $z$}^0$ given at the initial time $t^0$.

We capitalize on this fact by introducing a slightly different notation. 
First, use $t^i$ instead of $t^0$ to denote the \emph{initial} time.
Similarly use $\mbox{\boldmath $z$}^i$ to denote initial conditions by writing
\begin{equation}
\mbox{\boldmath $z$}^i = \mbox{\boldmath $z$}^0 = \mbox{\boldmath $z$}(t^i).
\end{equation}
Next, let $t^f$ be some \emph{final} time, and define final conditions
$\mbox{\boldmath $z$}^f$ by writing
\begin{equation}
\mbox{\boldmath $z$}^f = \mbox{\boldmath $z$}(t^f).
\end{equation}
Then, with this notation, (1.6) can be rewritten in the form
\begin{equation}
\mbox{\boldmath $z$}^f = \mbox{\boldmath $g$} (\mbox{\boldmath $z$}^i; t^i,t^f;\mbox{\boldmath $\lambda$}).
\end{equation}

We now view (1.9) as a {\em map} that sends the initial conditions $\mbox{\boldmath $z$}^i$ to the final conditions $\mbox{\boldmath $z$}^f$.  This map will be called the \emph{transfer map} between the times $t^i$ and $t^f$, and will be denoted by the symbol $\cal{M}$.  What we have emphasized is that a set of first-order differential equations of the form (1.5) can be integrated to produce a transfer map $\cal{M}$.  We express the fact that $\cal{M}$ sends $\mbox{\boldmath $z$}^i$ to $\mbox{\boldmath $z$}^f$ in symbols by writing 
\begin{equation}
{\mbox{\boldmath $z$}^f} = {\cal{M}}  {\mbox{\boldmath $z$}}^i,
\end{equation} 
and illustrate this relation by the picture shown in Figure 1. We also note that $\cal{M}$ is always invertible: Given $\mbox{\boldmath $z$}^f$, $t^f$, and $t^i$, we can always integrate (1.5) backward in time from the moment $t=t^f$ to the moment $t=t^i$ and thereby find  the initial conditions $\mbox{\boldmath $z$}^i$.  In the context of maps, our goal is to find a Taylor representation for $\cal M$.  If parameters are present, we may wish to have an expansion in some or all of them as well.

%Fig_141new
\begin{figure}[htp]
  \centering
\includegraphics*[height=2.1in,angle=0]{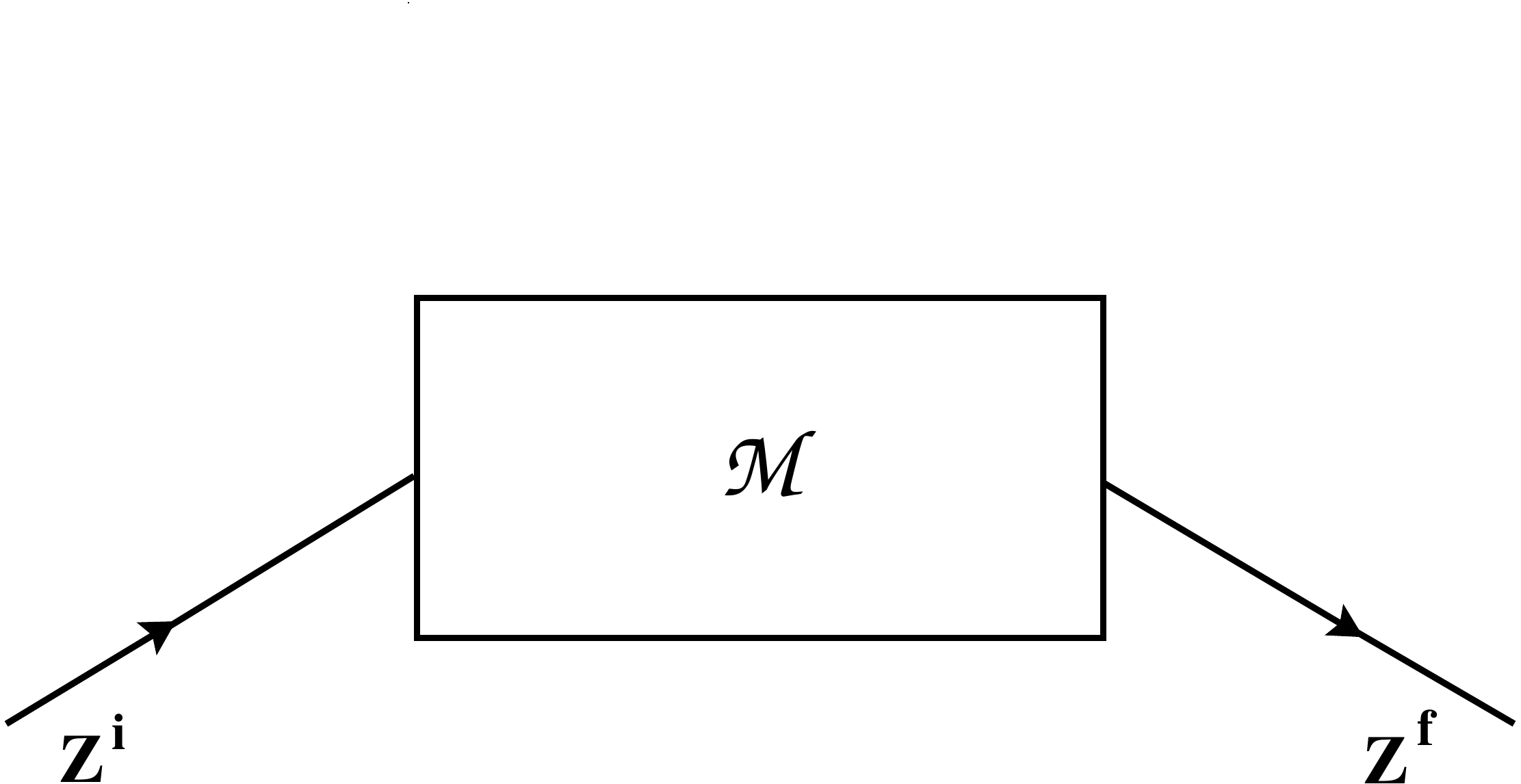}
  \caption{The transfer map $\cal{M}$ sends the initial conditions
           $\mbox{\boldmath $z$}^i$ to the final conditions
           $\mbox{\boldmath $z$}^f$.}
\end{figure}

The organization of this paper is as follows:  Section 2 derives the complete variational equations without and with dependence on some or all parameters.  Sections 3 and 4 describe their solution using forward and backward integration.  As an example, Section 5 treats the Duffing equation and describes the properties of an associated stroboscopic map $\cal{M}$.  Section 6 sets up the complete variational equations for the Duffing equation, including some parameter dependence, and studies some of the properties of the map obtained by solving these variational equations numerically. There we will witness the remarkable fact that a truncated Taylor map approximation to $\cal M$ can reproduce the infinite period-doubling Feigenbaum cascade and associated strange attractor exhibited by the exact $\cal M$.  Section 7 describes how the variational equations can be solved numerically.  A final section provides a concluding summary.

\section{Complete Variational Equations}
\setcounter{equation}{0}

This section derives the complete variational equations, first without parameter dependence, and then with parameter dependence.

\subsection{Case of No or Ignored Parameter Dependence}

Suppose the equations (1.1) do not depend on any parameters $\lambda_b$ or we do not wish to make expansions in them.  We may then suppress their appearance to rewrite (1.1) in the form
\begin{equation}
\dot{z}_a = f_a(z,t), \ \ a = 1,m.
\end{equation}
Suppose that $z^d(t)$ is some given {\em design} solution to these equations, and we wish to study solutions in the vicinity of this solution.  That is, we wish to make expansions about this solution.  Introduce deviation variables $\zeta_a$ by writing
\begin{equation}
z_a = z^d_a + \zeta_a .
\end{equation}
Then the equations of motion (2.1) take the form
\begin{equation}
\dot{z}^d_a + \dot{\zeta}_a = f_a (z^d + \zeta ,t).
\end{equation}
In accord with our hypothesis of analyticity, assume that the right side of (2.3) is analytic about $z^d$.  Then we may write the relation
\begin{equation}
f_a (z^d + \zeta ,t) = f_a (z^d,t) + g_a(z^d,t,\zeta )
\end{equation}
where each $g_a$ has a Taylor expansion of the form
\begin{equation}
g_a(z^d,t,\zeta ) = \sum_r g_a^r(t) G_r(\zeta ).
\end{equation}
Here the $G_r(\zeta )$ are the various monomials in the $m$ variables $\zeta_b$ labeled by an {\em index} $r$ using some convenient labeling scheme, and the $g^r_a$ are (generally) time-dependent coefficients which we call {\em forcing terms}.\footnote{Here and in what follows the quantities $g_a$ are not to be confused with those appearing in (1.3).}  By construction, all the monomials occurring in the right side of (2.5) have degree one or greater.  We note that the $g^r_a(t)$ are known once $z^d(t)$ is given.  

By assumption, $z^d$ is a solution of (2.3) and therefore satisfies the relations
\begin{equation}
\dot{z}^d_a = f_a(z^d,t).
\end{equation}
It follows that the deviation variables satisfy the equations of motion
\begin{equation}
\dot{\zeta}_a = g_a(z^d,t,\zeta ) = \sum_r g_a^r(t)G_r(\zeta ).
\end{equation}
These equations are evidently generalizations of the usual first-degree (linear) variational equations, and will be called the {\em complete variational} equations.

Consider the solution to the complete variational equations with {\em initial} conditions $\zeta^i_b$ specified at some initial time $t^i$.  Following Poincar\'{e}, we expect that this solution will be an analytic function of the initial conditions $\zeta^i_b$.  Also, since the right side of (2.7) vanishes when all $\zeta_b = 0$ [all the monomials $G_r$ in (2.7) have degree one or greater], $\zeta (t) = 0$ is a solution to (2.7).  It follows that the solution to the complete variational equations has a Taylor expansion of the form
\begin{equation}
\zeta_a(t) = \sum_r h^r_a(t) G_r(\zeta^i)
\end{equation}
where the $h^r_a(t)$ are functions to be determined, and again all the monomials $G_r$ that occur have degree one or greater.  When the quantities $h^r_a(t)$ are evaluated at some {\em final} time $t^f$, (2.8) provides a representation of the transfer map ${\cal M}$ about the design orbit in the Taylor form
\begin{equation}
\zeta^f_a = \zeta_a(t^f) = \sum_r h^r_a(t^f) G_r(\zeta^i).
\end{equation}

\subsection{Complete Variational Equations with Parameter Dependence}

What can be done if  we desire to have an expansion in parameters as well?  Suppose that there are $n$ such parameters, or that we wish to have expansions in $n$ of them.  The work of the previous section can be extended to handle this case by means of a simple trick:  View the $n$ parameters as additional {\em variables}, and ``augment" the set of differential equations by additional differential equations that ensure these additional variables remain constant.

In detail, suppose we label the parameters so that those in which we wish to have an expansion are $\lambda_1\cdots\lambda_n$.  Introduce $n$ additional variables $z_{m+1},\cdots z_\ell$ where $\ell=m+n$ by making the replacements
\begin{equation}
\lambda_b\rightarrow z_{m+b}, \ \ b=1,n.
\end{equation}
Next augment the equations (1.1) by $n$ more of the form
\begin{equation}
\dot{z}_a=0, \ \ a=m+1,\ell.
\end{equation}
By this device we can rewrite the equations (1.1) in the form
\begin{equation}
\dot{z}_a = f_a(z,t), \ \ a = 1,\ell
\end{equation}
with the understanding that
\begin{equation} 
f_a=f_a(z;t;\lambda^{\rm{rem}}),\ \ a=1,m,
\end{equation}
where $\lambda^{\rm{rem}}$ denotes the other {\em remaining} parameters, if any, and
\begin{equation}
f_a=0, \ \ a=m+1,\ell.
\end{equation}
For the first $m$ equations we impose, as before, the initial conditions
\begin{equation}
z_a(t^i)=z_a^i, \ \ a=1,m.
\end{equation}
For the remaining equations we impose the initial conditions
\begin{equation}
z_a(t^i)=\lambda_{a-m}, \ \ a=m+1,\ell.
\end{equation}
Note that the relations (2.14) then ensure the $z_a$ for $a>m$ retain these values for all $t$.

To continue, let $z^d(t)$ be some design solution.  Then, by construction,  we have the result
\begin{equation}
z^d_a(t)=\lambda^d_{a-m}=\lambda_{a-m}, \ \  a=m+1,\ell. 
\end{equation}
Again introduce deviation variables by writing 
\begin{equation}
z_a=z_a^d+\zeta_a \ \ a=1,\ell.
\end{equation}
Then the quantities $\zeta_a$ for $a>m$ will describe deviations in the parameter values.  Moreover, because we have assumed analyticity in the parameters as well, relations of the forms (2.4) and (2.5) will continue to hold except that the $G_r(\zeta)$ are now the various monomials in the $\ell$ variables $\zeta_b$.  Relations of the forms (2.6) and (2.7) will also hold with the provisos (2.13) and (2.14) and 
\begin{equation}
g_a^r(t)=0,\ \ a=m+1,\ell.
\end{equation}
Therefore, we will only need to integrate the equations of the forms (2.6) and (2.7) for $a\le m$.  Finally, relations of the form (2.9) will continue to hold for 
$a\le m$ supplemented by the relations
\begin{equation}
\zeta_a^f=\zeta_a^i, \ \ a=m+1,\ell.
\end{equation}
Since the $G_r(\zeta^i)$ now involve $\ell$ variables, the relations of the form (2.9) will provide an expansion of the final quantities $\zeta^f_a$ (for $a\le m$) in terms of the initial quantities $\zeta^i_a$ (for $a\le m$) and also the parameter deviations $\zeta_a^i$ with $a=m+1,\ell$.

\section{Solution of Complete Variational Equations Using Forward Integration}
\setcounter{equation}{0}

This section and the next describe two methods for the solution of the complete variational equations.  This section describes the method that employs integration forward in time, and is the conceptually simpler of the two methods.

\subsection{Method of Forward Integration}

To determine the functions $h^r_a$, let us insert the expansion (2.8) into both sides of (2.7).  With $r^{\prime \prime}$ as a dummy index, the left side becomes the relation
\begin{equation}
\dot{\zeta}_a = \sum_{r^{\prime \prime}} \dot{h}^{r^{\prime \prime}}_a(t) G_{r^{\prime \prime}}(\zeta^i).
\end{equation}
For the right side we find the intermediate result
\begin{equation}
\sum_r g^r_a(t) G_r(\zeta ) = \sum_r g^r_a(t) \ G_r\!\!\left( \sum_{r^{\prime}} h_1^{r^{\prime}}(t) G_{r^{\prime}}(\zeta^i), \cdots \sum_{r^{\prime}} h^{r^{\prime}}_m(t) G_{r^{\prime}}(\zeta^i)\right) .
\end{equation}
However, since the $G_r$ are monomials, there are relations of the form
\begin{equation}
G_r\!\!\left( \sum_{r^{\prime}} h_1^{r^{\prime}}(t) G_{r^{\prime}}(\zeta^i), \cdots \sum_{r^{\prime}} h^{r^{\prime}}_m(t) G_{r^{\prime}}(\zeta^i)\right) = \sum_{r^{\prime \prime}} U^{r^{\prime \prime}}_r (h^s_n) G_{r^{\prime \prime}} (\zeta^i),
\end{equation}
and therefore the right side of (2.7) can be rewritten in the form
\begin{equation}
\sum_r g^r_a(t) G_r(\zeta ) = \sum_{r^{\prime \prime}} \sum_r g^r_a(t) U^{r^{\prime \prime}}_r (h^s_n) G_{r^{\prime \prime}} (\zeta^i) .
\end{equation}
The notation $U^{r^{\prime \prime}}_r (h^s_n)$ is employed to indicate that these quantities might (at this stage of the argument) depend on all the $h^s_n$ with $n$ ranging from $1$ to $m$, and $s$ ranging over all possible values.

Now, in accord with (2.7), equate the right sides of (3.1) and (3.4) to obtain the relation
\begin{equation}
\sum_{r^{\prime \prime}} \dot{h}^{r^{\prime \prime}}_a (t) G_{r^{\prime \prime}}(\zeta^i) = \sum_{r^{\prime \prime}} \sum_r g^r_a(t) U^{r^{\prime \prime}}_r (h^s_n) G_{r^{\prime \prime}} (\zeta^i).
\end{equation}
Since the monomials $G_{r^{\prime \prime}} (\zeta^i)$ are linearly independent, we must have the result
\begin{equation}
\dot{h}^{r^{\prime \prime}}_a (t)  = \sum_r g^r_a(t) U^{r^{\prime \prime}}_r(h^s_n).
\end{equation}
We have found a set of differential equations that must be satisfied by the $h^r_a$.   Moreover, from (2.8) there is the relation
\begin{equation}
\zeta_a(t^i) = \sum_r h^r_a(t^i) G_r(\zeta^i) = \zeta^i_a.
\end{equation}
Thus, all the functions $h^r_a(t)$ have a known value at the initial time $t^i$, and indeed are mostly initially zero.  When the equations (3.6) are integrated {\em forward} from $t=t^i$ to $t=t^f$ to obtain the quantities $h^r_a(t^f)$, the result is the transfer map ${\cal M}$ about the design orbit in the Taylor form (2.9).

Let us now examine the structure of this set of differential equations.  A key observation is that the functions $U^{r^{\prime \prime}}_r(h^s_n)$ are {\em universal}.  That is, as (3.3) indicates, they describe certain {\em combinatorial} properties of monomials.  They depend only on the dimension $m$ of the system under study, and are the {\em same} for all such systems.  As (2.7) shows, what are peculiar to any given system are the forcing terms $g^r_a(t)$.

\subsection{Application of Forward Integration to the Two-variable Case}
To see what is going on in more detail, it is instructive to work out the first nontrivial case, that with $m=2$.  For two variables, all monomials in $(\zeta_1,\zeta_2)$ are of the form $(\zeta_1)^{j_1}(\zeta_2)^{j_2}$.  Here, to simplify notation, we have dropped the superscript $i$.  Table 1 below shows a convenient way of labeling such monomials, and for this labeling we write
\begin{equation}
G_r (\zeta ) = (\zeta_1)^{j_1} (\zeta_2)^{j_2}
\end{equation}
with
\begin{equation}
j_1=j_1(r) \ \ {\rm{and}} \ \  j_2=j_2(r).
\end{equation}

\begin{table}[h,t]
\caption{A labeling scheme for monomials in two variables.}
\begin{center}
\begin{tabular}{cccc}
$r$ & $j_1$ & $j_2$ & $D$ \\ \hline
1 & 1 & 0 & 1 \\
2 & 0 & 1 & 1 \\
3 & 2 & 0 & 2 \\
4 & 1 & 1 & 2 \\
5 & 0 & 2 & 2 \\
6 & 3 & 0 & 3 \\
7 & 2 & 1 & 3 \\
8 & 1 & 2 & 3 \\
9 & 0 & 3 & 3 \\
\end{tabular}
\end{center}
\end{table}

\noindent Thus, for example,
\begin{equation}
G_1 = \zeta_1,
\end{equation}
\begin{equation}
G_2 = \zeta_2,
\end{equation}
\begin{equation}
G_3 = \zeta_1^2,
\end{equation}
\begin{equation}
G_4 = \zeta_1\zeta_2,
\end{equation}
\begin{equation}
G_5 = \zeta_2^2, \ {\rm etc}.
\end{equation}
For more detail about monomial labeling schemes, see Section 7.

Let us now compute the first few $U^{r^{\prime \prime}}_r(h^s_n)$.  From (3.3) and (3.10) we find the relation
\begin{equation}
G_1 \!\!\left( \sum_{r^{\prime}} h^{r^{\prime}}_1 G_{r^{\prime}}(\zeta), \sum_{r^{\prime}} h^{r^{\prime}}_2 G_{r^{\prime}}(\zeta)\right) = \sum_{r^{\prime}} h^{r^{\prime}}_1 G_{r^{\prime}}(\zeta) = \sum_{r^{\prime \prime}} U^{r^{\prime \prime}}_1 G_{r^{\prime \prime}}(\zeta).
\end{equation}
It follows that there is the result
\begin{equation}
U^{r^{\prime \prime}}_1 = h^{r^{\prime \prime}}_1.
\end{equation}
Similarly, from (3.3) and (3.11), we find the result
\begin{equation}
U^{r^{\prime \prime}}_2 = h^{r^{\prime \prime}}_2.
\end{equation}
From (3.3) and (3.12) we find the relation
\begin{eqnarray}
G_3 &&\!\!\!\!\!\!\!\!\!\!\!\!\!\!\!\left( \sum_{r^{\prime}} h^{r^{\prime}}_1 G_{r^{\prime}}(\zeta), \sum_{r^{\prime}} h^{r^{\prime}}_2 G_{r^{\prime}}(\zeta)\right) = \left( \sum_{r^{\prime}} h^{r^{\prime}}_1 G_{r^{\prime}}(\zeta)\right)^2 \nonumber \\
&=& \sum_{s,t} h_1^sh^t_1 G_s(\zeta ) G_t(\zeta )=\sum_{r^{\prime \prime}}  U^{r^{\prime \prime}}_3 G_{r^{\prime \prime}}(\zeta).
\end{eqnarray}
Use of (3.18) and inspection of (3.10) through (3.14) yields the results
\begin{equation}
U^1_3 = 0,
\end{equation}
\begin{equation}
U^2_3 = 0,
\end{equation}
\begin{equation}
U^3_3 = (h^1_1)^2,
\end{equation}
\begin{equation}
U^4_3 = 2h^1_1h^2_1,
\end{equation}
\begin{equation}
U^5_3 = (h^2_1)^2.
\end{equation}
From (3.3) and (3.13) we find the relation
\begin{eqnarray}
G_4 &&\!\!\!\!\!\!\!\!\!\!\!\!\!\!\!\left( \sum_{r^{\prime}} h^{r^{\prime}}_1 G_{r^{\prime}}(\zeta), \sum_{r^{\prime}} h^{r^{\prime}}_2 G_{r^{\prime}}(\zeta)\right) = \left( \sum_{r^{\prime}} h_1^{r^{\prime}} G_{r^{\prime}}(\zeta )\right) \left( \sum_{r^{\prime}} h_2^{r^{\prime}} G_{r^{\prime}}(\zeta )\right) \nonumber \\
&=& \sum_{s,t}  h_1^s h_2^t G_s(\zeta ) G_t(\zeta ) = \sum_{r^{\prime \prime}} U^{r^{\prime \prime}}_4 G_{r^{\prime \prime}}(\zeta).
\end{eqnarray}
It follows that there are the results
\begin{equation}
U^1_4 = 0,
\end{equation}
\begin{equation}
U^2_4 = 0,
\end{equation}
\begin{equation}
U^3_4 = h^1_1h_2^1,
\end{equation}
\begin{equation}
U^4_4 = h^1_1h^2_2 + h^2_1h^1_2,
\end{equation}
\begin{equation}
U^5_4 = h^2_1 h^2_2.
\end{equation}
Finally, from (3.3) and (3.14), we find the results
\begin{equation}
U^1_5 = 0,
\end{equation}
\begin{equation}
U^2_5 = 0,
\end{equation}
\begin{equation}
U^3_5 = (h^1_2)^2,
\end{equation}
\begin{equation}
U^4_5 = 2h^1_2h^2_2,
\end{equation}
\begin{equation}
U^5_5 = (h^2_2)^2.
\end{equation}

Two features now become apparent.  As in Table 1, let $D(r)$ be the {\em degree} of the monomial with label $r$.  Then, from the examples worked out, and quite generally from (3.3), we see that there is the relation
\begin{equation}
U^{r^{\prime \prime}}_r = 0 \ {\rm when} \ D(r) > D(r^{\prime \prime}).
\end{equation}
It follows that the sum on the right side of (3.6) always terminates.  Second, for the arguments $h^s_n$ possibly appearing in $U^{r^{\prime \prime}}_r (h^s_n)$, we see that there is the relation
\begin{equation}
D(s) \leq D(r^{\prime \prime}).
\end{equation}
It follows, again see (3.6), that the right side of the differential equation for  any $h^{r^{\prime \prime}}_a$ involves only the $h^s_n$ for which (3.36) holds.    Therefore, to determine the coefficients $h^r_a(t^f)$ of the Taylor expansion (2.9) through terms of some degree $D$, it is only necessary to integrate a finite number of equations, and the right sides of these equations involve only the coefficients for this degree and lower.

For example, to continue our discussion of the case of two variables, the equations (3.6) take the explicit form
\begin{equation}
\dot{h}^1_1(t) = \sum^2_{r=1} g^r_1(t) U^1_r = g^1_1(t) h^1_1(t) + g^2_1(t) h^1_2(t),
\end{equation}
\begin{equation}
\dot{h}^1_2(t) = \sum^2_{r=1} g^r_2(t) U^1_r = g^1_2(t) h^1_1(t) + g^2_2(t) h^1_2(t),
\end{equation}
\begin{equation}
\dot{h}^2_1(t) = \sum^2_{r=1} g^r_1(t) U^2_r = g^1_1(t) h^2_1(t) + g^2_1(t) h^2_2(t),
\end{equation}
\begin{equation}
\dot{h}^2_2(t) = \sum^2_{r=1} g^r_2(t) U^2_r = g^1_2(t) h^2_1(t) + g^2_2(t) h^2_2(t),
\end{equation}
\begin{eqnarray}
\dot{h}^3_1(t) &=& \sum^5_{r=1} g^r_1(t) U^3_r \nonumber \\
&=& g^1_1(t) h^3_1(t) + g^2_1(t) h^3_2(t) + g^3_1(t) [h^1_1(t)]^2 \nonumber\\
& & +g^4_1(t)h^1_1(t) h^1_2(t) + g^5_1(t) [h^1_2(t)]^2,
\end{eqnarray}
\begin{eqnarray}
\dot{h}^3_2(t) &=& \sum^5_{r=1} g^r_2(t) U^3_r \nonumber \\
&=& g^1_2(t) h^3_1(t) + g^2_2(t) h^3_2(t) + g^3_2(t) [h^1_1(t)]^2 \nonumber\\
& & +g^4_2(t)h^1_1(t) h^1_2(t) + g^5_2(t) [h^1_2(t)]^2,
\end{eqnarray}
\begin{eqnarray}
\dot{h}^4_1(t) &=& \sum^5_{r=1} g^r_1(t) U^4_r \nonumber \\
&=& g^1_1(t) h^4_1(t) + g^2_1(t) h^4_2(t) + 2g^3_1(t) h^1_1(t) h^2_1(t) \nonumber\\
& & +g^4_1(t)[h^1_1(t) h^2_2(t) + h^2_1(t)h^1_2(t)] + 2g^5_1(t) h^1_2(t)h^2_2(t),
\end{eqnarray}
\begin{eqnarray}
\dot{h}^4_2(t) &=& \sum^5_{r=1} g^r_2(t) U^4_r \nonumber \\
&=& g^1_2(t) h^4_1(t) + g^2_2(t) h^4_2(t) + 2g^3_2(t) h^1_1(t) h^2_1(t) \nonumber\\
& &+ g^4_2(t)[h^1_1(t) h^2_2(t) + h^2_1(t)h^1_2(t)] + 2g^5_2(t) h^1_2(t)h^2_2(t),
\end{eqnarray}
\begin{eqnarray}
\dot{h}^5_1(t) &=& \sum^5_{r=1} g^r_1(t) U^5_r \nonumber \\
&=& g^1_1(t) h^5_1(t) + g^2_1(t) h^5_2(t) + g^3_1(t) [h^2_1(t)]^2 \nonumber\\
& &+ g^4_1(t)h^2_1(t) h^2_2(t) + g^5_1(t) [h^2_2(t)]^2,
\end{eqnarray}
\begin{eqnarray}
\dot{h}^5_2(t) &=& \sum^5_{r=1} g^r_2(t) U^5_r \nonumber \\
&=& g^1_2(t) h^5_1(t) + g^2_2(t) h^5_2(t) + g^3_2(t) [h^2_1(t)]^2 \nonumber\\
& &+ g^4_2(t)h^2_1(t) h^2_2(t) + g^5_2(t) [h^2_2(t)]^2, \ {\rm etc}.
\end{eqnarray}
And, from (3.7), we have the initial conditions
\begin{equation}
h^r_a(t^i) = \delta^r_a.
\end{equation}

We see that if we desire only the degree one terms in the expansion (2.8), then it is only necessary to integrate the equations (3.37) through (3.40) with the initial conditions (3.47).  A moment's reflection shows that so doing amounts to integrating the first-degree variational equations.  We also observe that if we desire only the degree one and degree two terms in the expansion (2.8), then it is only necessary to integrate the equations (3.37) through (3.46) with the initial conditions (3.47), etc.

\section{Solution of Complete Variational Equations Using Backward Integration}
\setcounter{equation}{0}

There is another method of determining the $h^r_a$ that is surprising, ingenious, and in some ways superior to that just described.  It involves integrating backward in time[5].

\subsection{Method of Backward Integration}
Let us rewrite (2.9) in the slightly more explicit form
\begin{equation}
\zeta^f_a = \sum_r h^r_a(t^i,t^f) G_r(\zeta^i)
\end{equation}
to indicate that there are two times involved, $t^i$ and $t^f$.  From this perspective, (3.6) is a set of differential equations for the quantities $(\partial /\partial t)h^r_a(t^i,t)$ that is to be integrated and evaluated at $t=t^f$.  An alternate procedure is to seek a set of differential equations for the quantities $(\partial /\partial \bar{t})h^r_a(\bar{t},t^f)$ that is to be integrated and evaluated at $\bar{t}=t^i$.

As a first step in considering this alternative, rewrite (4.1) in the form
\begin{equation}
\zeta^f_a = \sum_r h^r_a(\bar{t},t^f) G_r(\zeta (\bar{t})).
\end{equation}
Now reason as follows: \ If $\bar{t}$ is varied and at the same time the quantities $\zeta (\bar{t})$ are varied (evolve) so as to remain on the solution to (2.7) having final conditions $\zeta^f$, then the quantities $\zeta^f$ must remain {\em unchanged}.  Consequently, there is the differential equation result
\begin{equation}
0 = d\zeta^f_a/d\bar{t} = \sum_r [(\partial /\partial \bar{t}) h_a(\bar{t},t^f)] G_r(\zeta (\bar{t})) + \sum_r h^r_a (\bar{t},t^f) (d/d\bar{t}) G_r(\zeta (\bar{t})).
\end{equation}

Let us introduce the notation $\dot{h}^r_a(\bar{t},t^f)$ for $(\partial /\partial \bar{t}) h_a^r(\bar{t},t^f)$ so that the first term on the right side of (4.3) can be rewritten in the form
\begin{equation}
\sum_r [(\partial /\partial \bar{t}) h_a^r(\bar{t},t^f)] G_r(\zeta ) = \sum_r \dot{h}^r_a G_r(\zeta ).
\end{equation}
Next, begin working on the second term on the right side of (4.3) by replacing the summation index $r$ by the dummy index $r^{\prime}$,
\begin{equation}
\sum_r h^r_a(\bar{t},t^f) (d/d\bar{t}) G_r(\zeta (\bar{t})) = \sum_{r^{\prime}} h^{r^{\prime}}_a(\bar{t},t^f) (d/d\bar{t}) G_{r^{\prime}}(\zeta (\bar{t})).
\end{equation}
Now carry out the indicated differentiation using the chain rule and the relation (2.7) which describes how the quantities $\zeta$ vary along a solution,
\begin{equation}
(d/d\bar{t}) G_{r^{\prime}} (\zeta (\bar{t})) = \sum_b (\partial G_{r^{\prime}}/\partial \zeta_b) (d\zeta_b/d\bar{t}) = \sum_{br^{\prime \prime}} (\partial G_{r^{\prime}}/\partial \zeta_b) g^{r^{\prime \prime}}_b(\bar{t}) G_{r^{\prime \prime}} (\zeta (\bar{t})).
\end{equation}
Watch closely: \ Since the $G_r$ are simply standard monomials in the $\zeta$, there must be relations of the form
\begin{equation}
[(\partial /\partial \zeta_b) G_{r^{\prime}} (\zeta )] G_{r^{\prime \prime}} (\zeta ) = \sum_r C^r_{br^{\prime}r^{\prime \prime}} G_r(\zeta )
\end{equation}
where the $C^r_{br^{\prime}r^{\prime \prime}}$ are {\em universal constant coefficients} that describe certain combinatorial properties of monomials.  As a result, the second term on the right side of (4.3) can be written in the form
\begin{equation}
\sum_{r^{\prime}} h^{r^{\prime}}_a (\bar{t},t^f) (d/d\bar{t}) G_{r^{\prime}} (\zeta (\bar{t})) = \sum_r G_r(\zeta ) \sum_{br^{\prime}r^{\prime \prime}} C^r_{br^{\prime}r^{\prime \prime}} g^{r^{\prime \prime}}_b(\bar{t}) h^{r^{\prime}}_a (\bar{t},t^f).
\end{equation}
Since the monomials $G_r$ are linearly independent, the relations (4.3) through (4.8) imply the result
\begin{equation}
\dot{h}^r_a (\bar{t},t^f) = -\sum_{br^{\prime}r^{\prime \prime}} C^r_{br^{\prime}r^{\prime \prime}} g^{r^{\prime \prime}}_b(\bar{t}) h^{r^{\prime}}_a (\bar{t},t^f).
\end{equation}

This result is a set of differential equations for the $h^r_a$ that are to be integrated from $\bar{t} = t^f$ {\em back} to $\bar{t} = t^i$.  Also, evaluating (4.2) for $\bar{t} = t^f$ gives the results
\begin{equation}
\zeta^f_a = \sum_r h^r_a (t^f,t^f) G_r(\zeta^f_a),
\end{equation}
from which it follows that (with the usual polynomial labeling) the $h^r_a$ satisfy the final conditions
\begin{equation}
h^r_a(t^f,t^f) = \delta^r_a.
\end{equation}
Therefore the solution to (4.9) is uniquely defined.  Finally, it is evident from the definition (4.7) that the coefficients $C^r_{br^{\prime}r^{\prime \prime}}$ satisfy the relation
\begin{equation}
C^r_{br^{\prime}r^{\prime \prime}} = 0 \ {\rm unless} \ [D(r^{\prime})-1] + D(r^{\prime \prime}) = D(r).
\end{equation}
Therefore, since $D(r^{\prime \prime}) \geq 1$ in (4.9), it follows from (4.12) that the only $h^{r^{\prime}}_a$ that occur on the right side of (4.9) are those that satisfy
\begin{equation}
D(r^{\prime}) \leq D(r).
\end{equation}
Similarly, the only $g^{r^{\prime \prime}}_b$ that occur are those that satisfy
\begin{equation}
D(r^{\prime \prime}) \leq D(r).
\end{equation}
Therefore, as before, to determine the coefficients $h^r_a$ of the Taylor expansion (2.9) through terms of some degree $D$, it is only necessary to integrate a finite number of equations, and the right sides of these equations involve only the coefficients for this degree and lower.

Comparison of the differential equation sets (3.6) and (4.9) shows that the latter has the remarkable property of being {\em linear} in the unknown quantities $h^r_a$.  This feature means that the evaluation of the right side of (4.9) involves only the retrieval of certain universal constants $C^r_{br^{\prime}r^{\prime \prime}}$ and straight-forward multiplication and addition.  By contrast, the use of (3.6) requires evaluation of the fairly complicated {\em nonlinear} functions $U^{r^{\prime \prime}}_r(h^s_n)$.  Finally, it is easier to insure that a numerical integration procedure is working properly for a set of linear differential equations than it is for a nonlinear set.

The only complication in the use of (4.9) is that the equations must be integrated backwards in $\bar{t}$.  Correspondingly the equations (2.6) for the design solution must also be integrated backwards since they supply the required quantities $g^r_a$ through use of (2.4) and (2.5).  This is no problem if the final quantities $z^d(t^{\rm fin})$ are known.  However if only the initial quantities $z^d(t^{\rm in})$ are known, then the equations (2.6) for $z^d$ must first be integrated forward in time to find the final quantities $z^d(t^{\rm fin})$.  

\subsection{The Two-variable Case Revisited}
For clarity, let us also apply this second method to the two-variable case.  Table 2 shows the nonzero values of $C^r_{br^{\prime}r^{\prime \prime}}$ for $r \in [1,5]$ obtained using (3.10) through (3.14) and (4.7).  Note that the rules (4.12) hold.  Use of this Table shows that in the two-variable case the equations (4.9) take the form
\begin{equation}
\dot{h}^1_1(\bar{t},t^f) = - \ g^1_1(\bar{t}) h^1_1(\bar{t},t^f) - g^1_2(\bar{t}) h^2_1(\bar{t},t^f),
\end{equation}
\begin{equation}
\dot{h}^1_2(\bar{t},t^f) = - \ g^1_1(\bar{t}) h^1_2(\bar{t},t^f) - g^1_2(\bar{t}) h^2_2(\bar{t},t^f),
\end{equation}
\begin{equation}
\dot{h}^2_1(\bar{t},t^f) = - \ g^2_1(\bar{t}) h^1_1(\bar{t},t^f) - g^2_2(\bar{t}) h^2_1(\bar{t},t^f),
\end{equation}
\begin{equation}
\dot{h}^2_2(\bar{t},t^f) = - \ g^2_1(\bar{t}) h^1_2(\bar{t},t^f) - g^2_2(\bar{t}) h^2_2(\bar{t},t^f),
\end{equation}
\begin{equation}
\dot{h}^3_1(\bar{t},t^f) = - \ 2g^1_1(\bar{t}) h^3_1(\bar{t},t^f) - g^3_1(\bar{t}) h^1_1(\bar{t},t^f) -g^1_2(\bar{t}) h^4_1(\bar{t},t^f) - g^3_2(\bar{t}) h^2_1(\bar{t},t^f),
\end{equation}
\begin{equation}
\dot{h}^3_2(\bar{t},t^f) = - \ 2g^1_1(\bar{t}) h^3_2(\bar{t},t^f) - g^3_1(\bar{t}) h^1_2(\bar{t},t^f) -g^1_2(\bar{t}) h^4_2(\bar{t},t^f) - g^3_2(\bar{t}) h^2_2(\bar{t},t^f),
\end{equation}
\begin{eqnarray}
\dot{h}^4_1(\bar{t},t^f) &=& - \ g^1_1(\bar{t}) h^4_1(\bar{t},t^f) - 2g^2_1(\bar{t}) h^3_1(\bar{t},t^f) -g^4_1(\bar{t}) h^1_1(\bar{t},t^f) \nonumber \\
& & -2g^1_2(\bar{t}) h^5_1(\bar{t},t^f) - g^2_2(\bar{t}) h^4_1(\bar{t},t^f) - g^4_2(\bar{t}) h^2_1(\bar{t},t^f),
\end{eqnarray}
\begin{eqnarray}
\dot{h}^4_2(\bar{t},t^f) &=& - \ g^1_1(\bar{t}) h^4_2(\bar{t},t^f) - 2g^2_1(\bar{t}) h^3_2(\bar{t},t^f) -g^4_1(\bar{t}) h^1_2(\bar{t},t^f) \nonumber \\
& &- 2g^1_2(\bar{t}) h^5_2(\bar{t},t^f) - g^2_2(\bar{t}) h^4_2(\bar{t},t^f) - g^4_2(\bar{t}) h^2_2(\bar{t},t^f),
\end{eqnarray}
\begin{equation}
\dot{h}^5_1(\bar{t},t^f) = - \ g^2_1(\bar{t}) h^4_1(\bar{t},t^f) - g^5_1(\bar{t}) h^1_1(\bar{t},t^f) - 2g^2_2(\bar{t}) h^5_1(\bar{t},t^f) - g^5_2(\bar{t}) h^2_1(\bar{t},t^f),
\end{equation}
\begin{equation}
\dot{h}^5_2(\bar{t},t^f) = - \ g^2_1(\bar{t}) h^4_2(\bar{t},t^f) - g^5_1(\bar{t}) h^1_2(\bar{t},t^f) - 2g^2_2(\bar{t}) h^5_2(\bar{t},t^f) - g^5_2(\bar{t}) h^2_2(\bar{t},t^f), \ {\rm etc.}
\end{equation}
As advertised, the right sides of (4.15) through (4.24) are indeed simpler than those of (3.37) through (3.46).

\begin{table}[htp]
\caption{Nonzero values of $C^r_{br^{\prime}r^{\prime \prime}}$ for $r\in [1,5]$ in the two-variable case.}
\begin{center}
\begin{tabular}{ccccc}
$r$ & $b$ & $r^{\prime}$ & $r^{\prime \prime}$ & $C$ \\ \hline
1 & 1 & 1 & 1 & 1 \\
1 & 2 & 2 & 1 & 1 \\
2 & 1 & 1 & 2 & 1 \\
2 & 2 & 2 & 2 & 1 \\
3 & 1 & 1 & 3 & 1 \\
3 & 1 & 3 & 1 & 2 \\
3 & 2 & 2 & 3 & 1 \\
3 & 2 & 4 & 1 & 1 \\
4 & 1 & 1 & 4 & 1 \\
4 & 1 & 3 & 2 & 2 \\
4 & 1 & 4 & 1 & 1 \\
4 & 2 & 2 & 4 & 1 \\
4 & 2 & 4 & 2 & 1 \\
4 & 2 & 5 & 1 & 2 \\
5 & 1 & 1 & 5 & 1 \\
5 & 1 & 4 & 2 & 1 \\
5 & 2 & 2 & 5 & 1 \\
5 & 2 & 5 & 2 & 1 \\
\end{tabular}
\end{center}
\end{table}

\section{Duffing Equation Example}
\setcounter{equation}{0}
\subsection{Introduction}
As an example application, this section studies some aspects of the {\em Duffing} equation. The behavior of the driven Duffing oscillator, like that of generic nonlinear systems, is enormously complicated.  Consequently, we will be able to only touch on some of the highlights of this fascinating problem.

Duffing's equation describes the behavior of a periodically driven damped {\em nonlinear} oscillator governed by the equation of motion
\begin{equation}
  \ddot{x} + a\dot{x} + bx + cx^3 = d \cos (\Omega t + \psi).
\end{equation} 
Here $\psi$ is an arbitrary phase factor that is often set to zero.  For our purposes it is more convenient to set 
\begin{equation}
\psi = \pi/2.
\end{equation}
Evidently any particular choice of $\psi$ simply results in a shift of the origin in time, and this shift has no physical consequence since the left side of (5.1) is independent of time.

 We assume $b,c>0$, which is the case of a positive hard spring restoring force.\footnote{Other authors consider other cases, particularly the `double well' case $b<0$ and $c>0$.}  We make these assumptions because we want the Duffing oscillator to behave like an ordinary harmonic oscillator when the amplitude is small, and we want the motion to be bounded away from infinity when the amplitude is large.  Then, by a suitable choice of time and length scales that introduces new variables $q$ and ${\tau}$, the equation of motion can be brought to the form
\begin{equation}
  \ddot{q} + 2\beta \dot{q} + q + q^3 = -\epsilon \sin \omega \tau ,
\end{equation}
where now a dot denotes $d/d\tau$ and we have made use of (5.2). In this form it is evident that there are 3 free parameters: $\beta$, $\epsilon$, and $\omega$.

\subsection{Stroboscopic Map}
While the Duffing equation is nonlinear, it does have the simplifying feature that the driving force is periodic with period
\begin{equation}
T = 2\pi /\omega .
\end{equation}
Let us convert (5.3) into a pair of first-order equations by making the definition
\begin{equation}
p = \dot{q},
\end{equation}
with the result
\[
\dot{q} = p,
\]
\begin{equation}
\dot{p} = -2\beta p - q - q^3 - \epsilon \sin \omega \tau .
\end{equation}
Let $q^0,p^0$ denote initial conditions at $\tau = 0$, and let $q^1,p^1$ be the final conditions resulting from integrating the pair (5.6) one full period to the time $\tau = T$.  Let $\cal{M}$ denote the transfer map that relates $q^1,p^1$ to $q^0,p^0$.  Then, using the notation $z=(q,p)$, we may write 
\begin{equation}
z^1 = {\cal{M}} z^0.
\end{equation}

Suppose we now integrate for a second full period to find $q^2,p^2$.  Since the right side of (5.6) is periodic, the rules for integrating from $\tau = T$ to $\tau = 2T$ are the same as the rules for integrating from $\tau = 0$ to $\tau = T$.  Therefore we may write
\begin{equation}
  z^2 = {\cal{M}}z^1 = {\cal{M}}^2z^0,
\end{equation}
and in general 
\begin{equation}
  z^{n+1} = {\cal{M}}z^n = {{\cal{M}}}^{n+1}z^0.
\end{equation}
We may regard the quantities $z^n$ as the result of viewing the motion in the light provided by a stroboscope that flashes at the times\footnote{Note that, with the choice (5.2) for $\psi$, the driving term described by the right side of (5.3) vanishes at the stroboscopic times $\tau^n$.}
\begin{equation}
  \tau^n = nT.
\end{equation}
Because of the periodicity of the right side of the equations of motion, the rule for sending $z^n$ to $z^{n+1}$ over the intervals between successive flashes is always the same, namely $\cal{M}$.  For these reasons $\cal{M}$ is called a \emph{stroboscopic map}.  Despite the explicit time dependence in the equations of motion, because of periodicity we have been able to describe the long-term motion by the repeated application of a single fixed map.

\subsection{Feigenbaum Diagram Overview}

One way to study a map and analyze its properties, in this case the Duffing stroboscopic map, is to find its fixed points.  When these fixed points are found, one can then display how they appear, move, and vanish as various parameters are varied.  Such a display is often called a {\em Feigenbaum} diagram. 
This subsection will present selected Feigenbaum diagrams, including the infinite period doubling cascade and its associated strange attractor, for the stroboscopic map obtained by high-accuracy  numerical integration of the equations of motion (5.6).  They will be made by observing the behavior of fixed points as the driving frequency $\omega$ is varied.  For simplicity, the damping parameter will be held constant at the value $\beta=0.10$.  Various sample values will be used for the driving strength $\epsilon$.\footnote{Of course, one can also make Feigenbaum diagrams in which some other parameter, say $\epsilon$, is varied while the others, including $\omega$, are held fixed.}

Let us begin with the case of small driving strength.  When the driving strength is small, we know from energy considerations that the steady-state response will be small, and therefore the behavior of the steady-state solution will be much like that of the driven damped {\em linear} harmonic oscillator.  That is, for small amplitude motion, the $q^3$ term in (5.3) will be negligible compared to the other terms.  We also know that, because of damping, there will be only {\em one} steady-state solution, and therefore $\cal M$ has only {\em one} fixed point $z^f$ such that
\begin{equation}
{\cal M}z^f=z^f.
\end{equation}
Finally, again because of damping, we know that this fixed point is {\em stable}.  That is, if $\cal M$ is applied repeatedly to a point near $z^f$, the result is a sequence of points that approach ever closer to $z^f$.  For this reason a stable fixed point is also called an {\em attractor}.

\subsubsection{A Simple Feigenbaum Diagram}

Figure 2 shows the values of $q_f(\omega)$ and for the case $\epsilon=0.150$, and Figure 3 shows 
 $p_f(\omega)$.  In the figures the phase-space axes are labeled as $q_\infty$ and $p_\infty$ to indicate that what are being displayed are steady-state values reached after a large number of applications of $\cal M$.  As anticipated, we observe from Figures 2 and 3 that the response is much like the resonance response of the driven damped linear harmonic oscillator.\footnote{It was the desire for $q_\infty$ to exhibit a resonance-like peak as a function of $\omega$ that dictated the choice (5.2) for $\psi$.}  Note that the coefficient of $q$ in (5.12) is 1, and therefore at small amplitudes, where $q^3$ can be neglected, the Duffing oscillator described by (5.3) has a natural frequency near 1.  Correspondingly, Figure 2 displays a large response when the driving frequency has the value $\omega\simeq1$.  Observe, however, that the response, while similar, is not exactly like that of the driven damped linear harmonic oscillator.  For example, the resonance peak at $\omega\simeq1$ is slightly tipped to the right, and there is also a small peak for $\omega\simeq1/3$.

%Fig_147e
\begin{figure}[htp]
  \centering
  \includegraphics*[height=4.5in,angle=0]{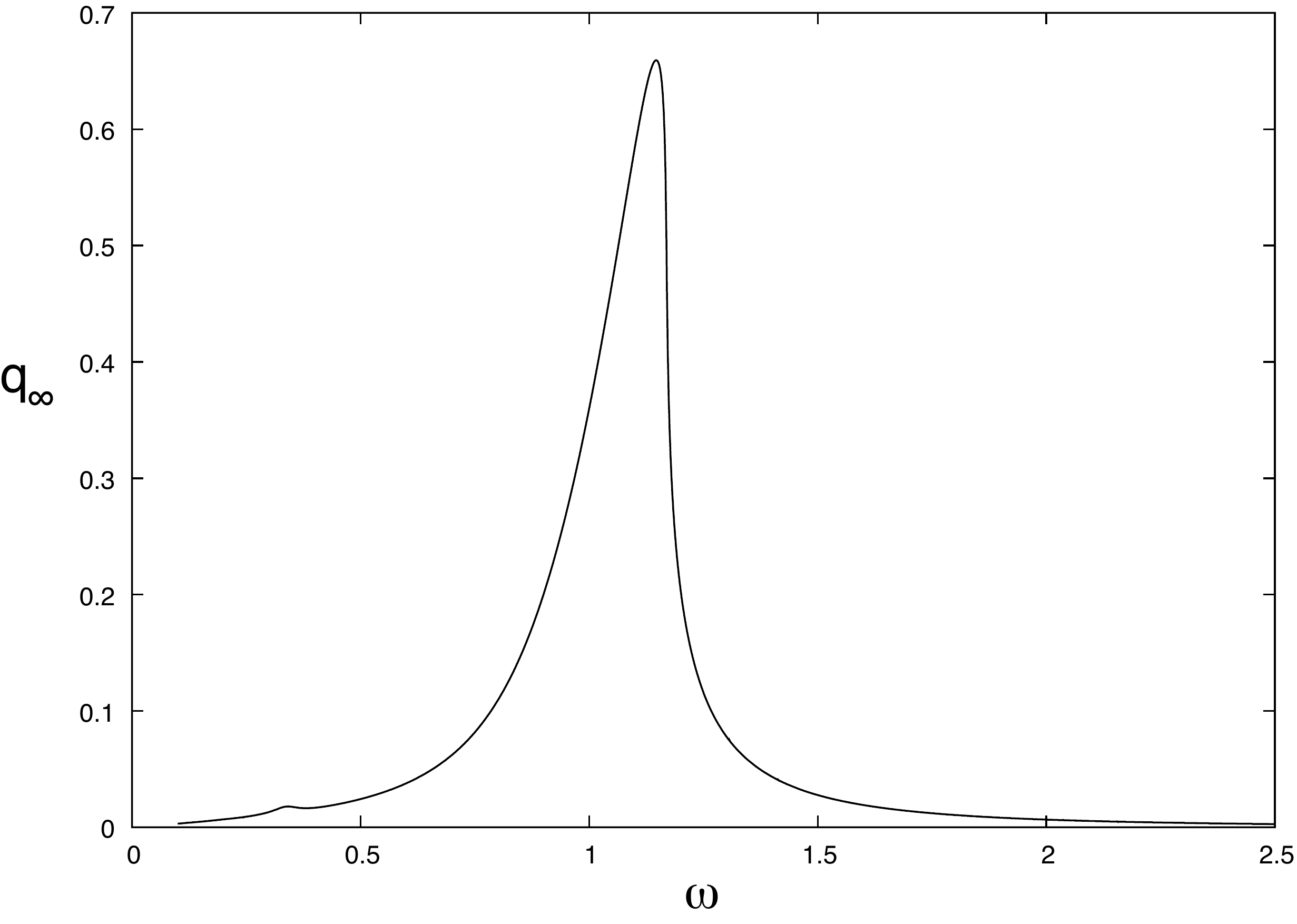}
  \caption{Feigenbaum diagram showing limiting values  $q_{\infty}$ as a function of $\omega$ (when $\beta = 0.1$ and $\epsilon = .15$) for the stroboscopic Duffing map.}
\end{figure}

%Fig_145e
\begin{figure}[htp]
  \centering
  \includegraphics*[width=0.7\textwidth]{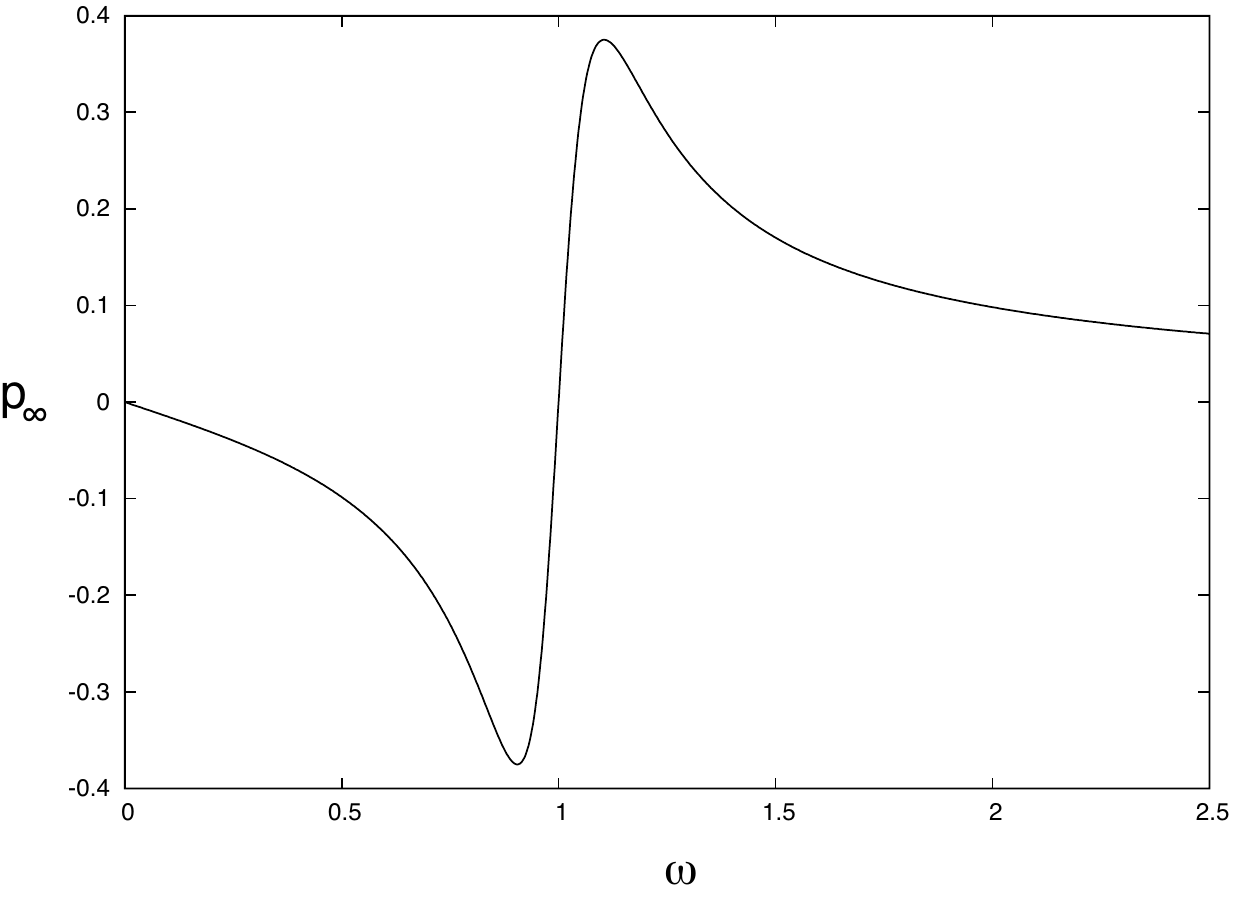}
  \caption{Feigenbaum diagram showing limiting values $p_{\infty}$ as a function
           of $\omega$ (when $\beta = 0.1$ and $\epsilon = .15$) for the
           stroboscopic Duffing map.}
\end{figure}

Our strategy for further exploration will be to increase the value of the driving strength $\epsilon$, all the while observing the stroboscopic Duffing map Feigenbaum diagram as a function of $\omega$.  We hasten to add the disclaimer that the driven Duffing oscillator displays an enormously rich behavior that varies widely with the parameter values $\beta$, $\epsilon$, $\omega$, and we shall be able to give a brief summary of some of it.  Also, for brevity, we shall generally only display $q_f(\omega)$.

\subsubsection{Saddle-Node (Blue-Sky) Bifurcations}

Figure 4 shows the $q$ Feigenbaum diagram for the case of somewhat larger driving strength, $\epsilon=1.50$.  For this driving strength the resonance peak, which previously occurred at $\omega\simeq1$, has shifted to a higher frequency and taken on a more complicated structure.  There are now also noticeable resonances at lower frequencies, with the most prominent one being at $\omega\simeq1/2$. 

Examination of Figure 4 shows that for $\omega\le1.5$ there is a single stable fixed point whose trail is shown in black.  Then, as $\omega$ is increased, a pair of fixed points is born at $\omega\simeq1.8$.\footnote{Actually, in the analytic spirit of Poincar\'{e}, these fixed points also exist for smaller values of $\omega$, but are then complex.  They first become purely real, and therefore physically apparent, when $\omega\simeq1.8$.}  One of them is stable.  The other, whose trail as $\omega$ is varied is shown in red, is {\em unstable}.  That is, if ${\cal M}$ is applied repeatedly to a point near this fixed point, the result is a sequence of points that move ever farther away from the fixed point.  For this reason an unstable fixed point is also called a {\em repellor}.

This appearance of two fixed points out of nowhere is called a saddle-node {\em bifurcation} or a {\em blue-sky} bifurcation, and the associated Fiegenbaum diagram is then sometimes called a bifurcation diagram.\footnote{Strictly speaking, a Feigenbaum diagram displays only the trails of stable fixed points while a bifurcation diagram displays the trails of all fixed points.}  The original stable fixed point persists as $\omega$ is further increased so that over some $\omega$ range there are 3 fixed points.  Then, as $\omega$ is further increased, the original fixed point and the unstable fixed point move until they meet and annihilate when  
$\omega\simeq2.6$.\footnote{Actually, they are not destroyed, but instead become complex and therefore disappear from view.}  This disappearance is called an inverse saddle-node or inverse blue-sky bifurcation.  Finally, for still larger $\omega$ values there is again only one fixed point, and it is stable. 

%Fig_149g
\begin{figure}[htp]
  \centering
  \includegraphics*[width=5in,angle=0]{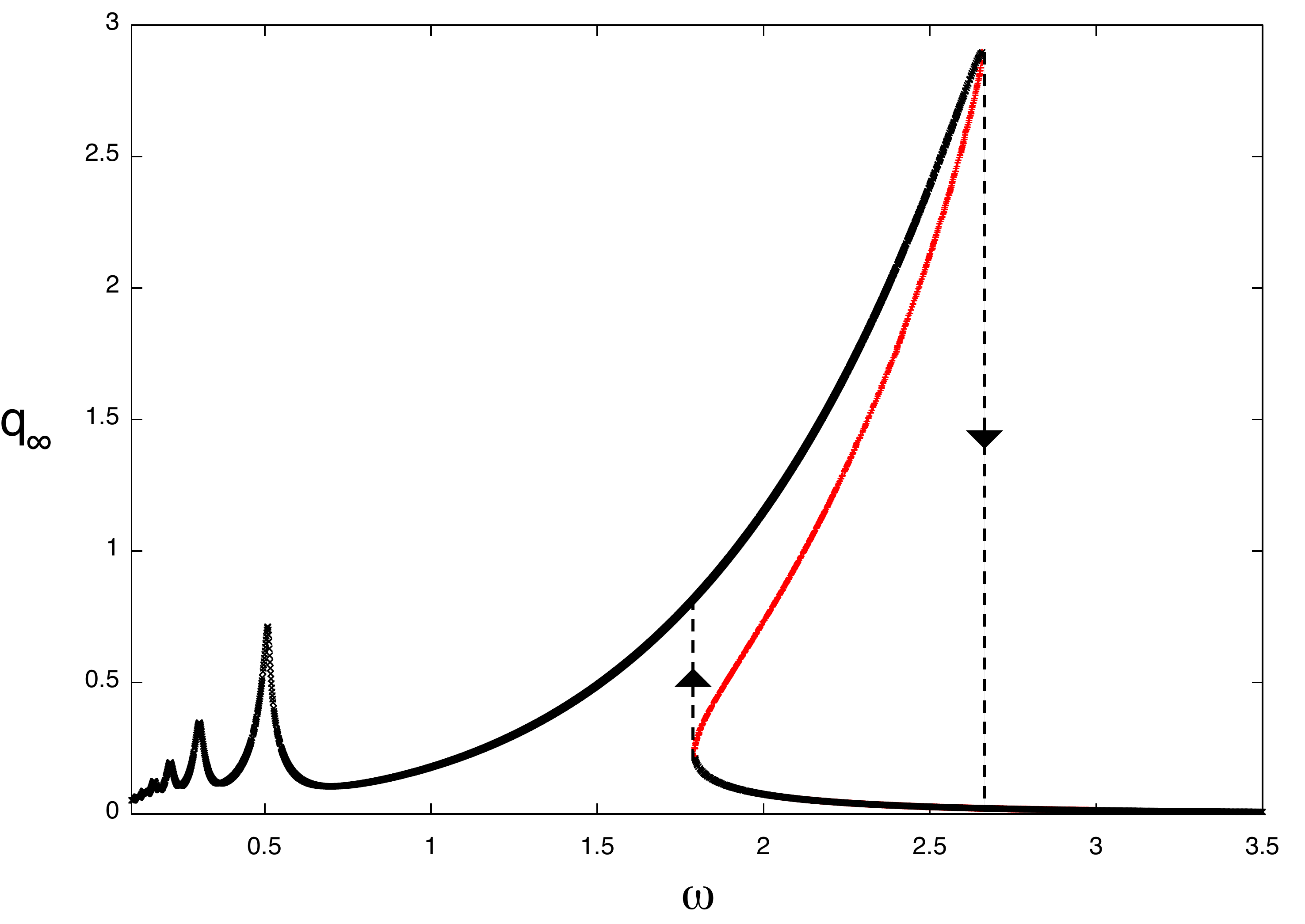}
  \caption{Feigenbaum/bifurcation diagram showing limiting values  $q_{\infty}$ as a function of $\omega$ (when $\beta = 0.1$ and $\epsilon = 1.5$) for the stroboscopic Duffing map.  Trails of the stable fixed points are shown in black.  Also shown, in red, is the trail of the unstable fixed point. Finally, jumps in the steady-state amplitude are illustrated by vertical dashed lines at $\omega \simeq 1.8$ and $\omega \simeq 2.6$.}
\end{figure}

The appearance and disappearance of stable-unstable fixed-point pairs, as $\omega$ is varied, has a striking dynamical consequence. 
Suppose, for example in the case of Figure 4, that the driving frequency $\omega$ is below the value $\omega \simeq 1.8$ where the saddle-node bifurcation occurs.  Then there is only one fixed point, and it is attracting.  Now suppose $\omega$ is {\em slowly} increased.  Then, since the fixed-point solution is attracting, the solution for the slowly increasing $\omega$ case will remain near this solution.  See the upper black trail in Figure 4.  This ``tracking" will continue until $\omega$ reaches the value  $\omega \simeq 2.6$ where the inverse saddle-node bifurcation occurs.  At this value the fixed point being followed disappears.  Consequently, since the one remaining fixed point is also an attractor, the solution evolves very quickly to that of the remaining fixed point.  It happens that the oscillation amplitude associated with this fixed point is much smaller, and therefore there appears to be a sudden jump in oscillation amplitude to a smaller value.  Now suppose $\omega$ is slowly decreased from a value above the value $\omega \simeq 2.6$ where the inverse saddle-node bifurcation occurs.  Then the solution will remain near that of the fixed point lying on the bottom black trail in Figure 4.  This tracking will continue until $\omega$ reaches the value $\omega \simeq 1.8$ where the fixed point being followed disappears.  Again, since the remaining fixed point is attracting, the solution will now evolve to that of the remaining fixed point.  The result is a jump to a larger oscillation amplitude.  Evidently the steady-state oscillation amplitude exhibits {\em hysteresis} as $\omega$ is slowly varied back and forth over an interval that begins below the value where the first saddle-node bifurcation occurs and ends at a value above that where the inverse saddle-node bifurcation occurs.

\subsubsection{Pitchfork Bifurcations}

Let us continue to increase $\epsilon$. Figure 5 shows that a qualitatively new feature appears when $\epsilon$ is near 2.2: a {\em bubble} is formed between the major resonant peak (the one that has saddle-node bifurcated) and the subresonant peak immediately to its left.  To explore the nature of this bubble, let us make $\epsilon$ still larger, which, we anticipate, will result in the bubble becoming larger. Figure 6 shows the Feigenbaum diagram in the case $\epsilon = 5.5$.  Now the major resonant peak and the subresonant peak have moved to larger $\omega$ values. Correspondingly, the bubble between them has also moved to larger $\omega$ values.  Moreover, it is larger, yet another smaller bubble has formed, and the subresonant peak between them has also undergone a saddle-node bifurcation.  For future use, we will call the major resonant peak the {\em first} or {\em leading} saddle-node bifurcation, and we will call the subresonant peak between the two bubbles the {\em second} saddle-node bifurcation, etc.  Also, we will call the bubble just to the left of the first saddle-node bifurcation the {\em first} or {\em leading} bubble, and the next bubble will be called the {\em second} bubble, etc.

We also note that three short trails have appeared in Figure 6 just to the right of $\omega=4$.  They correspond to a period-three bifurcation followed shortly thereafter by an inverse bifurcation.   Actually, much closer examination shows that there are six trails consisting of three closely-spaced pairs.  Each pair comprises a stable and an unstable fixed point of the map ${\cal {M}}^3$.  They are not fixed points of $\cal M$ itself, but rather are sent into each other in cyclic fashion under the action of $\cal M$.

%Fig_1414e
\begin{figure}[htp]
  \centering
  \includegraphics*[width=0.75\textwidth]{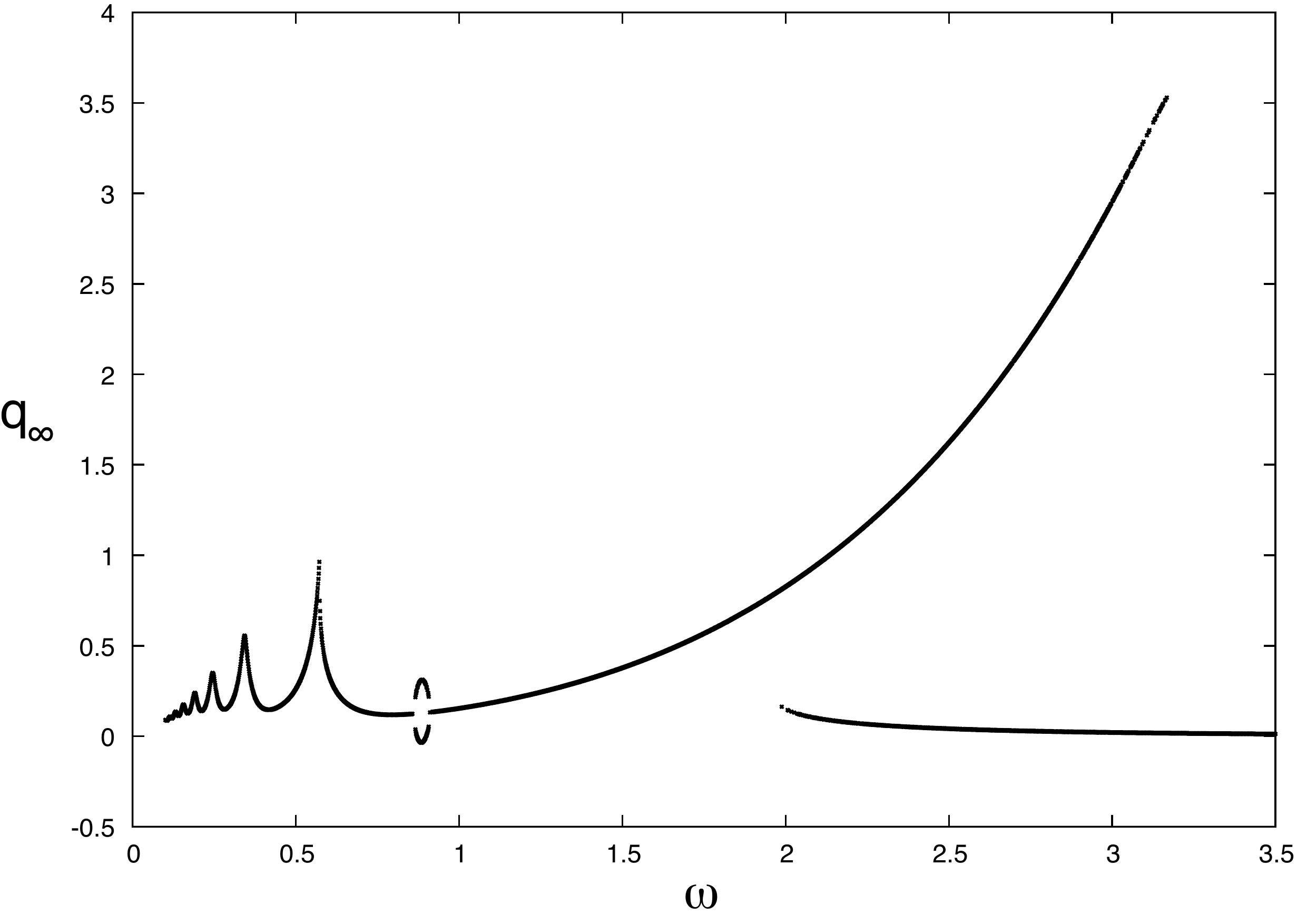}
  \caption{Feigenbaum diagram showing limiting values  $q_{\infty}$ as a function of $\omega$ (when $\beta = 0.1$ and $\epsilon = 2.2$) for the stroboscopic Duffing map. It displays that a bubble has now formed at $\omega\approx.8$.}
\end{figure}

%Fig_1415e
\begin{figure}[htp]
  \centering
  \includegraphics*[width=0.75\textwidth]{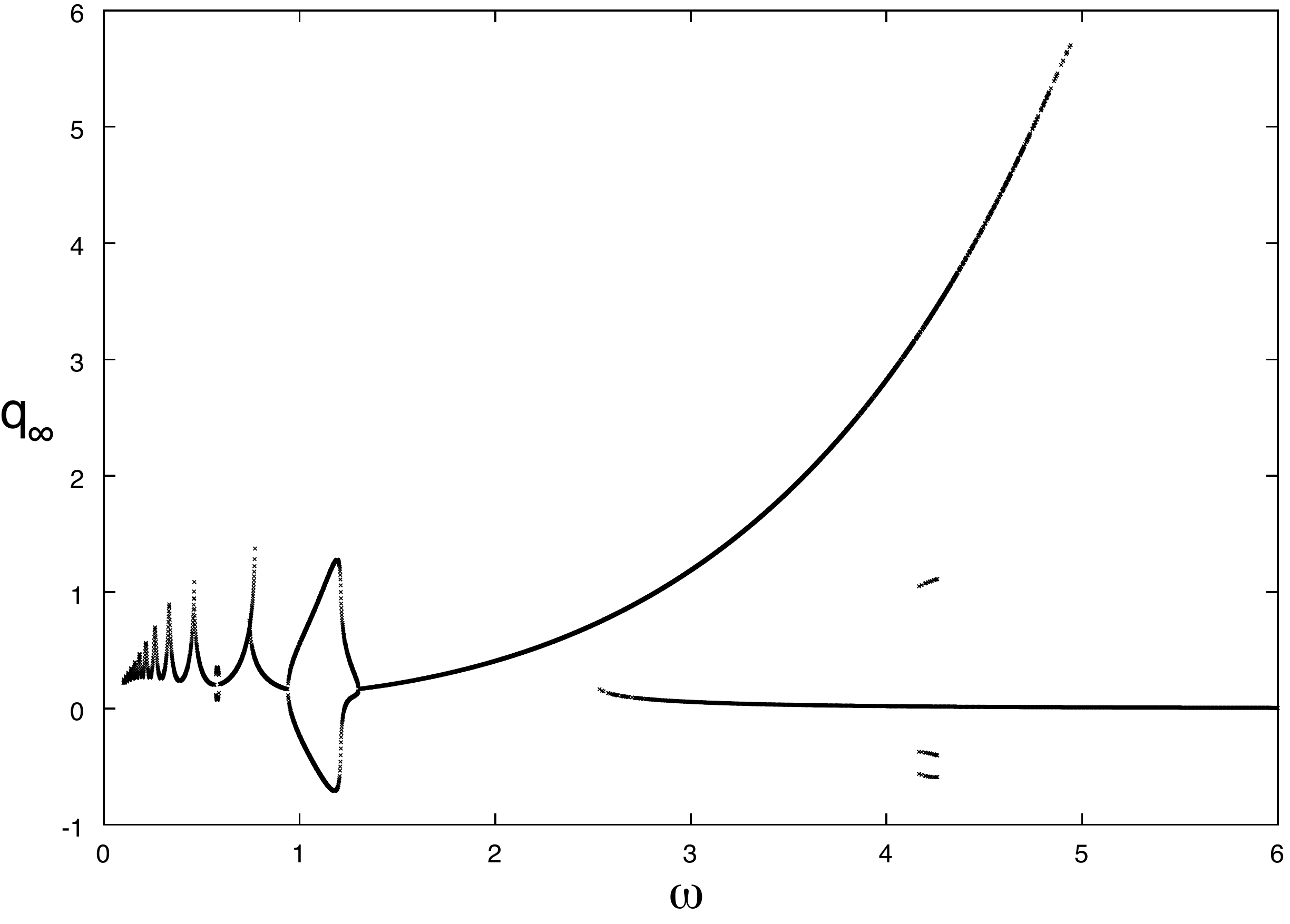}
  \caption{Feigenbaum diagram showing limiting values  $q_{\infty}$ as a function of $\omega$ (when $\beta = 0.1$ and $\epsilon = 5.5$) for the stroboscopic Duffing map. The first bubble has grown, a second smaller bubble has formed to its left, and the sub-resonant peak between them has saddle-node bifurcated to become the second saddle-node bifurcation.}
\end{figure}

Figure 7 shows the larger (leading) bubble in Figure 6 in more detail and with the addition of red lines indicating the trails of unstable fixed points. It reveals that the bubble describes the {\em simultaneous} bifurcation of a single fixed point into three fixed points.  Two of these fixed points are stable and the third, whose $q$ coordinate as a function of $\omega$ are shown as a red line, is unstable.  What happens is that, as $\omega$ is increased, a {\em single} stable fixed point becomes a {\em triplet} of fixed points, two of which are stable and one of which is unstable. This is called a {\em pitchfork} bifurcation. Then, as $\omega$  is further increased, these three fixed points again merge, in an inverse pitchfork bifurcation, to form what is again a single stable fixed point. 

%Fig_1417h
\begin{figure}[htp]
  \centering
   \includegraphics*[width=5in,angle=0]{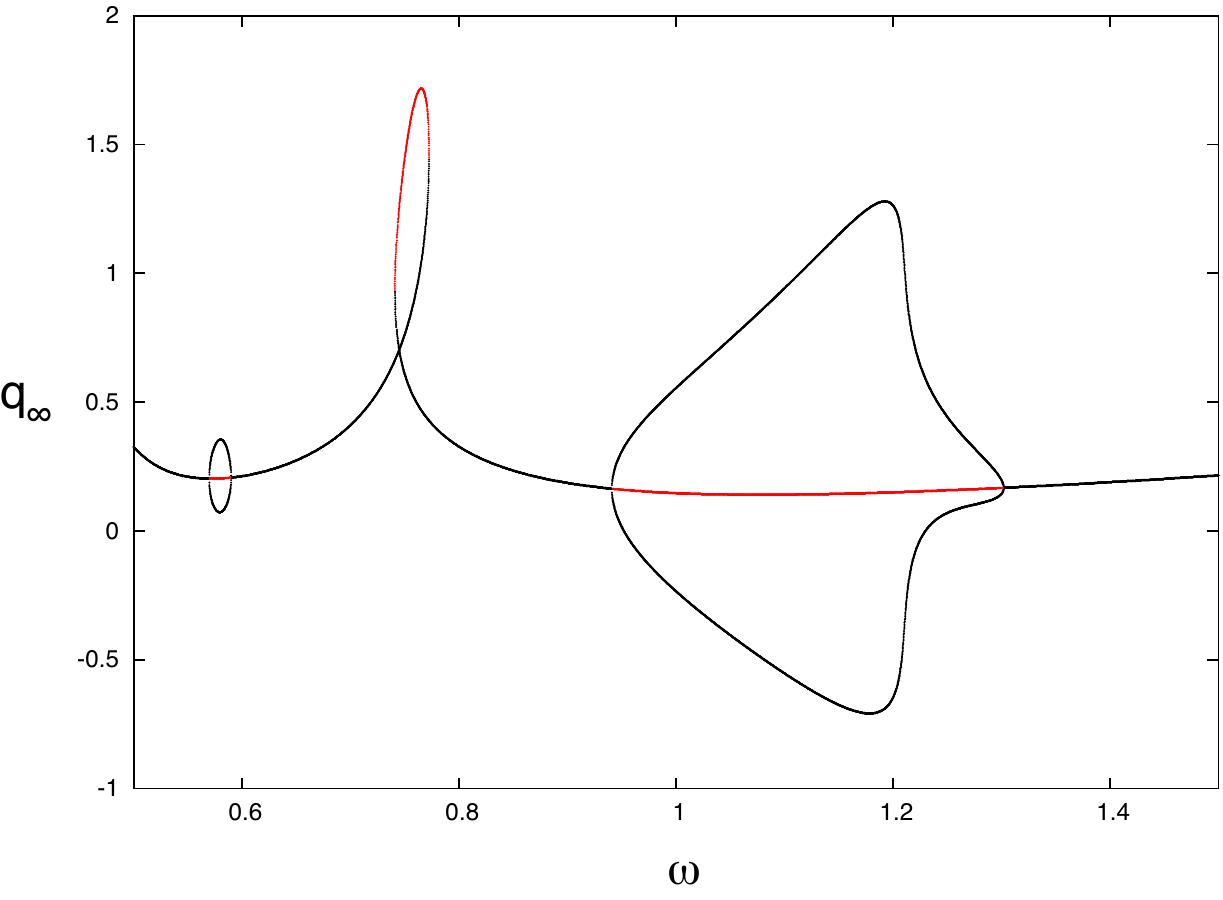}
  \caption{An enlargement of Figure 6 with the addition of red lines indicating the trails of unstable fixed points.}
\end{figure}

\newpage
\subsubsection{A Plethora of Bifurcations and Period Doubling Cascades}

We end our study of the Duffing equation by increasing $\epsilon$ from its earlier value $\epsilon=5.5$ to much larger values.  First we will set $\epsilon=22.125$.  Based on our experience so far, we might anticipate that the Feigenbaum diagram would become much more complicated.  That is indeed the case.  Figure 8 displays  $q_\infty$  when $\beta=0.1$ and $\epsilon=22.125$, as a function of $\omega$,  for the range $\omega \in (0,12)$. Evidently the behavior of the attractors for the stroboscopic Duffing map, which is what is shown in Figure 8, is extremely complicated.  There are now a great many fixed points both of $\cal M$ itself and various powers of $\cal M$.

Of particular interest to us are the two areas around $\omega=.8$ and $\omega=1.25$.  They contain what has become of the first two bubbles in Figure 7, and are shown in greater magnification in Figure 9.  What has happened is that bubbles have formed within bubbles, and bubbles have formed within these bubbles, etc. to form a cascade.  However, these interior bubbles are not the result of pitchfork bifurcations, but rather the result of {\em period-doubling} bifurcations.  For example, the bifurcation that creates the first bubble at $\omega\simeq1.2$ is a pitchfork bifurcation.  But the successive bifurcations within the bubble are period-doubling bifurcations.  In a period-doubling bifurcation a fixed point that is initially stable becomes unstable as $\omega$ is increased.  When this happens, simultaneously two stable fixed points of ${\cal {M}}^2$ are born.  They are not fixed points of $\cal M$ itself, but rather are sent into each other under the action of $\cal M$.  Hence the name ``period doubling".  The map $\cal M$ must be applied twice to send such a fixed point back into itself.  In the next period doubling, fixed points of ${\cal {M}}^4$ are born, etc.  However we note that, as $\omega$ increases,  the sequence of period-doubling bifurcations only occurs a {\em finite} number of times and then undoes itself.

Remarkably, when $\epsilon$ is just somewhat larger, {\em infinite} sequences of period doubling cascades can occur.  Figure 10 shows what happens when $\epsilon=25$.

%Fig_1429e
\begin{figure}[htp]
  \centering
    \includegraphics*[height=5in,angle=90]{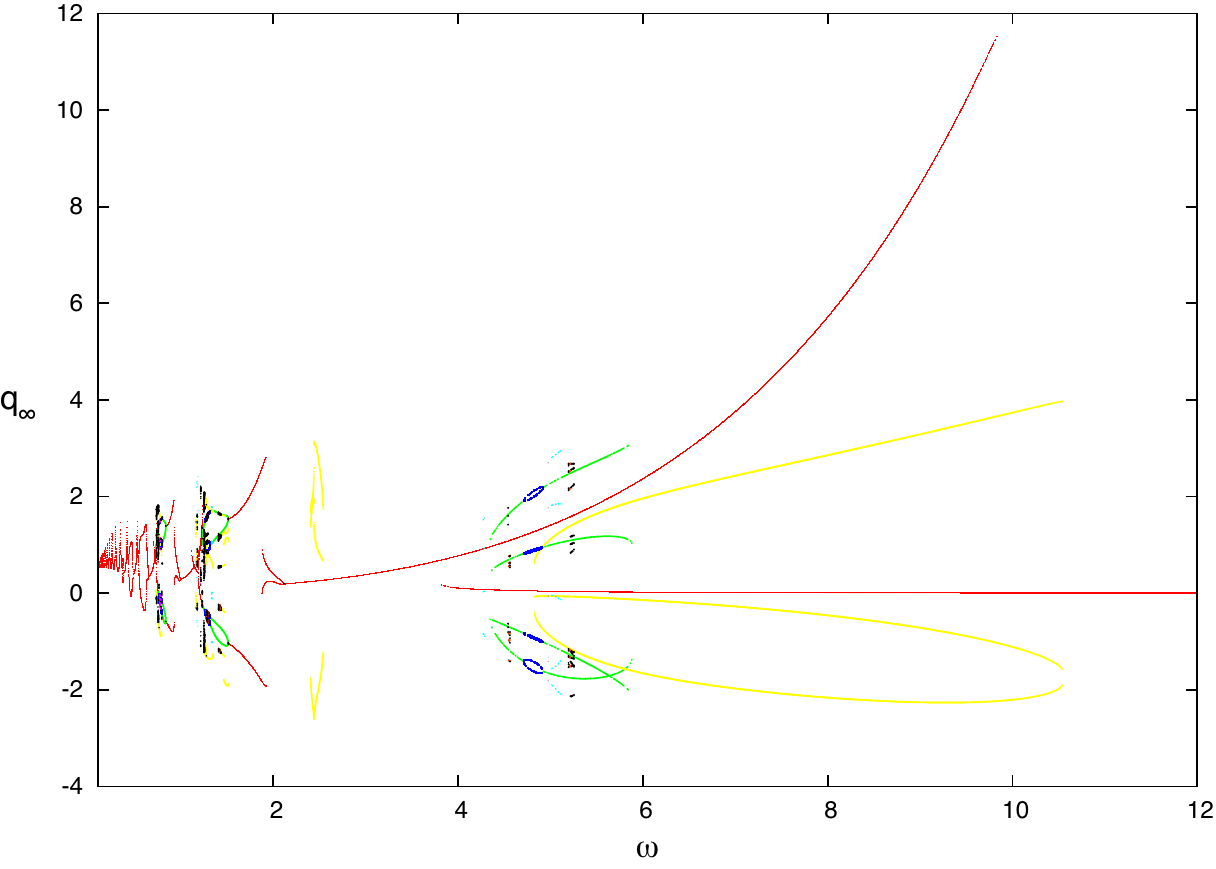}
  \caption{Feigenbaum diagram showing limiting values $q_{\infty}$ as a function of $\omega$ (when $\beta = 0.1$ and $\epsilon = 22.125$) for the stroboscopic Duffing map.}
\end{figure}

%Fig_1430e
\begin{figure}[htp]
  \centering
  \includegraphics*[height=4in,angle=0]{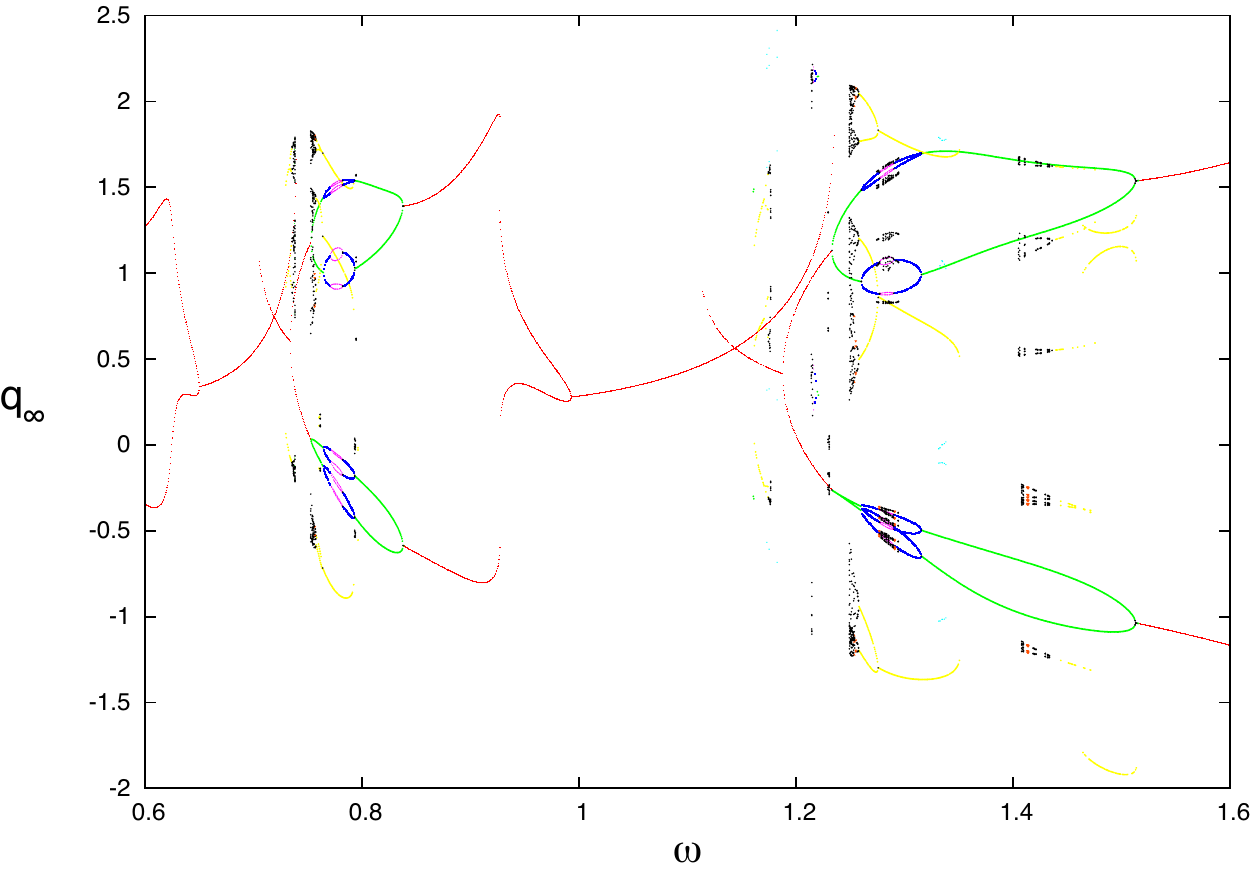}

  \caption{Enlarged portion of the Feigenbaum diagram of Figure 8 displaying  limiting values $q_{\infty}$ as a function of $\omega$ (when $\beta = 0.1$ and $\epsilon = 22.125$) for the stroboscopic Duffing map.  It shows part of the  first bubble at the far right, the second bubble, and part of a third bubble at the far left. Examine the first and second bubbles.  Each initially consists of two stable period-one fixed points.  Each also contains the beginnings of  period-doubling cascades. These cascades do not complete, but rather remerge to again result in a pair of stable period-one fixed points.  There are also many higher-period fixed points and their associated cascades.}
\end{figure}

%Fig_1431e
\begin{figure}[htp]
  \centering
  \includegraphics*[height=5in,angle=90]{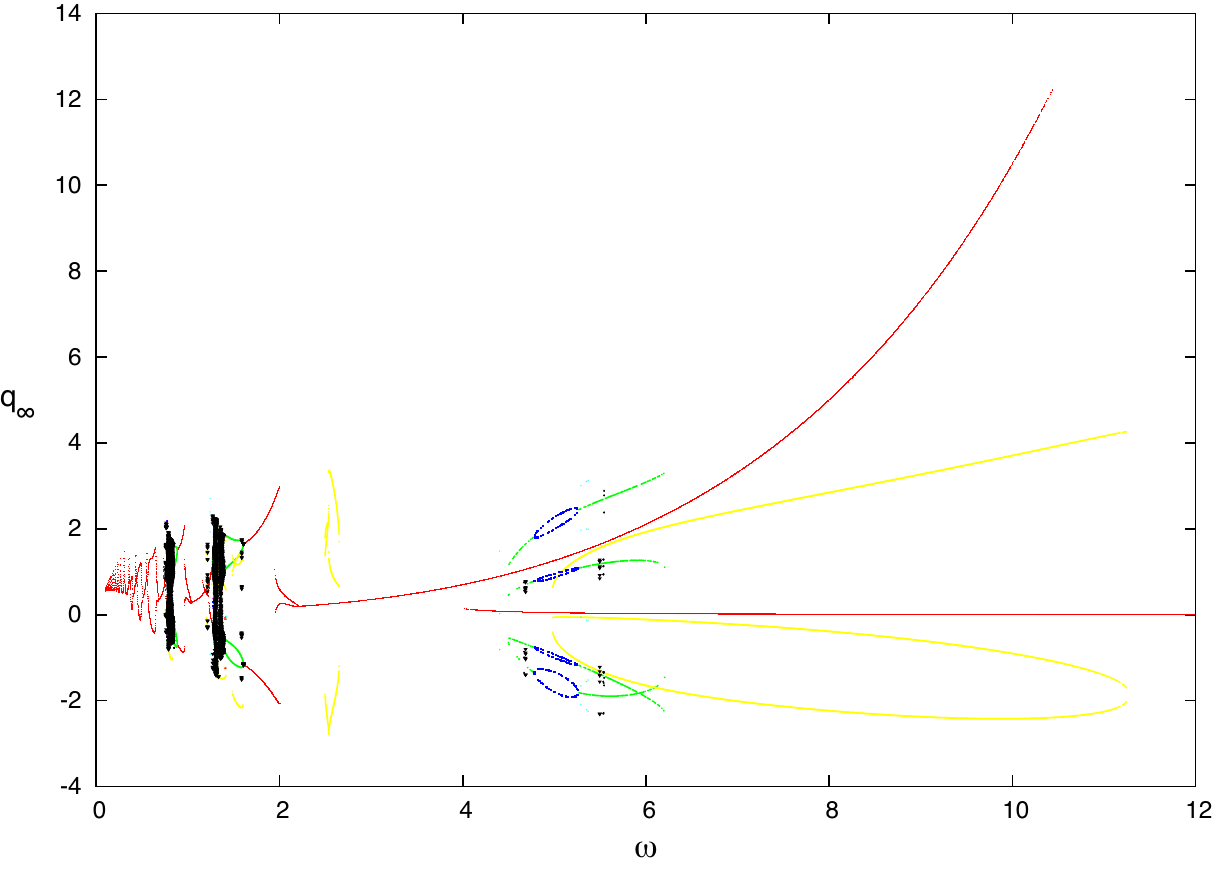}
  \caption{Feigenbaum diagram showing limiting values $q_{\infty}$ as a function of $\omega$ (when $\beta = 0.1$ and $\epsilon = 25$) for the stroboscopic Duffing map.}
\end{figure}

\subsubsection{More Detailed View of Infinite Period Doubling Cascades}

To display the infinite period doubling cascades in more detail, Figure 11 shows an enlargement of part of Figure 10.  And Figures 12 and 13 show successive enlargements of parts of the first bubble in Figure 10.  From Figure 11 we see that the first bubble forms as a result of a pitchfork bifurcation just to the right of $\omega=1.2$, and from Figures 12 and 13 we see that the first period doubling bifurcation occurs in the vicinity of $\omega=1.268$.  From Figure 13 it is evident that successive period-doublings occur an infinite number of times to ultimately produce a chaotic region when $\omega$ exceeds $\omega\simeq1.29$.

%Fig_1432f
\begin{figure}[htp]
  \centering
  \includegraphics*[height=4in,angle=0]{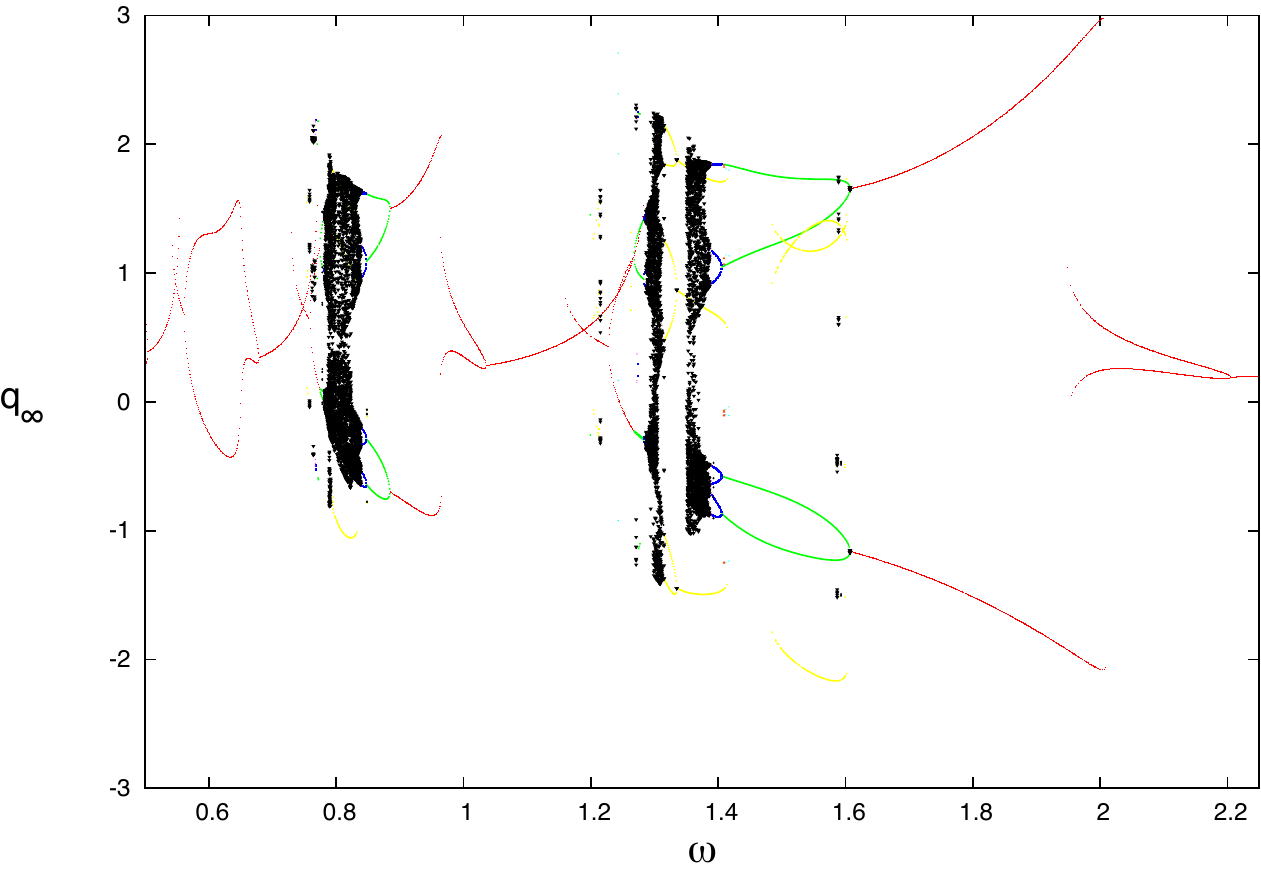}
  \caption{Enlargement of a portion Figure 10 showing the first, second, and third bubbles. The period-doubling cascades in each of the first and second bubbles complete. Then they undo themselves as $\omega$ is further increased. There is no period doubling in the third bubble when $\epsilon=25$.}
\end{figure}

%Fig_1433f
\begin{figure}[htp]
  \centering
  \includegraphics*[height=3.5in]{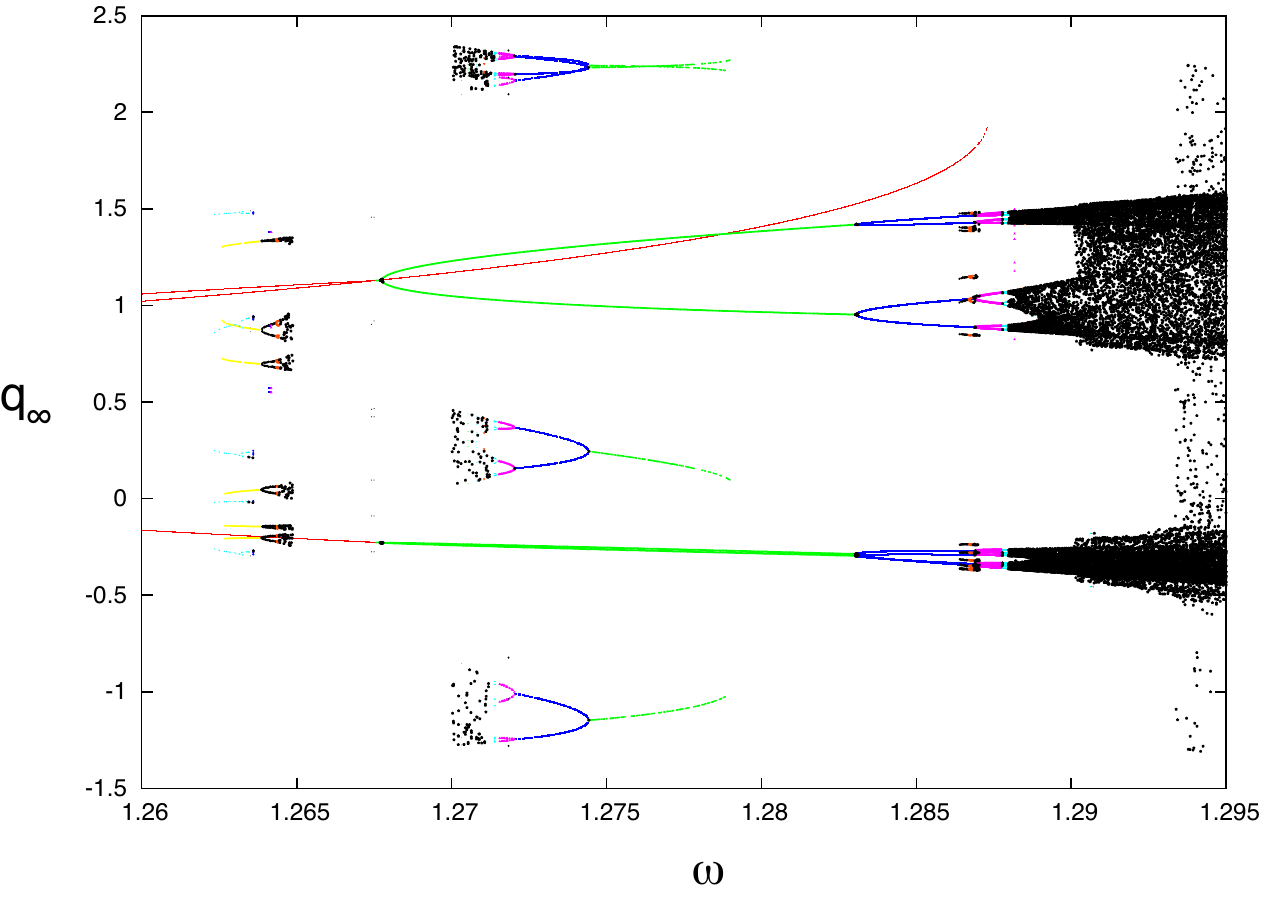}
  \caption{Detail of part of the first bubble in Figure 11  showing upper and lower infinite period-doubling cascades. Part of the trail of the stable fixed point associated with the second saddle-node bifurcation accidentally appears to overlay the upper period doubling bifurcation. Finally, there are numerous cascades and remergings associated with higher-period fixed points.}
\end{figure}

%Fig_1434f
\begin{figure}[htp]
  \centering
  \includegraphics*[height=3.5in]{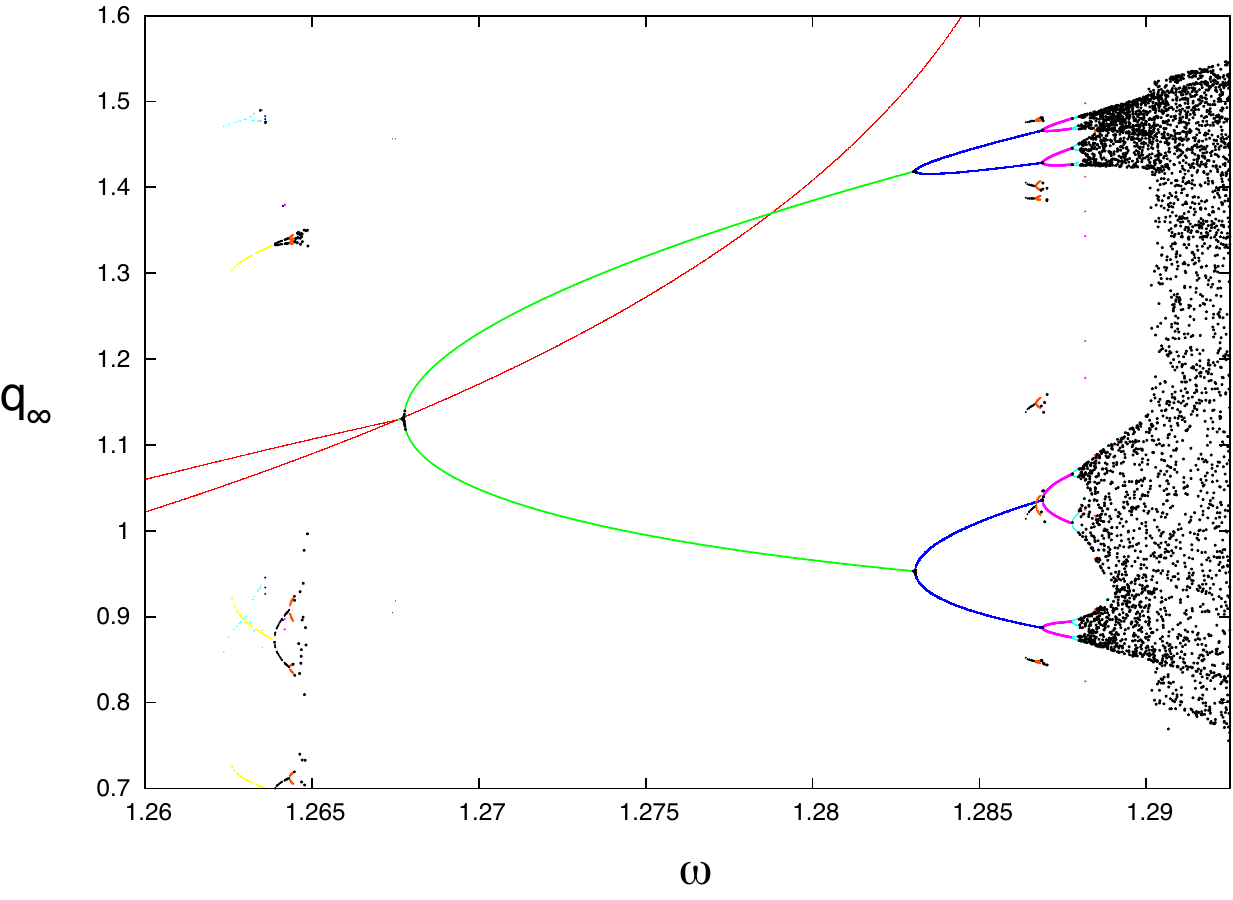}
  \caption{Detail of part of the upper cascade in Figure 12 showing an infinite period-doubling cascade, followed by chaos, for what was initially a stable period-one fixed point.}
\end{figure}

\subsubsection{Strange Attractor}

As evidence that the behavior in this region is chaotic, Figures 14 and 15 show portions of the {\em full} phase space, the $q,p$ plane, when $\omega=1.2902$.  Note the evidence for fractal structure.  The points appear to lie on a {\em strange attractor}.

%Fig_1435g
\begin{figure}[htp]
  \centering
   \includegraphics*[height=3.5in]{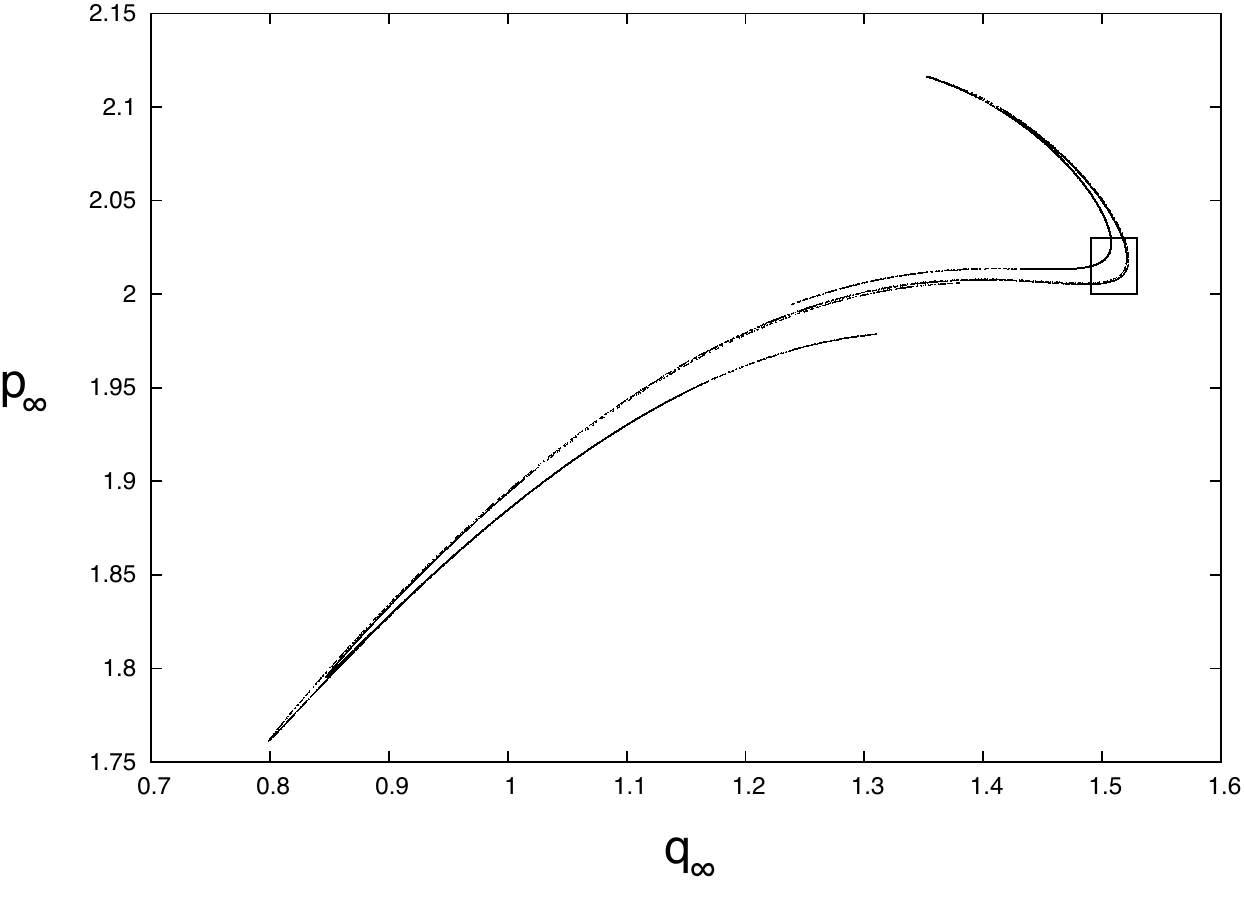}
  \caption{Limiting values of $q_{\infty},p_{\infty}$ for the stroboscopic Duffing map when $\omega=1.2902$ (and $\beta=.1$ and $\epsilon=25$). They appear to lie on a   strange attractor.}
\end{figure}

%Fig_1436f
\begin{figure}[htp]
  \centering
  \includegraphics*[height=3.5in]{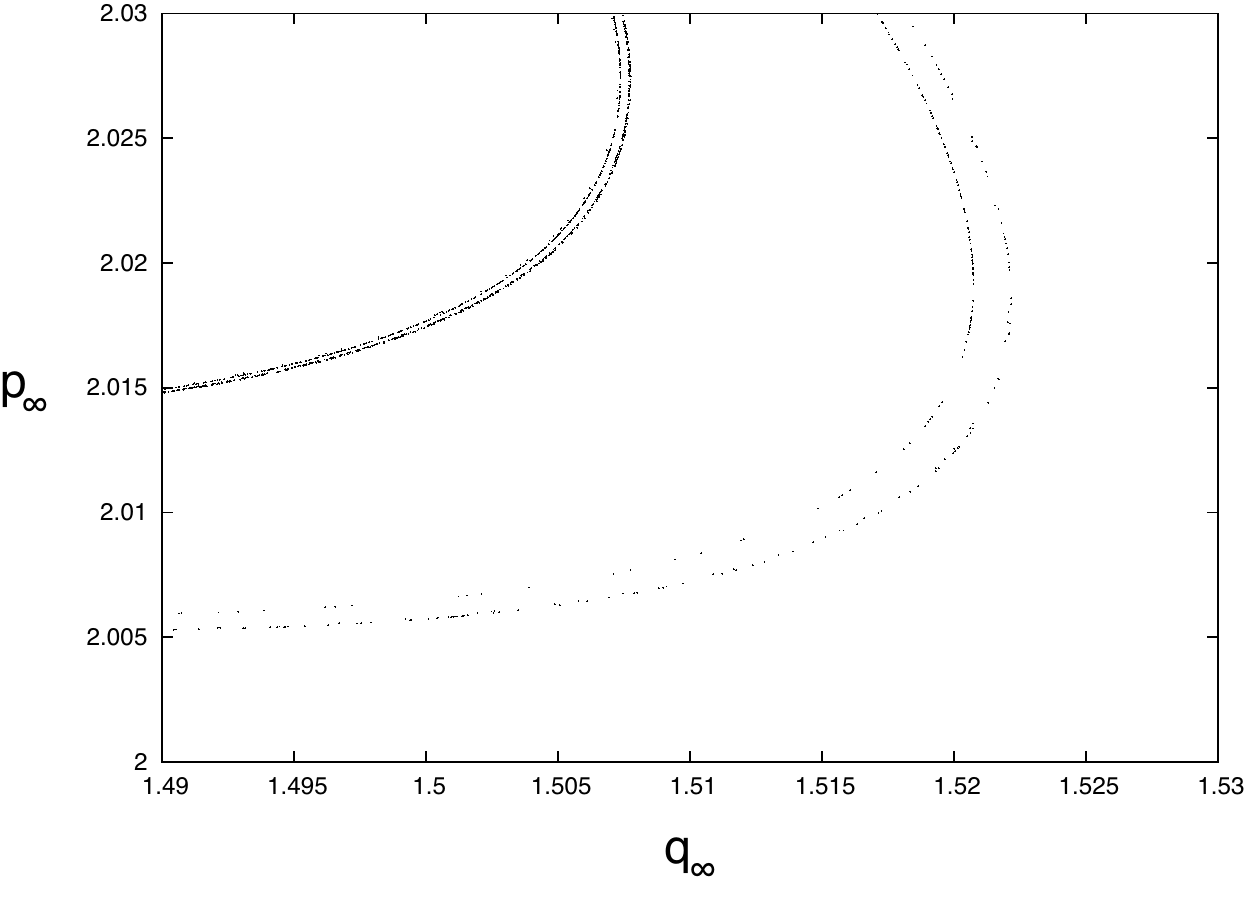}
  \caption{Enlargement of boxed portion of Figure 14 illustrating the
           beginning of self-similar fractal structure.}
\end{figure}

\newpage

\section{Polynomial Approximation to Duffing Stroboscopic Map}
\setcounter{equation}{0}

In this section we will find the complete variational equations for the Duffing equation including dependence on the parameter $\omega$.  The two remaining parameters, $\beta$ and $\epsilon$, will remain fixed.  We will then study the properties of the resulting polynomial approximation to the Duffing stroboscopic map obtained by truncating its Taylor expansion.

\subsection{Complete Variational Equations with Selected Parameter Dependence}

To formulate the complete variational equations we could proceed as in Section 2.2 by setting $\omega=z_3$ and $z_3=\omega^d+\zeta_3$.  So doing would require expanding the function $\sin(\omega\tau)=\sin[(\omega^d+\zeta_3)\tau]$ as a power series in $\zeta_3$.  Such an expansion is, of course, possible, but it leads to variational equations with a large number of forcing terms $g_a^r$ since the expansion of $\sin[(\omega^d+\zeta_3)\tau]$ contains an infinite number of terms.

In the case of Duffing's equation this complication can be avoided by a further change of variables.  Recall that the first change of variables brought the Duffing equation to the form (5.3), which we now slightly rewrite as
\begin{equation}
q^{\prime \prime} + 2\beta q^\prime + q + q^3 = 
-\epsilon \sin \omega \tau 
\end{equation}
where $d/d\tau$ is now denoted by a prime.
Next make the further change of variables
\begin{equation}
q = \omega Q,
\end{equation}
\begin{equation}
\omega = 1/\sigma ,
\end{equation}
\begin{equation}
\omega \tau = t.
\end{equation}
When this is done, there are the relations
\begin{equation}
q^\prime =\omega^2\dot{Q}
\end{equation}
and
\begin{equation}
q^{\prime \prime} = \omega^3 \ddot{Q}
\end{equation}
where now a dot denotes $d/dt$.  [Note that the variable $t$ here is different from that in (5.1).]
Correspondingly, Duffing's equation takes the form
\begin{equation}
\ddot{Q} + 2\beta \sigma \dot{Q} + \sigma^2Q + Q^3 = -\epsilon \sigma^3 \sin t.
\end{equation}
This equation can be converted to a first-order set of the form (2.1) by writing
\begin{equation}
Q = z_1
\end{equation}
and
\begin{equation}
\dot{Q} = z_2.
\end{equation}

Following the method of Section 2.2, we augment the first-order equation set associated with (6.7) by adding the equation
\begin{equation}
\dot{\sigma} = 0.
\end{equation}
Then we may view $\sigma$ as a variable, and (6.10) guarantees that this variable remains a constant.  Taken together, (6.7) and (6.10) may be converted to a first-order triplet of the form (2.12) by writing (6.8), (6.9), and
\begin{equation}
\sigma = z_3.
\end{equation}
Doing so gives the system
\begin{equation}
\dot{z}_1 = z_2,
\end{equation}
\begin{equation}
\dot{z}_2 = -2\beta z_3z_2 -z^2_3z_1-z^3_1-\epsilon z^3_3 \sin t,
\end{equation}
\begin{equation}
\dot{z}_3 = 0,
\end{equation}
and we see that there are the relations
\begin{equation}
f_1(z,t) = z_2,
\end{equation}
\begin{equation}
f_2(z,t) = - \ 2\beta z_3z_2 -z^2_3z_1-z^3_1-\epsilon z^3_3 \sin t,
\end{equation}
\begin{equation}
f_3(z,t) = 0.
\end{equation}
Note that, with the change of variables just made, the stroboscopic map is obtained by integrating the equations of motion from $t=0$ to $t=2\pi$.

As before, we introduce deviation variables using (2.2) and carry out the steps (2.3) through (2.9).  In particular, we write
\begin{equation}
z_3=z_3^d+\zeta_3=\sigma^d+\zeta_3.
\end{equation}
This time we are working with monomials in the three variables $\zeta_1$, $\zeta_2$, and $\zeta_3$.  [That is, $a$ ranges from 1 to 3 in (2.12).]  They are conveniently labeled using the indices $r$ given in Table 3 below.  With regard to the expansions (2.4) and (2.5), we find the results
\begin{equation}
f_1(z^d+\zeta ,t) = z^d_2 + \zeta_2,
\end{equation}
\begin{eqnarray}
f_2(z^d+\zeta ,t) &=& - \ 2\beta (z^d_3 + \zeta_3)(z^d_2+\zeta_2) - (z^d_3+\zeta_3)^2(z^d_1+\zeta_1) \nonumber \\
& &- (z^d_1+\zeta_1)^3 - \epsilon (z^d_3+\zeta_3)^3 \sin t\nonumber \\
%constant terms
&=&[-2\beta z_2^dz_3^d-z_1^d(z_3^d)^2-(z_1^d)^3
-\epsilon (z_3^d)^3\sin t]\nonumber \\
%linear terms
& &-[3(z_1^d)^2+(z_3^d)^2]\zeta_1-2\beta z_3^d\zeta_2
-[2\beta z_2^d+2z_1^dz_3^d+3\epsilon(z_3^d)^2\sin t ]\zeta_3 \nonumber \\
%quadratic terms
& &-2\beta\zeta_2\zeta_3-(z_1^d+3\epsilon z_3^d)\zeta_3^2-z_3^d\zeta_1\zeta_2-3z_1^d\zeta_1^2 \nonumber \\
%cubic terms
& &-\zeta_1^3-\zeta_1\zeta_3^2-\epsilon(\sin t)\zeta_3^3,
\end{eqnarray}
\begin{equation}
f_3(z^d+\zeta ,t) = 0.
\end{equation}
Note the right sides of (6.19) through (6.21) are at most cubic in the deviation variables $\zeta_a$.  Therefore, from Table 3, we see that the index $r$ for the $g^r_a$ should range from 1 through 19.  
It follows that for Duffing's equation (with $\sigma$ parameter expansion) the only nonzero forcing terms are given by the relations
\begin{equation}
g^2_1 = 1,
\end{equation}
\begin{equation}
g^1_2 = -3(z^d_1)^2 - (z^d_3)^2,
\end{equation}
\begin{equation}
g^2_2 = -2\beta z^d_3,
\end{equation}
\begin{equation}
g^3_2 = -2\beta z^d_2 - 2z^d_1z^d_3 - 3\epsilon (z^d_3)^2 \sin t,
\end{equation}
\begin{equation}
g^4_2 = -3z^d_1,
\end{equation}
\begin{equation}
g^6_2 = -2z^d_3,
\end{equation}
\begin{equation}
g^8_2 = -2\beta ,
\end{equation}
\begin{equation}
g^9_2 = -z^d_1 - 3\epsilon z^d_3 \sin t,
\end{equation}
\begin{equation}
g^{10}_2 = -1,
\end{equation}
\begin{equation}
g^{15}_2 = -1,
\end{equation}
\begin{equation}
g^{19}_2 = -\epsilon\sin t.
\end{equation}

\begin{table}[h,t]
\caption{A labeling scheme for monomials in three variables.}
\begin{center}
\begin{tabular}{ccccc}
$r$ & $j_1$ & $j_2$ & $j_3$ & $D$ \\ \hline
1 & 1 & 0 & 0 & 1 \\ 
2 & 0 & 1 & 0 & 1 \\
3 & 0 & 0 & 1 & 1 \\
4 & 2 & 0 & 0 & 2 \\
5 & 1 & 1 & 0 & 2 \\
6 & 1 & 0 & 1 & 2 \\
7 & 0 & 2 & 0 & 2 \\
8 & 0 & 1 & 1 & 2 \\
9 & 0 & 0 & 2 & 2 \\
10 & 3 & 0 & 0 & 3 \\
11 & 2 & 1 & 0 & 3 \\
12 & 2 & 0 & 1 & 3 \\
13 & 1 & 2 & 0 & 3 \\
14 & 1 & 1 & 1 & 3 \\
15 & 1 & 0 & 2 & 2 \\
16 & 0 & 3 & 0 & 3 \\
17 & 0 & 2 & 1 & 3 \\
18 & 0 & 1 & 2 & 3 \\
19 & 0 & 0 & 3 & 3 \\
\end{tabular}
\end{center}
\end{table}

\newpage
\subsection{Performance of Polynomial Approximation}

Let ${\cal{M}}_8$ denote the $8^{\rm{th}}$-order polynomial map (with parameter dependence) approximation to the stroboscopic Duffing map $\cal M$.  Provided the relevant phase-space region is not too large, we have found that ${\cal{M}}_8$  reproduces all the features, described in Section 5.3, of the exact map[6].  (The phase-space region must lie within the convergence domain of the Taylor expansion.)  This reproduction might not be too surprising in the cases of elementary bifurcations such as saddle-node and pitchfork bifurcations.  What is more fascinating, as we will see, is that  ${\cal{M}}_8$ also reproduces the infinite period doubling cascade and associated strange attractor.

\subsubsection{Infinite Period Doubling Cascade}

Figure 16 shows the partial Feigenbaum diagram for the map ${\cal{M}}_8$ in the case that $\beta=0.1$ and $\epsilon=25$.  The Taylor expansion is made about the point indicated by the {\em black dot}.  This point has the coordinates
\begin{equation}
q_{\rm{bd}}=1.26082,{\;}p_{\rm{bd}}=2.05452,{\;}\omega_{\rm{bd}}=1.285.
\end{equation}
It was selected to be an unstable fixed point of $\cal M$, but that is not essential.  Any nearby expansion point would have served as well.
Note the remarkable resemblance of Figures 13 and 16.

We have referred to Figure 16 as a {\em partial} Feigenbaum diagram because it shows only $q_\infty$ and not $p_\infty$.  In order to give a complete picture, Figure 17 displays them both.

%Fig_22p12p24
\begin{figure}[htp]
  \centering
  \includegraphics*[width=4.5in]{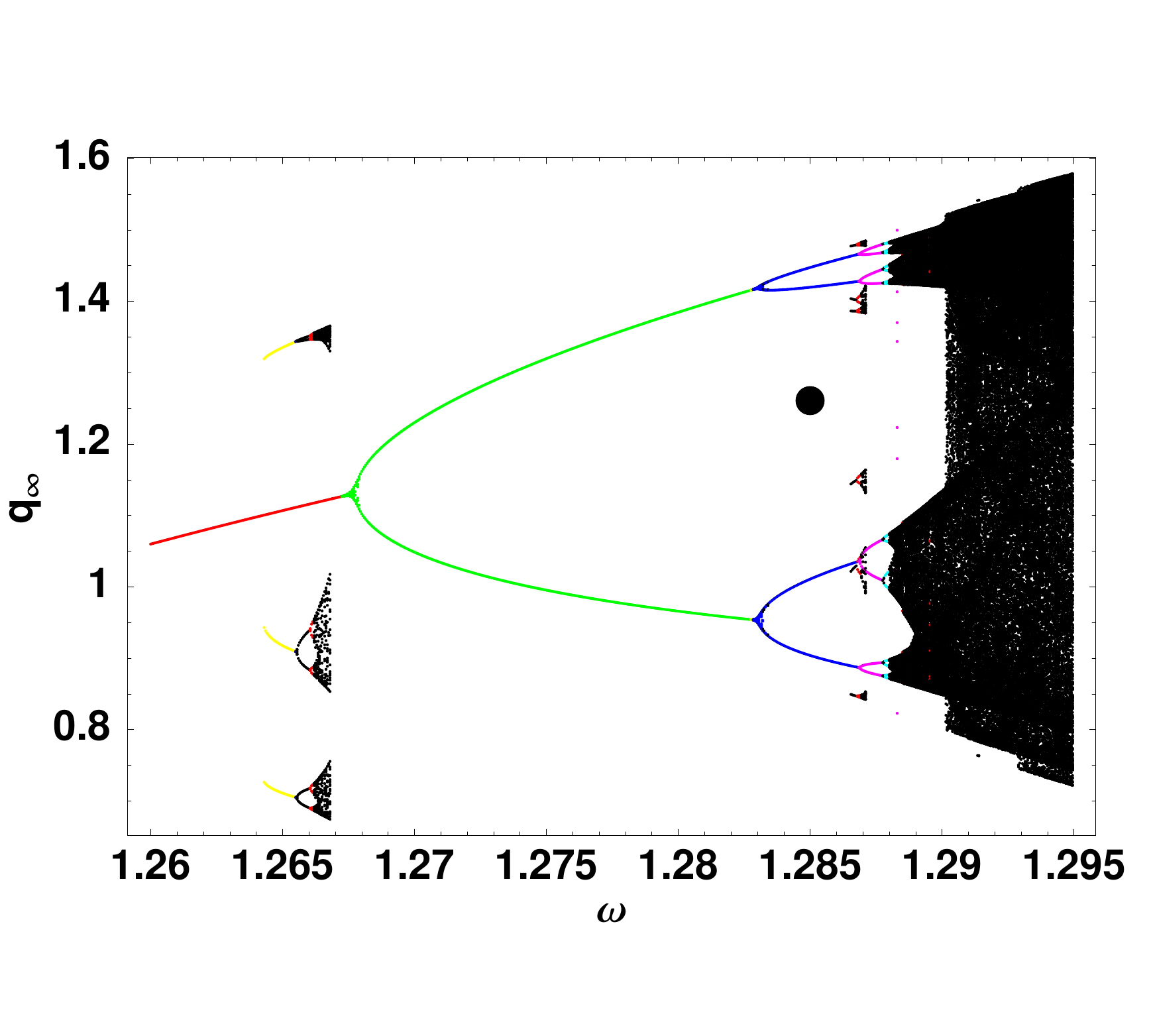}
  \caption{Partial Feigenbaum diagram for the map  ${\cal {M}}_8$. The black dot marks the point about which ${\cal M}$ is expanded to yield ${\cal {M}}_8$}.
\end{figure}

%Fig_22p12p25
\begin{figure}[htp]
  \centering
  \includegraphics*[width=4.5in]{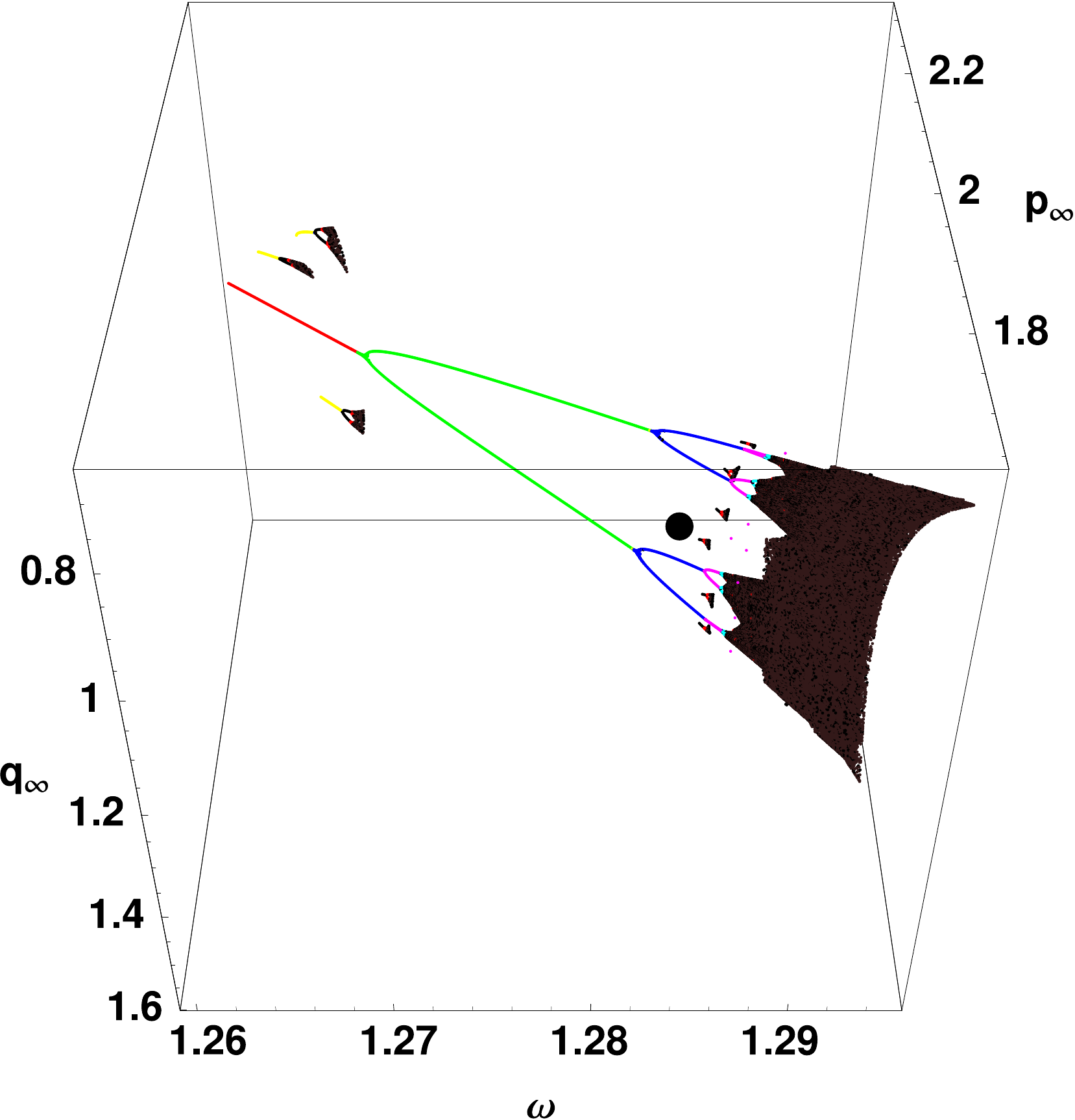}
  \caption{Full Feigenbaum diagram for the map  ${\cal {M}}_8$. The black dot again marks the expansion point.}
\end{figure}

\newpage
\subsubsection{Strange Attractor}

As displayed in Figures 18 through 21 ${\cal {M}}_8$, like $\cal M$,  appears to have a strange attractor.  Note the remarkable agreement between Figures 14 and 15 for $\cal M$ and their ${\cal {M}}_8$ counterparts, Figures 18 and 19.  In the case of ${\cal {M}}_8$ we have been able to obtain additional {\em enlargements}, Figures 20 and 21, further illustrating a self-similar fractal structure.  Analogous  figures are more difficult to obtain for the exact map $\cal M$ due to the excessive numerical integration time required.  By contrast the map ${\cal {M}}_8$, because it is a simple polynomial, is easy to evaluate repeatedly.
 
%Fig_22p12p26
\begin{figure}[htp]
  \centering
  \includegraphics*[height=4.5in]{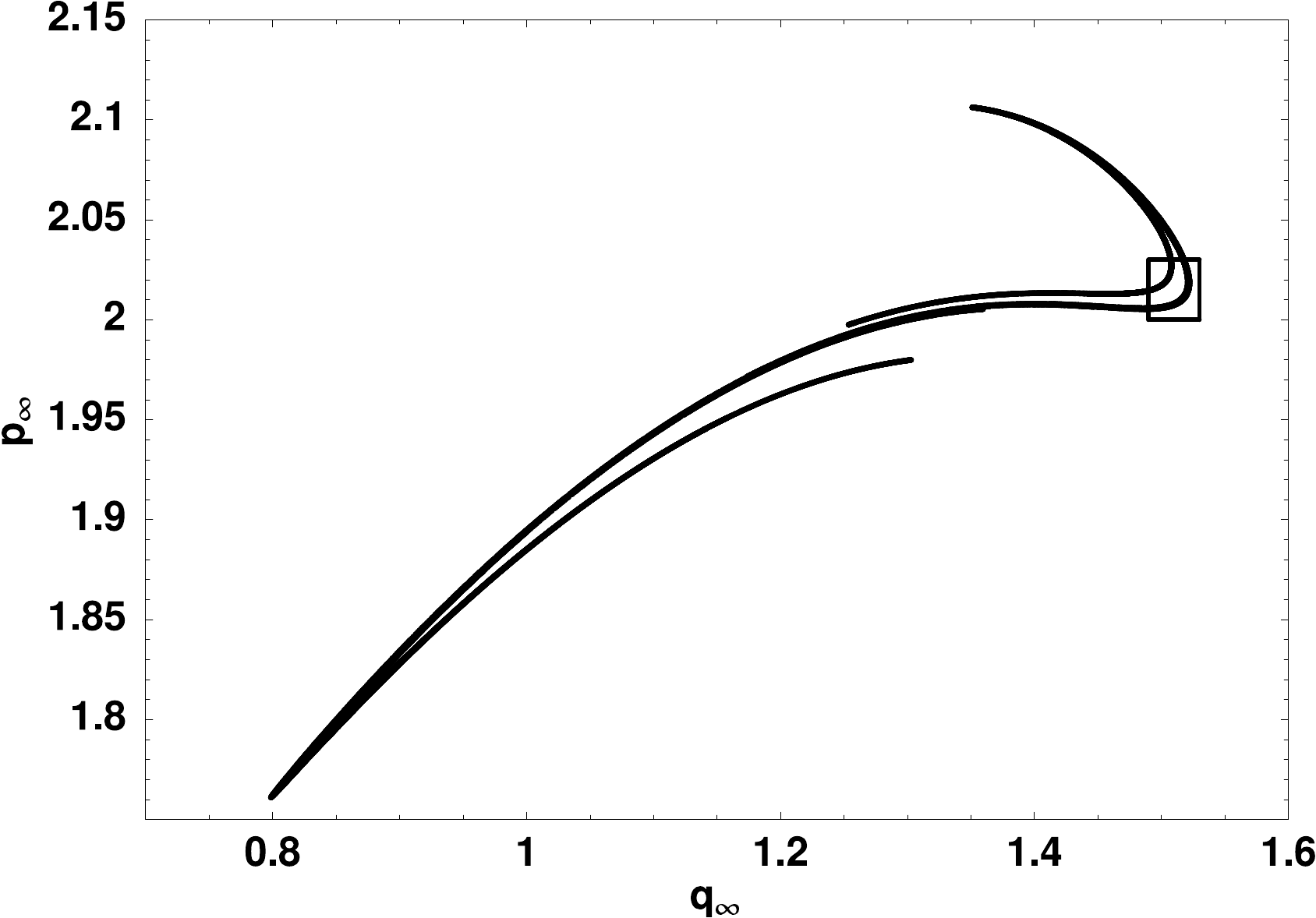}
  \caption{Limiting values of $q_{\infty},p_{\infty}$ for the map ${\cal {M}}_8$ when $\omega=1.2902$. They appear to lie on a   strange attractor.}
\end{figure}

%Fig_22p12p27
\begin{figure}[htp]
  \centering
  \includegraphics*[width=5.5in]{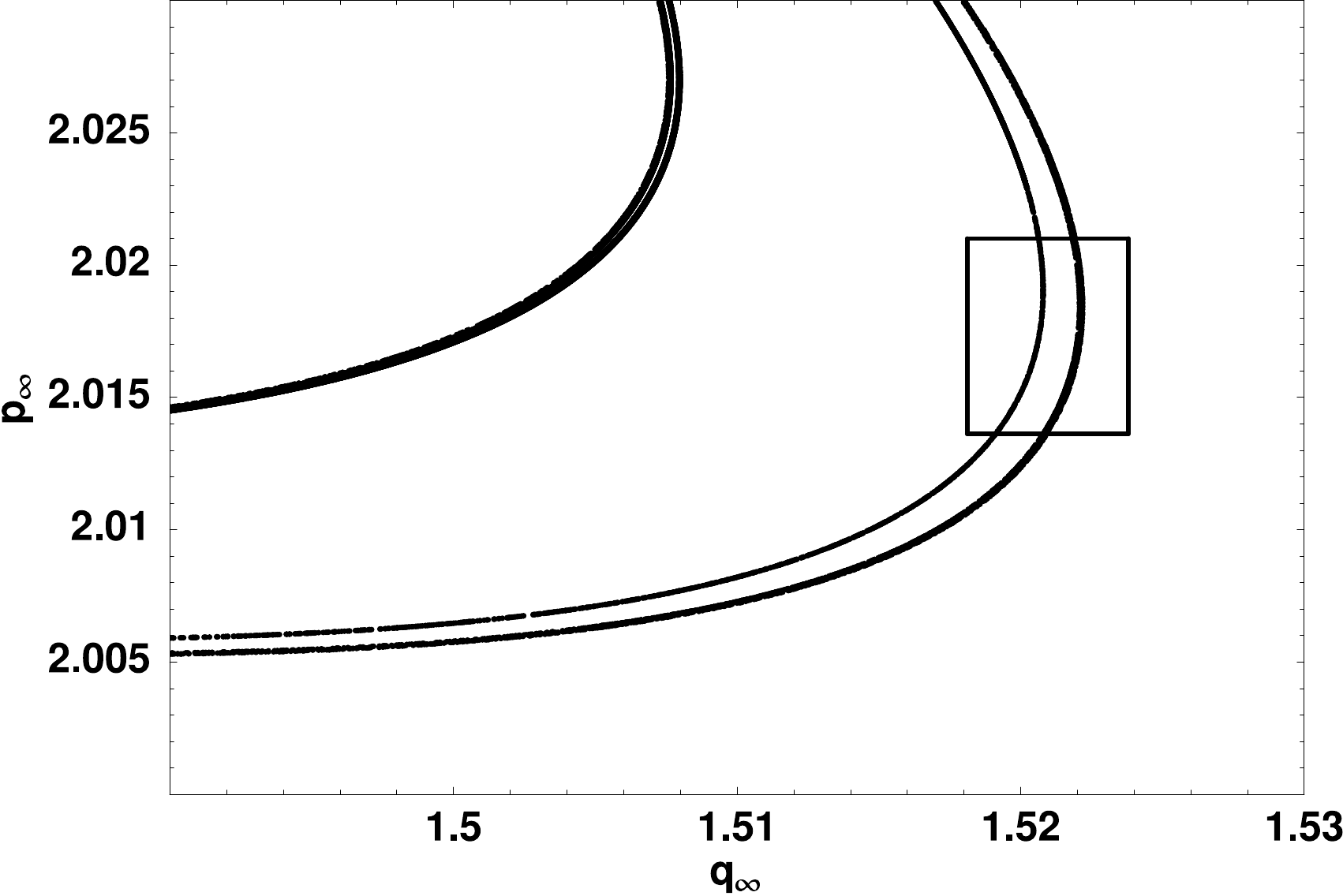}
  \caption{Enlargement of boxed portion of Figure 18 illustrating the
           beginning of self-similar fractal structure.}
\end{figure}

%Fig_22p12p28
\begin{figure}[htp]
  \centering
  \includegraphics*[height=5.5in]{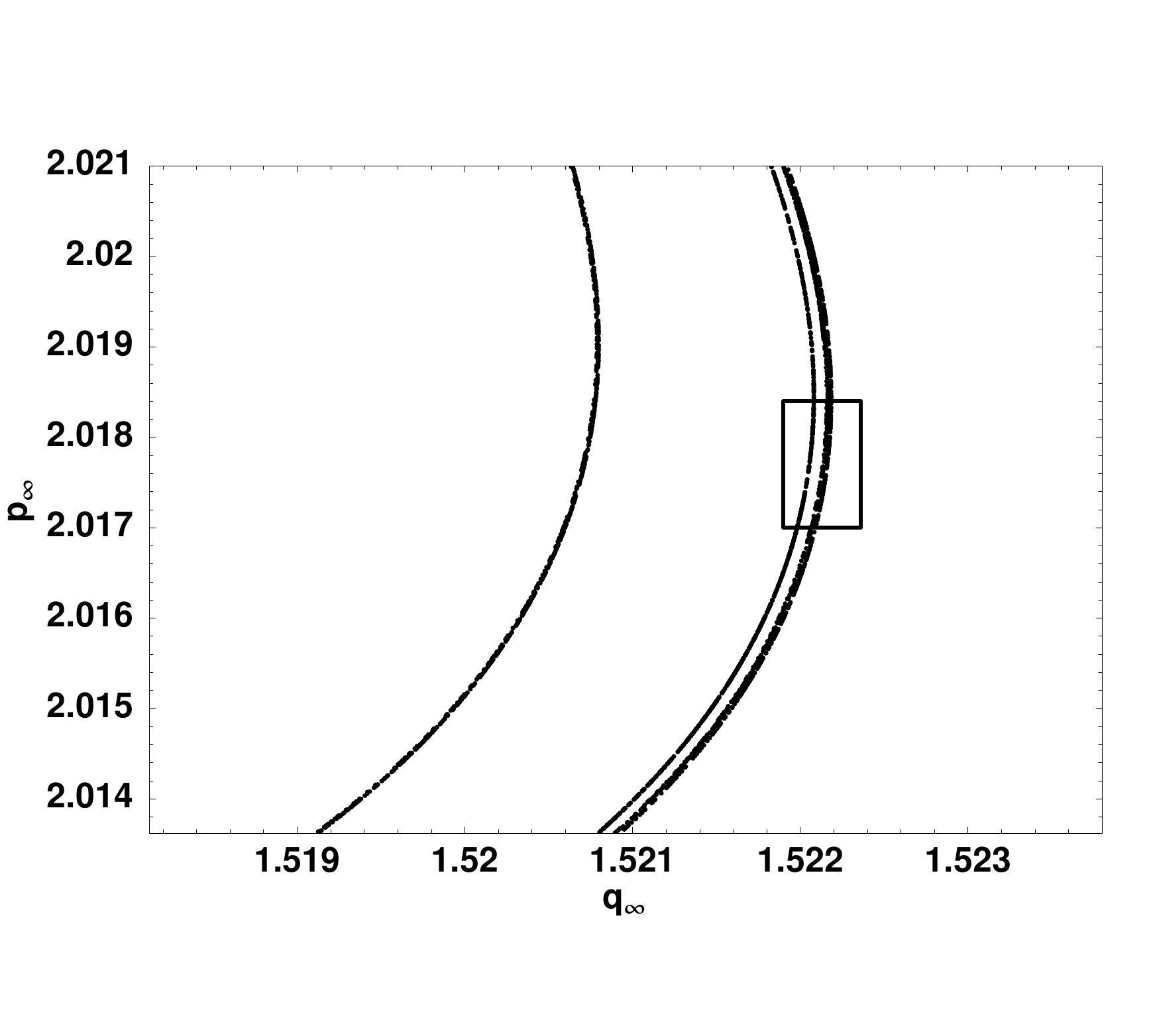}
  \caption{Enlargement of boxed portion of Figure 19 illustrating the
           continuation of self-similar fractal structure.}
\end{figure}

%Fig_22p12p29
\begin{figure}[htp]
  \centering
  \includegraphics*[height=5.5in]{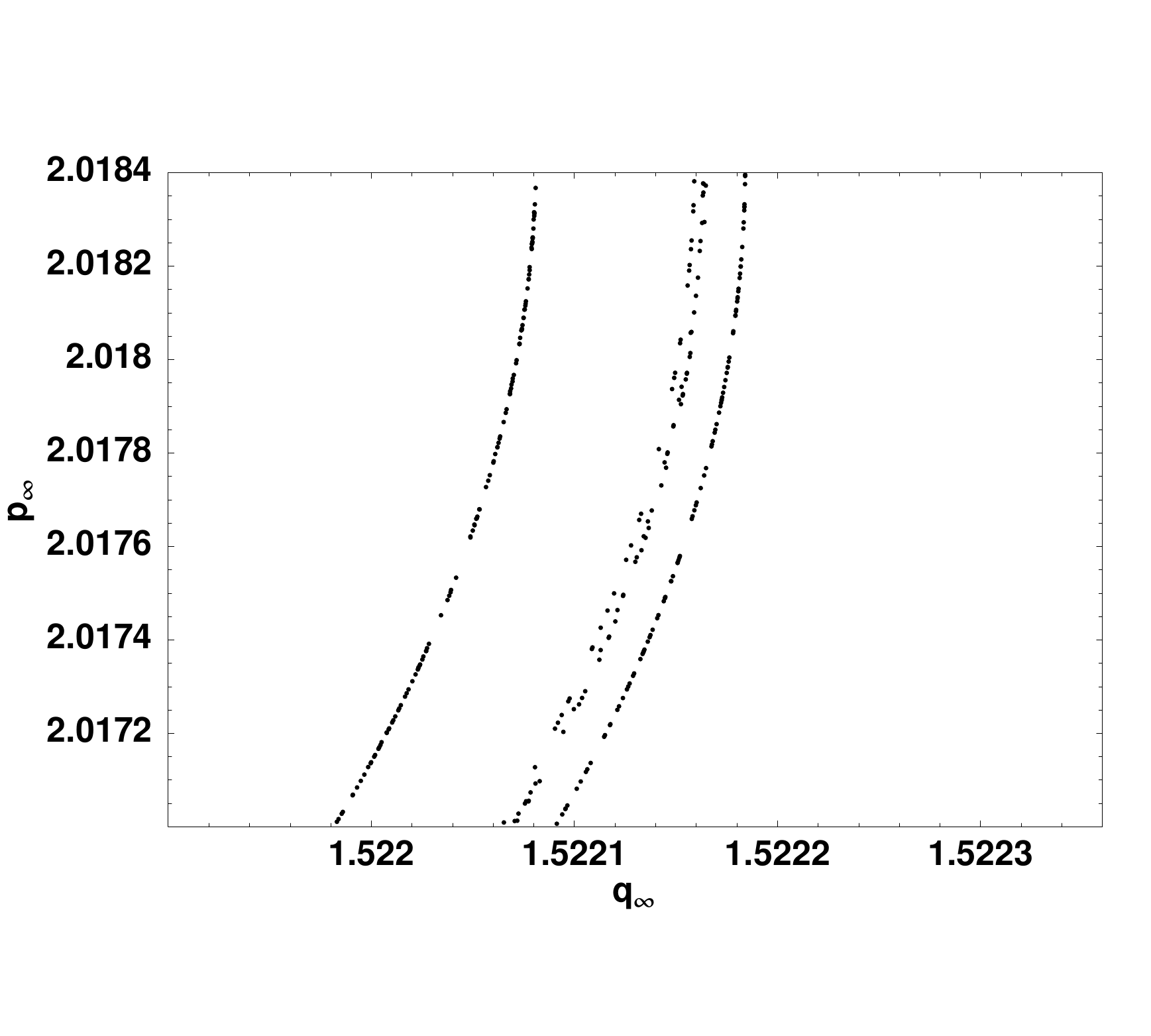}
  \caption{Enlargement of boxed portion of Figure 20 illustrating the
           further continuation of self-similar fractal structure.}
\end{figure}

\newpage

\section{Numerical Implementation}
\setcounter{equation}{0}
The forward integration method (Section~3.1) can be implemented by a code employing the tools of
{\em automatic differentiation} (AD) described by Neidinger [7].\footnote{Some authors refer to AD as {\em truncated power series algebra} (TPSA) since AD algorithms arise from manipulating multivariable truncated power series.  Other authors refer to AD as {\em Differential Algebra} (DA).}  In this approach arrays of Taylor coefficients of various functions are referred to as AD variables or {\em pyramids} since, as will be seen, they have a hyper-pyramidal structure.  Generally the first entry in the array will be the value of the function about some expansion point, and the remaining entries will be the higher-order Taylor coefficients about the expansion point and truncated beyond some specified order.  Such truncated Taylor expansions are also commonly called {\em jets}.

In our application elements in these arrays will be addressed and manipulated with the aid of scalar indices associated with look-up tables generated at run time.  We have also replaced the original APL implementation of Neidinger with a code written in the language of {\it Mathematica} (Version 6, or 7) [8].  Where necessary, for those unfamiliar with the details of {\it Mathematica}, we will explain the consequences of various {\it Mathematica} commands.  
The inputs to the code are the right sides (RS) of (1.1),
or (2.1).  Other input parameters are the number of variables  $m$, the desired order of the Taylor map $p$, and the initial conditions $(z_a^d)^i$ for the design-solution equation (2.3).

Various AD tools for describing and manipulating pyramids are outlined in Section 7.1. There we show how pyramid operations are encoded in the case of polynomial RS, as needed for the Duffing equation.  For brevity, we omit the cases of rational, fractional power, and transcendental RS.   These cases can also be handled using various methods based on functional identities and known Taylor coefficients, or the differential equations that such functions obey along with certain recursive relations [7].
In Section 7.2, based on the work of Section (7.1), we in effect obtain and integrate numerically the set  of differential equations (3.6) in pyramid form, i.e.  valid for any map order and any number of variables.  
Section 7.3 treats the specific case of the Duffing equation. A final Section 7.4 describes in more detail  the relation between integrating equations for pyramids and the complete variational equations.

\subsection{AD tools}
This section describes how arithmetic expressions representing $f_a(\bmv z,t)$, the right sides of (1.1)  where $\bmv z$ denotes the dependent variables, are replaced with expressions for arrays (pyramids) of Taylor coefficients.  These pyramids in turn constitute the input to our code. Such an ad-hoc replacement, according to the problem at hand, as opposed to operator overloading where the kind of operation depends on the type of its argument, is also the approach taken in [7,9,10].

Let $u,v,w$ be general arithmetic expressions, i.e. scalar-valued functions of $\bmv z$. They contain various arithmetic operations such as addition, multiplication $(*)$, and raising to a power $(\wedge)$.  (They may also entail the computation of various transcendental functions such as the sine function, etc.  However, as stated earlier, for simplicity we will omit these cases.) The arguments of these operations may be a constant,  a single variable or multiple variables $z_a$, or even some other expression. The  idea of AD is to redefine the arithmetic operations in such a way (see Definition 1), that all functions $u,v,w$ can be consistently  replaced with the arrays of coefficients of their Taylor expansions. For example, by redefining the usual product of numbers ($*$) and introducing the pyramid operation ${\tt PROD}$,  $u*v$ is replaced with {\tt PROD[U,V]}. 

We use upper typewriter font for pyramids ({\tt U,V,\dots}) and for operations on pyramids ({\tt PROD, POW, \dots}). Everywhere, equalities written in typewriter fonts have equivalent {\it Mathematica} expressions. That is, they have associated realizations in {\it Mathematica} and directly correspond to various operations and commands in {\it Mathematica}. In effect, our code operates entirely on pyramids.  However, as we will see, any pyramid  expression contains, as its first entry, its usual arithmetic  counterpart.

We begin with a description of our method of monomial labeling. In brief, we list all monomials in a polynomial in some sequence, and label them by where they occur in the list.  Next follow Definition~1 and the recipes for encoding operations on pyramids. Subsequently, by using Definition~2, which simply states the rule by which an arithmetic expressions is replaced with its pyramid counterpart, we show  how a general expression can be encoded by using only the pyramid of a constant and of a single variable. 

\subsubsection{Labeling Scheme}
A monomial $G_{\bmv j}(\bmv z )$ in $m$ variables is of the form 
\begin{equation}
G_{\bmv j}(\bmv z )=(z_1)^{j_1} (z_2)^{j_2}\cdots (z_m)^{j_m}.
\end{equation}
Here we have introduced an exponent vector ${\bmv j}$ by the rule
\begin{equation}
{\bmv j}=(j_1,j_2,\cdots j_m).
\end{equation}
Evidently $\bmv j$ is an $m$-tuple of non-negative integers.  The degree of $G_{\bmv j}(\bmv z )$, denoted by $|\bmv j|$, is given by the sum of exponents,
\begin{equation}
|\bmv j|=j_1+j_2+\cdots + j_m.
\end{equation}
The set of all exponents for monomials in $m$ variables with degree less than or equal to $p$ will be denoted by $\Gamma_m^p$,
\begin{equation}
\Gamma_m^p=\{\bmv j \; |\; |\bmv j| \leq p \}.
\end{equation}
It can be shown that this set has $L(m,p)$ entries with $L(m,p)$ given given by a binomial coefficient,
\begin{equation}
L(m,p)=\binom{p+m}{p}.
\end{equation}
In [6] this quantity is called $S_0(m,p)$.
Assuming that  $m$ and $p$ are fixed input variables, 
we will often write $\Gamma$ and $L$.  With this notation, 
a Taylor series expansion (about the origin) of a scalar-valued function $u$ of $m$ variables $\bmv z = (z_1,z_2,\dots z_m)$, truncated beyond terms of degree $p$, can be written in the form
\begin{equation}
u(\bmv z)=\sum_{\bmv j \;\in\; \Gamma_m^p }{\tt U}(\bmv j)\; G_{\bmv
j}(\bmv z ).
\end{equation}
Here, for now, ${\tt U}$ simply denotes an array of numerical coefficients.  When employed in code that has symbolic manipulation capabilities,  each ${\tt U}(\bmv j)$ may also be a symbolic quantity.

To proceed, what is needed is some way of listing monomials systematically.  With such a list, as already mentioned, we may assign a label $r$ to each monomial based on where it appears in the list.  A summary of labeling methods, and an analysis of storage requirements, may be found in [6,9].  Here we describe one of them that is particularly useful for our purposes.  

The first step is to {\em order} the monomials or, equivalently, the exponent vectors.  One possibility is {\em lexicographic} ({\em lex}) order.  Consider two exponent vectors $\bmv j$ and $\bmv k$.  Let $\bmv j - \bmv k$ be the vector whose entries are obtained by component-wise subtraction,
\begin{equation}
\bmv j - \bmv k=(j_1-k_1,j_2-k_2,\cdots j_m-k_m).
\end{equation}
We say that the exponent vector $\bmv j$ is lexicographically greater than the exponent vector $\bmv k$, and write $\bmv j >_{\rm{lex}}\bmv k$, if the {\em left-most nonzero} entry in $\bmv j - \bmv k$ is {\em positive}.
Thus, for example in the case of monomials in three variables $(z_1,z_2,z_3)$ with exponents $(j_1,j_2,j_3)$, we have the ordering 
\begin{equation}
(1,0,0)>_{\rm{lex}}(0,1,0)>_{\rm{lex}}(0,0,1)
\end{equation}
and 
\begin{equation}
(4,2,1)>_{\rm{lex}}(4,2,0)>_{\rm{lex}}(2,5,1).
\end{equation}

For our purposes we have found it convenient to label the monomials 
in such a way that monomials of a given degree $D=|\bmv j|$ occur together.  One possibility is to employ {\em  graded} lexicographic ({\em glex}) ordering.  If $\bmv j$ and $\bmv k$ are two exponent vectors, we say that $\bmv j>_{\rm{glex}}\bmv k$ if either $|\bmv j|>|\bmv k|$, or $|\bmv j|=|\bmv k|$ and $\bmv j>_{\rm{lex}}\bmv k$.  

Table 4 shows a list of monomials in three variables.  As one goes down the list, first the monomial of degree $D=0$ appears, then the monomials of degree $D=1$, etc.  Within each group of monomials of fixed degree the individual monomials appear in descending lex order.  Note that Table 4 is similar to Table 3 except that it begins with the monomial of degree 0.

\newpage
\begin{table}[h,t]
\caption{A labeling scheme for monomials in three variables.}
\begin{center}
\begin{tabular}{ccccc}
$r$ & $j_1$ & $j_2$ & $j_3$ & $D$ \\ \hline
1 & 0 & 0 & 0 & 0\\
2 & 1 & 0 & 0 & 1\\
3 & 0 & 1 & 0 & 1\\
4 & 0 & 0 & 1 & 1\\
5 & 2 & 0 & 0 & 2\\
6 & 1 & 1 & 0 & 2\\
7 & 1 & 0 & 1 & 2\\
8 & 0 & 2 & 0 & 2\\
9 & 0 & 1 & 1 & 2\\
10 & 0 & 0 & 2 & 2\\
11 & 3 & 0 & 0 & 3\\
12 & 2 & 1 & 0 & 3\\
13 & 2 & 0 & 1 & 3\\
14 & 1 & 2 & 0 & 3\\
15 & 1 & 1 & 1 & 3\\
16 & 1 & 0 & 2 & 3\\
17 & 0 & 3 & 0 & 3 \\
18 & 0 & 2 & 1 & 3\\
19 & 0 & 1 & 2 & 3\\
20 & 0 & 0 & 3 & 3\\
$.$ & $.$ & $.$ & $.$ & $.$\\
$.$ & $.$ & $.$ & $.$ & $.$\\
$.$ & $.$ & $.$ & $.$ & $.$\\
28 & 1 & 2 & 1 & 4\\
$.$ & $.$ & $.$ & $.$ & $.$\\
$.$ & $.$ & $.$ & $.$ & $.$\\
$.$ & $.$ & $.$ & $.$ & $.$\\
\end{tabular}
\end{center}
\end{table}

We give the name {\em modified glex sequencing} to a monomial listing of the kind shown in Table 4.  This is the labeling scheme we will use.  Other possible  listings include ascending true glex order in which monomials appear in ascending lex order within each group of degree $D$, and lex order for the whole monomial list as in [7]. 

With the aid of the scalar index $r$ the relation (7.6) can be rewritten in the form
\begin{equation}
u(\bmv z)=\sum_{r=1}^{L(m,p)} {\tt U}(r) G_r(\bmv z),
\end{equation}
because (by construction and with fixed $m$) for each positive integer $r$ there is a unique exponent ${\bmv j}(r)$, and for each ${\bmv j}$ there is a unique $r$.  Here $\tt U$ may be viewed as a vector with entries ${\tt U(r)}$, and $G_r(\bmv z)$ denotes $G_{{\bmv j}(r)}(\bmv z)$.

Consider, in an $m$-dimensional space, the points defined by the vectors ${\bmv j}\in\Gamma^p_m$.  See (7.4).  Figure 22 displays them in the case $m=3$ and $p=4$.  Evidently they form a grid that lies on the surface and interior of what can be viewed as an $m$-dimensional {\em pyramid} in $m$-dimensional space.  At each grid point there is an associated coefficient ${\tt U(r)}$.  Because of its association with this pyramidal structure, we will refer to the entire set of coefficients in (7.6) or (7.10) as the {\it pyramid} ${\tt U}$ of $u(\bmv z)$.

%no eps file 
\begin{figure}[h]
\begin{center}
\includegraphics[angle=0,width=.35\textwidth]{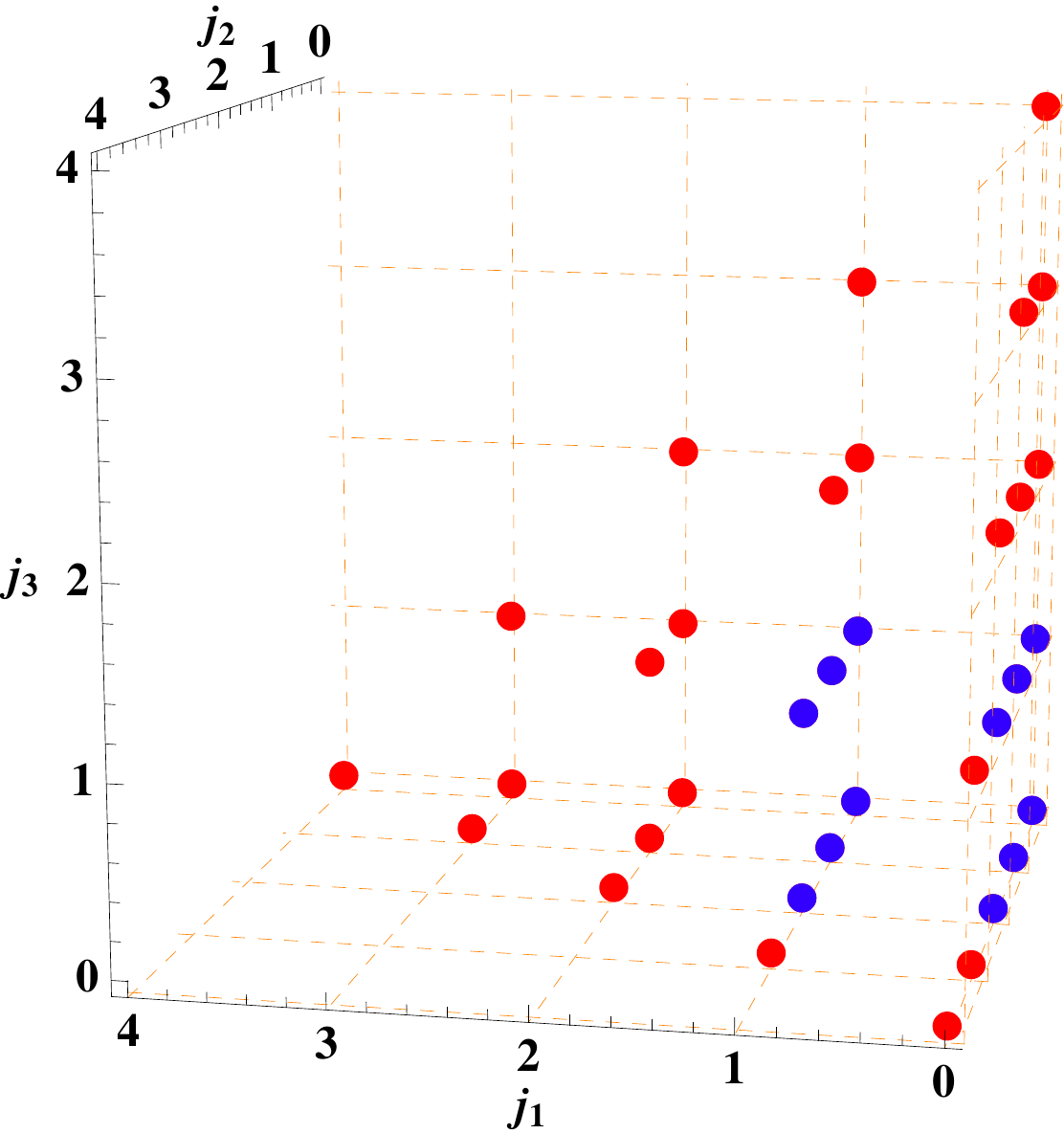}
\caption{\label{fig:pyram} A grid of points representing the set $\Gamma^4_3$.  For future reference a subset of $\Gamma^4_3$, called a {\em box}, is shown in blue.} 
\end{center}
\end{figure}

\subsubsection{Implementation of Labeling Scheme}

We have seen that use of modified glex sequencing, for any specified number of variables $m$, provides a labeling rule such that for each positive integer $r$ there is a unique exponent ${\bmv j}(r)$, and for each ${\bmv j}$ there is a unique $r$.  That is, there is a invertible function $r({\bmv j})$ that provides a 1-to-1 correspondence between the positive integers and the exponent vectors ${\bmv j}$.  To proceed further, it would be useful to have this function and its inverse in more explicit form.

First, we will learn that there is a formula for $r({\bmv j})$, which we will call the {\em Giorgilli} formula [6].
For any specified $m$ the exponent vectors ${\bmv j}$ take the form (7.2) where all the entries $j_i$ are positive integers or zero.   Begin by defining the integers
\begin{equation}
n(\ell ;j_1,\cdots ,j_m) = \ell -1 + \sum^{\ell -1}_{k=0} j_{m-k}
\end{equation}
for $\ell \in \{ 1,2,\cdots m\}$.  Then, to the general monomial $G_{\bmv j}({\bmv z})$ or exponent vector ${\bmv j}$, assign the label
\begin{equation}
r({\bmv j}) = r(j_1,\cdots j_m) = 
1+\sum^m_{\ell =1} \ {\rm Binomial} \ [n(\ell ;j_1,\cdots j_m),\ell ].
\end{equation}
Here the quantities
\begin{equation}
{\rm Binomial} \ [n,\ell ] = \left( \begin{array}{c} n \\
\ell \end{array}\right) = \left\{ \begin{array}{ccc} \frac{n!}{\ell !(n-\ell
)!} & , & 0 \leq \ell \leq n \\
0 & , & {\rm otherwise}\end{array}\right\}
\
\end{equation}
denote the usual binomial coefficients.  It can be shown that this formula reproduces the results of modified glex sequencing [6].

Below is simple {\em Mathematica} code that implements the Giorgilli formula in the case of three variables, and evaluates it for selected exponents $\bm j$.  Observe that these evaluations agree with results in Table 4.
\begin{eqnarray}
&&{\tt Gfor[j1\_, j2\_, j3\_] := (}\nonumber\\
&&{\tt s1 = j3;{\;}s2 = 1 + j3 + j2;{\;}s3 = 2 + j3 + j2 + j1;}\nonumber\\
&&{\tt t1 = Binomial[s1, 1];{\;}t2 = Binomial[s2, 2];{\;}t3 = Binomial[s3, 3];}\nonumber\\
&&{\tt r = 1 + t1 + t2 + t3;{\;}r}\nonumber\\
&&{\tt )}\nonumber\\
&&{\tt Gfor[0, 0, 0]}\nonumber\\
&&{\tt Gfor[1, 0, 0]}\nonumber\\
&&{\tt Gfor[2, 0, 1]}\nonumber\\
&&{\tt Gfor[1, 2, 1]}\nonumber\\
&&{    1}\nonumber\\
&&{    2}\nonumber\\
&&{    13}\nonumber\\
&&{    28}
\end{eqnarray}

Second, for the inverse relation, we have found it convenient to introduce a  rectangular matrix associated with the set $\Gamma_m^p$.  By abuse of notation, it will also be called $\Gamma$.  It has $L(m,p)$ rows and $m$ columns with entries
\begin{equation}
\Gamma_{r,a}=j_a(r).
\end{equation}
For example, looking a Table 4, we see (when $m=3$) that $\Gamma_{1,1}=0$ and $\Gamma_{17,2}=3$.  Indeed, if the first and last columns of Table 4 are removed, what remains (when $m=3$) is the matrix $\Gamma_{r,a}$.  In the language of [6], $\Gamma$ is a {\em look up table} that, given $r$, produces the associated ${\bmv j}$.  In our {\em Mathematica} implementation $\Gamma$ is the matrix ${\tt GAMMA}$ with elements ${\tt GAMMA[[r,a]]}$.  

The matrix ${\tt GAMMA}$ is constructed using the {\em Mathematica}
code illustrated below,
\begin{eqnarray}
&&{\tt Needs[\texttt{"}Combinatorica`\texttt{"}];}\nonumber\\
&&{\tt m=3;p=4;}\nonumber\\
&&{\tt GAMMA = Compositions[0, m];}\nonumber\\
&&{\tt Do[GAMMA = Join[GAMMA, Reverse[Compositions[d, m]]], \{d, 1, p, 1\}];}\nonumber\\
&&{\tt L=Length[GAMMA]}\nonumber\\
&&{\tt r=17; a=2;}\nonumber\\
&&{\tt GAMMA[[r]]}\nonumber\\
&&{\tt GAMMA[[r,a]]}\nonumber\\
&&{    35\nonumber}\\
&&{    \{0,3,0\}}\nonumber\\
&&{    3}
\end{eqnarray}
It employs the {\em Mathematica} commands \p{Compositions}, \p{Reverse}, and \p{Join}.  

We will first describe the ingredients of this code and illustrate the function of each:
\begin{itemize}
\item The command ${\tt Needs[\texttt{"}Combinatorica`\texttt{"}];}$ loads a combinatorial package.
\item The command \p{Compositions[i, m]} produces, as a list of arrays (a rectangular array), all {\em compositions} (under addition) of the integer $i$ into $m$ integer parts.  Furthermore, the compositions appear in {\em ascending} lex order.  
For example, the command \p{Compositions[0, 3]} produces the single row
\begin{eqnarray}
&&0\ \ 0\ \ 0
\end{eqnarray}
As a second example, the command \p{Compositions[1, 3]} produces the rectangular array
\begin{eqnarray}
&&0\ \ 0\ \ 1 \nonumber\\
&&0\ \ 1\ \ 0 \nonumber\\
&&1\ \ 0\ \ 0
\end{eqnarray}
As a third example, the command \p{Compositions[2, 3]} produces the rectangular array
\begin{eqnarray}
&&0\ \ 0\ \ 2 \nonumber\\
&&0\ \ 1\ \ 1 \nonumber\\
&&0\ \ 2\ \ 0 \nonumber\\
&&1\ \ 0\ \ 1 \nonumber\\
&&1\ \ 1\ \ 0 \nonumber\\
&&2\ \ 0\ \ 0 
\end{eqnarray}
\item The command \p{Reverse} acts on the list of arrays, and reverses the order of the list while leaving the arrays intact.  For example, the nested sequence of commands \p{Reverse[Compositions[1, 3]]} produces the rectangular array
\begin{eqnarray}
&&1\ \ 0\ \ 0\nonumber\\
&&0\ \ 1\ \ 0\nonumber\\
&&0\ \ 0\ \ 1
\end{eqnarray}
As a second example, the nested sequence of commands \p{Reverse[Compositions[2, 3]]} produces the rectangular array
\begin{eqnarray}
&&2\ \ 0\ \ 0\nonumber\\
&&1\ \ 1\ \ 0\nonumber\\
&&1\ \ 0\ \ 1\nonumber\\
&&0\ \ 2\ \ 0\nonumber\\
&&0\ \ 1\ \ 1\nonumber\\
&&0\ \ 0\ \ 2
\end{eqnarray}
Now the compositions appear in {\em descending} lex order.
\item Look, for example, at Table 4.  We see that the exponents $j_a$ for the $r=1$ entry 
are those appearing in (7.17).  Next, exponents for the $r=2$ through $r=4$ entries are those appearing in (7.20).  Following them, the exponents for the $r=5$ through $r=10$ entries, are those appearing in (7.21), etc.  Evidently, to produce the exponent list of Table 4, what we must do is successively {\em join} various lists.  That is what the {\em Mathematica} command \p{Join} can accomplish.
\end{itemize}

We are now ready to describe how ${\tt GAMMA}$ is constructed:
\begin{itemize}
\item   The second line in 
(7.16) sets the values of $m$ and $p$.  They are assigned the values $m=3$ and $p=4$ for this example, which will construct ${\tt GAMMA}$ for the case of Table 4.  The third line in (7.16) initially sets ${\tt GAMMA}$ to a row of $m$ zeroes.  The fourth  line is a \p{Do} loop that successively redefines ${\tt GAMMA}$ by generating and joining to it successive descending lex order compositions. The net result is the exponent list of Table 4.  

\item The quantity $L=L(m,p)$ is obtained by applying the {\em Mathematica} command \p{Length} to the the rectangular array ${\tt GAMMA}$.  

\item The last  6 lines of (7.16) illustrate that $L$ is computed properly and that the command ${\tt GAMMA[[r,a]]}$
accesses the array ${\tt GAMMA}$ in the desired fashion.  Specifically, in this example, we find from (7.5) that $L(3,4)=35$ in agreement with the {\em Mathematica} output for $L$.  Moreover, ${\tt GAMMA[[17]]}$ produces the exponent array $\{0,3,0\}$, in agreement with the $r=17$ entry in Table 4, and ${\tt GAMMA[[17,2]]}$ produces $\Gamma_{17,2}=3$, as expected.
\end{itemize}

\subsubsection{Pyramid Operations: Addition and Multiplication}
Here we {\it derive} the pyramid operations in terms of $\bmv j$-vectors by using the ordering previously described, and provide scripts to {\it encode} them in the $r$-representation (7.10). 
 \newtheorem{mydef}{Definition}
\begin{mydef}
Suppose that $w(\bm z)$ arises from carrying out various {\it arithmetic operations} on $u(\bmv z)$ and $v(\bmv z)$, and the pyramids \p{U} and  \p{V} are known. The corresponding pyramid operation on \p{U} and \p{V} is so defined that it yields the pyramid \p{W} of $w(\bm z)$.
\end{mydef}
Here we assume that $u,v,w$ are polynomials such as (7.6).

We begin with the operations of scalar multiplication and addition, which are easy to implement.  If 
\begin{equation}
w(\bmv z) = c\; u(\bmv z),
\end{equation}
then
\begin{equation}
{\tt W}(r)=c\;{\tt U}(r),
\end{equation}
and we write
\begin{equation}
{\tt W}=c\;{\tt U }.
\end{equation}
If
\begin{equation}
w(\bmv z) = u(\bmv z)+ v(\bmv z),
\end{equation}
then
\begin{equation}
{\tt W}(r)={\tt U}(r)+{\tt V}(r),
\end{equation}
and we write
\begin{equation}
{\tt W}={\tt U}+{\tt V}.
\end{equation}
In both cases all operations are performed coordinate-wise (as for vectors).

Implementation of scalar multiplication and addition is easy in {\em Mathematica} because, as the example below illustrates, it has built in vector routines.  There we define two vectors, multiply them by scalars, and add the resulting vectors.
\begin{eqnarray}
&&{\tt Unprotect[V];}\nonumber\\
&&{\tt U=\{1,2,3\};}\nonumber\\
&&{\tt V=\{4,5,6\};}\nonumber\\
&&{\tt W=.1U+.2V}\nonumber\\
&&     \{.9,1.2,1.5\}
\end{eqnarray}
Since \p{V} is a ``protected" symbol in the {\em Mathematica} language, and, for purposes of illustration, we wish to use it as an ordinary vector variable, it must first be unprotected as in line 1 above.  The last line shows that the {\em Mathematica} output is indeed the desired result.

The operation of polynomial multiplication is more involved.  Now we have the relation
\begin{equation}
w(\bmv z) = u(\bmv z)*v(\bmv z),
\end{equation}
and we want to encode
\begin{equation}
{\tt W} = {\tt PROD[U,V]}.
\end{equation}
Let us write $u(\bmv x)$ in the form (7.6), but with a change of dummy indices, so that it has the representation
\begin{equation}
u(\bmv z)=\sum_{\bmv i \;\in\; \Gamma_m^p }{\tt U}(\bmv i)\; G_{\bmv
i}(\bmv z ).
\end{equation}
Similarly, write $v(\bmv z)$ in the form
\begin{equation}
v(\bmv z)=\sum_{\bmv j \;\in\; \Gamma_m^p }{\tt V}(\bmv j)\; G_{\bmv
j}(\bmv z ).
\end{equation}
Then, according to Leibniz, there is the result
\begin{equation}
u(\bmv z)*v(\bmv z)=
\sum_{\bmv i \;\in\; \Gamma_m^p }{\;}\sum_{\bmv j \;\in\; \Gamma_m^p }
{\tt U}(\bmv i){\tt V}(\bmv j)G_{\bmv
i}(\bmv z )*G_{\bmv j}(\bmv z ).
\end{equation}
From (7.1) we observe that
\begin{eqnarray}
G_{\bmv i}(\bmv z )*G_{\bmv j}(\bmv z )&=&
(z_1)^{i_1} (z_2)^{i_2}\cdots (z_m)^{i_m}*(z_1)^{j_1} (z_2)^{j_2}\cdots (z_m)^{j_m}\nonumber\\
&=&(z_1)^{i_1+j_1} (z_2)^{i_2+j_2}\cdots (z_m)^{i_m+j_m}=
G_{{\bmv i}+{\bmv j}}(\bmv z ).
\end{eqnarray}
Therefore, we may also write
\begin{equation}
u(\bmv z)*v(\bmv z)=
\sum_{\bmv i \;\in\; \Gamma_m^p }{\;}\sum_{\bmv j \;\in\; \Gamma_m^p }
{\tt U}(\bmv i){\tt V}(\bmv j)G_{{\bmv i}+{\bmv j}}(\bmv z ).
\end{equation}
Now we see that there are two complications.  First, there may be terms on the right side of (7.35) whose degree is higher than $p$ and therefore need not be computed.  Second, there are generally many terms on the right side of (7.35) that contribute to a given monomial term in $w(\bmv z) = u(\bmv z)*v(\bmv z)$.  Suppose we write
\begin{equation}
w(\bmv z)=\sum_{\bmv k} {\tt W}(\bmv k)\; G_{\bmv
k}(\bmv z ).
\end{equation}
Upon comparing (7.35) and (7.36) we conclude that
\begin{equation}
 {\tt W}(\bmv k)=
\sum_{{\bmv i}+ {\bmv j}={\bmv k}}
{\tt U}(\bmv i){\tt V}(\bmv j)=\sum_{\bmv j \le  \bmv k } 
{\tt U}({\bmv k-\bmv j})  {\tt V} ({\bmv j}).
\end{equation}
Here, by $\bmv j \le  \bmv k $, we mean that the sum ranges over all ${\bmv j}$ such that $j_a\le k_a$ for all $a\in[1,m]$.  That is,
\begin{equation}
\bmv j \le  \bmv k{\;} \Leftrightarrow{\;}
j_a\le k_a{\;}{\rm{for}}{\;}{\rm{all}}{\;}a\in[1,m].
\end{equation}

Evidently, to implement the relation (7.37) in terms of $r$ labels, we need to describe the exponent relation $\bmv j \le  \bmv k $ in terms of $r$ labels.  Suppose ${\bmv k}$ is some exponent vector with label $r({\bmv k})$ as, for example, in Table 4.  Introduce the notation
\begin{equation}
k=r({\bmv k}).
\end{equation}
This notation may be somewhat confusing because $k$ is not the norm of the vector ${\bmv k}$, but rather the label associated with 
${\bmv k}$.  However, this notation is very convenient.  Now, given 
a label $k$, we can find ${\bmv k}$.  Indeed, from (7.15), we have the result
\begin{equation}
k_a=\Gamma_{k,a}.
\end{equation}
Having found ${\bmv k}$,
we define a set of exponents $B_k$ by the rule
\begin{equation}
B_k=\{{\bmv j}|\bmv j \le  \bmv k\}.
\end{equation}
This set of exponents is called the $k^{\rm{th}}$ {\em box}.  For example (when $m=3$), suppose $k=28$.  Then we see from Table 4 that ${\bmv k}$ = $(1,2,1)$.  Table 5 lists, in modified glex order, all the vectors in $B_{28}$, i.e. all vectors ${\bmv j}$ such that $\bmv j \le  (1,2,1)$.  These are the vectors shown in blue in Figure 22.  Finally, with this notation, we can rewrite (7.37) in the form
\begin{equation}
 {\tt W}({\bmv k})=
\sum_{\bmv j \in B_k }  {\tt U}({\bmv k-\bmv j})  {\tt V} ({\bmv j}).
\end{equation}
 
\begin{table}[h,t]
\caption{The vectors in $B_{28}=\{{\bmv j}|{\bmv j}\le(1,2,1)\}$.}
\begin{center}
\begin{tabular}{ccccc}
$r$ & $j_1$ & $j_2$ & $j_3$ & $D$ \\ \hline
1 & 0 & 0 & 0 & 0\\
2 & 1 & 0 & 0 & 1\\
3 & 0 & 1 & 0 & 1\\
4 & 0 & 0 & 1 & 1\\
6 & 1 & 1 & 0 & 2\\
7 & 1 & 0 & 1 & 2\\
8 & 0 & 2 & 0 & 2\\
9 & 0 & 1 & 1 & 2\\
14 & 1 & 2 & 0 & 3\\
15 & 1 & 1 & 1 & 3\\
18 & 0 & 2 & 1 & 3\\
28 & 1 & 2 & 1 & 4\\
\end{tabular}
\end{center}
\end{table}

What can be said about the vectors $({\bmv k-\bmv j})$ as $\bmv j$ ranges over $B_\ell$?  Table 6 lists, for example, the vectors 
$\bmv j\in B_{28}$ and the associated vectors $\bmv i$ with
${\bmv i}=({\bmv k-\bmv j})$.  Also listed are the labels $r(\bmv j)$ and $r(\bmv i)$.  Compare columns 2,3,4, which specify the $\bmv j\in B_{28}$, with columns 5,6,7, which specify the associated ${\bmv i}$ vectors.  We see that every vector that appears in the $\bmv j$ list also occurs somewhere in the $\bmv i$ list, and vice versa.  This to be expected because the operation of multiplication is commutative: we can also write (7.37) in the form
\begin{equation}
 {\tt W}({\bmv k})=
\sum_{\bmv j \in B_k }  {\tt U}({\bmv j})  {\tt V} ({\bmv k-\bmv j}).
\end{equation}
We also observe the more remarkable feature that 
the two lists are {\em reverses} of each other: running down the $\bmv j$ list gives the same vectors as running up the $\bmv i$ list, and vice versa.  This feature is a consequence of our ordering procedure.

As indicated earlier, what we really want is a version of (7.37) that involves labels instead of exponent vectors.  Looking at Table 6, we see that this is easily done.  We may equally well think of $B_k$ as containing a collection of labels $r({\bmv j})$, and we may introduce a {\em reversed} array $Brev_k$ of {\em complementary} labels $r^c({\bmv j})$ where
\begin{equation}
r^c({\bmv j})=r({\bmv i}).
\end{equation}
That is, for example, $B_{28}$ would consist of the first column of Table 6 and $Brev_{28}$ would consist of the last column of Table 6.
Finally, we have already introduced $k$ as being the label associated with ${\bmv k}$.  We these understandings in mind, we may rewrite (7.37) in the label form
\begin{equation}
 {\tt W}(k)=
\sum_{r \in B_k }  {\tt U}(r^c)  {\tt V} (r)=\sum_{r \in B_k }  {\tt U}(r)  {\tt V} (r^c).
\end{equation}
This is the rule ${\tt W} = {\tt PROD[U,V]}$ for multiplying pyramids.  In the language of [6], $B_k$ and $Brev_k$ are {\em look back tables} that, given a $k$, look back to find all monomial pairs with labels $r,r^c$ which produce, when multiplied, the monomial with label $k$.

\begin{table}[h,t]
\caption{The vectors ${\bmv j}$ and ${\bmv i}= ({\bmv k-\bmv j})$ for
${\bmv j}\in B_{28}$ and $k_a=\Gamma_{28,a}$.}
\begin{center}
\begin{tabular}{cccccccc}
$r({\bmv j})$&$j_1$&$j_2$& $j_3$&$i_1$&$i_2$&$i_3$&$r({\bmv i})$ \\ 
\hline
1 & 0 & 0 & 0 & 1&2&1&28\\
2 & 1 & 0 & 0 & 0&2&1&18\\
3 & 0 & 1 & 0 & 1&1&1&15\\
4 & 0 & 0 & 1 & 1&2&0&14\\
6 & 1 & 1 & 0 & 0&1&1&9\\
7 & 1 & 0 & 1 & 0&2&0&8\\
8 & 0 & 2 & 0 & 1&0&1&7\\
9 & 0 & 1 & 1 & 1&1&0&6\\
14 & 1 & 2 & 0 & 0&0&1&4\\
15 & 1 & 1 & 1 & 0&1&0&3\\
18 & 0 & 2 & 1 & 1&0&0&2\\
28 & 1 & 2 & 1 & 0&0&0&1\\
\end{tabular}
\end{center}
\end{table}

\subsubsection{Implementation of Multiplication}

The code shown below in (7.46) illustrates how $B_k$ and $Brev_k$ are constructed using {\em Mathematica}. 
\begin{eqnarray}
&&{\tt JSK[list\_, K\_] :=}\nonumber\\
&&{\tt Position[Apply[And,Thread[\texttt{\#}1\texttt{<=\#2\&[\#,K]]]\&
/@}\; list, True]//Flatten;}
\nonumber\\  
&&{\tt B = Table[JSK[GAMMA, GAMMA[[k]]], \{k, 1, L\}];}\nonumber\\
&&{\tt Brev = Reverse\; \texttt{/@}\; B;}
\end{eqnarray}
As before, some explanation is required.  The main tasks are to implement the $\bmv j \le  \bmv k $ operation (7.38) and then to employ this implementation.  We will begin by implementing the $\bmv j \le  \bmv k $ operation.  Several steps are required, and each of them is described briefly below:

\begin{itemize}
\item  When {\em Mathematica} is presented with a statement of the 
form $j<=k$, with $j$ and $k$ being {\em integers}, it replies with the answer True or the answer False.  (Here $j<=k$ denotes $j\le k$.)  Two sample {\em Mathematica} runs are shown below:
\begin{eqnarray}
&&\texttt{3\;<=\;4}\nonumber\\
&&{\rm{True}}
\end{eqnarray}
\begin{eqnarray}
&&\texttt{5\;<=\;4}\nonumber\\
&&{\rm{False}}
\end{eqnarray}

\item A {\em Mathematica} function can be constructed that does the same thing.  It takes the form
\begin{eqnarray}
&&{\tt \texttt{\#1\;<=\;\#2\;\&\;}[j,k]}
\end{eqnarray}
Here the symbols {\tt \#1} and {\tt \#2} set up two {\em slots} and the symbol {\tt \&} means the operation to its left is to be regarded as a function and is to be applied to the arguments to its right by inserting the arguments into the slots.  Below is a short {\em Mathematica} run illustrating this feature.
\begin{eqnarray}
&&{\tt j=3; k=4;}\nonumber\\
&&{\tt \texttt{\#1\;<=\;\#2\;\&\;}[j,k]}\nonumber\\
&&{\rm{True}}
\end{eqnarray}
Observe that the output of this run agrees with that of (7.47).

\item  The same operation can be performed on pairs of {\em arrays} (rather than pairs of numbers) in such a way that corresponding entries from each array are compared, with the output then being an array of True and False answers.  This is done using the {\em Mathematica} command \p{Thread}.  Below is a short {\em Mathematica} run illustrating this feature.
\begin{eqnarray}
&&{\tt j=\{1,2,3\};k=\{4,5,1\};}\nonumber\\
&&{\tt Thread[ \texttt{\#1\;<=\;\#2\;\&\;}[j,k]]}\nonumber\\
&&\{{\rm{True}},{\rm{True}},{\rm{False}}\}
\end{eqnarray}
Note that the first two answers in the output array are True because 
the statements $1\le4$ and $2\le5$ are true.  The last answer in the output array is False because the statement $3\le1$ is false. 
 
\item Suppose, now, that we are given two arrays $\bm j$ and $\bm k$ and we want to determine if ${\bm j}\le {\bm k}$ in the sense of (7.38).  This can be done by {\em applying} the logical \p{And} operation (using the {\em Mathematica} command \p{Apply}) to the True/False output array described above.  Below is a short {\em Mathematica} run illustrating this feature. 
\begin{eqnarray}
&&{\tt j=\{1,2,3\};k=\{4,5,1\};}\nonumber\\
&&{\tt Apply[And,Thread[\texttt{\#1\;<=\;\#2{\;}\&{\;}}[j,k]]]}\nonumber\\
&&{\rm{False}}
\end{eqnarray}
Note that the output answer is False because at least one of the entries in the output array in (7.51) is False.  The output answer would be True if, and only if, all entries in the output array in (7.51) were True.

\item Now that the ${\bm j}\le {\bm k}$ operation has been defined for two exponent arrays, we would like to construct a related operator/function, to be called \p{JSK}.  (Here the letter $\tt S$ stands for {\em smaller than or equal to}.)  It will depend on the exponent array $\bm k$, and its task will be to search a list of exponent arrays to find those $\bm j$ within it that satisfy ${\bm j}\le {\bm k}$.  The first step in this direction is to slightly modify the function appearing in (7.52).  Below is a short {\em Mathematica} run that specifies this modified function and illustrates that it has the same effect.
\begin{eqnarray}
&&{\tt j=\{1,2,3\};k=\{4,5,1\};}\nonumber\\
&&{\tt Apply[And,Thread[ \texttt{\#1\;<=\;\#2\;\&\;}[\texttt{\#},k]]]\;\texttt{\&}\;}[j]\nonumber\\
&&{\rm{False}}
\end{eqnarray}
Comparison of the functions in (7.52) and (7.53) reveals that what has been done is to replace the argument $j$ in (7.52) by a slot {\tt \#}, then follow the function by the character {\tt \&}, and finally add the symbols ${\tt [j]}$.  What this modification does is to redefine the function in such a way that it acts on what follows the second {\tt \&}.

\item The next step is to extend the function appearing in (7.53) so that it acts on a list of exponent arrays.  To do this, we replace the symbols ${\tt [j]}$ by the symbols {\tt /@} ${\tt list}$.  The symbols {\tt /@} indicate that what stands to their left is to act on what stands to their right, and what stands to their right is a list of exponent arrays.  The result of this action will be a list of True/False results with one result for each exponent array in the list.  Below is a short {\em Mathematica} run that illustrates how the further modified function acts on lists.
\begin{eqnarray}
&&{\tt k=\{4,5,1\};}\nonumber\\
&&{\tt ja=\{3,4,1\};jb=\{1,2,3\};jc=\{1,2,1\};}\nonumber\\
&&{\tt list=\{ja,jb,jc\};}\nonumber\\
&&{\tt Apply[And,Thread[\texttt{\#1\;<=\;\#2\;\&\;}[ \texttt{\#},k]]]\; \texttt{\&\;/@}\;list
}\nonumber\\
&&\{\rm{True},\rm{False},\rm{True}\}
\end{eqnarray}
Observe that the output answer list is $\{\rm{True},\rm{False},\rm{True}\}$ because $\{3,4,1\}\le\{4,5,1\}$ is True, $\{1,2,3\}\le\{4,5,1\}$ is False, and $\{1,2,1\}\le\{4,5,1\}$ is True.

\item What we would really like to know is where the True items are in the list, because  that will tell us where the $\bm j$ that satisfy ${\bm j}\le {\bm k}$ reside.  This can be accomplished by use of the {\em Mathematica} command \p{Position} in conjunction with the result  True.
Below is a short {\em Mathematica} run that illustrates how this works.
\begin{eqnarray}
&&{\tt k=\{4,5,1\};}\nonumber\\
&&{\tt ja=\{3,4,1\};jb=\{1,2,3\};jc=\{1,2,1\};}\nonumber\\
&&{\tt list=\{ja,jb,jc\};}\nonumber\\
&&{\tt Position[Apply[And,Thread[\texttt{\#1\;<=\;\#2\;\&\;}[\texttt{\#},k]]]\;\texttt{\&\;/@}\;list,True]}
\nonumber\\
&&\{\{1\},{\;}\{3\}\}
\end{eqnarray}
Note that the output is an array of positions in the list for which ${\bm j}\le {\bm k}$.  There is, however, still one defect.  Namely, the output array is an array of single-element subarrays, and we would like it to be simply an array of location numbers.  This defect can be remedied by appending  the {\em Mathematica} command \p{Flatten}, preceded by {\tt //},  to the instruction string in (7.55).  The short {\em Mathematica} run below illustrates this modification.
\begin{eqnarray}
&&{\tt k=\{4,5,1\};}\nonumber\\
&&{\tt ja=\{3,4,1\};jb=\{1,2,3\};jc=\{1,2,1\};}\nonumber\\
&&{\tt list=\{ja,jb,jc\};}\nonumber\\
&&{\tt Position[Apply[And,Thread[\texttt{\#1\;<=\;\#2\;\&\;}[\texttt{\#},k]]]\;\texttt{\&\;/@}\;list,True]\texttt{//}Flatten}
\nonumber\\
&&\{1,{\;}3\}
\end{eqnarray}
Now the output is a simple array containing the positions in the list for which ${\bm j}\le {\bm k}$.

\item The last step is to employ the ingredients in (7.56) to define the operator ${\tt JSK[list,k]}$.  The short {\em Mathematica} run below illustrates how this can be done.
\begin{eqnarray}
&&{\tt k=\{4,5,1\};}\nonumber\\
&&{\tt ja=\{3,4,1\};jb=\{1,2,3\};jc=\{1,2,1\};}\nonumber\\
&&{\tt list=\{ja,jb,jc\};}\nonumber\\
&&{\tt JSK[list\_,k\_]\;:=}\nonumber\\
&&{\tt Position[Apply[And,Thread[\texttt{\#1\;<=\;\#2\;\&\;}[\texttt{\#},k]]]\;\texttt{\&\;/@}\;list,True]\texttt{//}Flatten;}\nonumber\\
&&{\tt JSK[list,\;k]}\nonumber\\
&&\{1,{\;}3\}
\end{eqnarray}

Lines 4 and 5 above define the operator ${\tt JSK[list,k]}$, line 6 invokes it, and line 7 displays its output, which agrees with the output of (7.56).

\item With the operator ${\tt JSK[list,k]}$ in hand, we are prepared to construct tables $B$ and $Brev$ that will contain the $B_k$ and the $Brev_k$.  The short {\em Mathematica} run below illustrates how this can be done.
\begin{eqnarray}
&&{\tt  B=Table[JSK[GAMMA, GAMMA[[k]]], \{k, 1, L, 1\}] ;}\nonumber\\
&&{\tt  Brev=Reverse\;\texttt{/@}\; B;}\nonumber\\
&&{\tt  B[[8]]}\nonumber\\
&&{\tt  Brev[[8]]}\nonumber\\
&&{\tt  B[[28]]}\nonumber\\
&&{\tt  Brev[[28]]}\nonumber\\
&&\{1,3,8\}\nonumber\\
&&\{8,3,1\}\nonumber\\
&&\{1,2,3,4,6,7,8,9,14,15,18,28\}\nonumber\\
&& \{28,18,15,14,9,8,7,6,4,3,2,1\}
\end{eqnarray}

The first line employs the {\em Mathematica} command \p{Table} in combination with an implied Do loop to produce a two-dimensional array $\tt B$.  Values of $k$ in the range $[1,L]$ are selected sequentially.  For each $k$ value the associated exponent array ${\bmv k}(k)={\tt GAMMA[[k]]}$ is obtained.  The operator \p{JSK} then searches the full \p{GAMMA} array to find the list of $r$ values associated with the ${\bmv j}\le{\bmv k}$.  All these $r$ values are listed in a row.  Thus, the array $\tt B$ consists of list of $L$ rows, of varying width.  The rows are labeled by $k\in[1,L]$, and in each row are the $r$ values associated with the ${\bmv j}\le{\bmv k}$.  In the second line the {\em Mathematica} command \p{Reverse} is applied to $\tt B$ to produce a second array called ${\tt Brev}$.  Its rows are the reverse of those in $\tt B$.  For example, as the {\em Mathematica} run illustrates,   ${\tt B[[8]]}$, which is the 8$^{th}$ row of $\tt B$, contains the list $\{1,3,8\}$, and ${\tt Brev[[8]]}$ contains the list $\{8,3,1\}$.  Inspection of the $r=8$ monomial in Table 4, that with exponents $\{0,2,0\}$, shows that it has the monomials with exponents \{0,0,0\}, \{0,1,0\}, and \{0,2,0\} as factors.  And further inspection of Table 4 shows that the exponents of these factors have the $r$ values $\{1,3,8\}$.  Similarly ${\tt B[[28]]}$, which is the 28$^{th}$ row of $B$, contains the same entries that appear in the first column of Table 6.  And ${\tt Brev[[28]]}$, which is the 28$^{th}$ row of $Brev$, contains the same entries that appear in the last column of Table 6.  
\end{itemize}

Finally, we need to explain how the arrays $B$ and $Brev$ can be employed to carry out polynomial multiplication.  This can be done using the {\em Mathematica} dot product command:
\begin{itemize}

\item  The exhibit below shows a simple {\em Mathematica} run that illustrates the use of the dot product command.
\begin{eqnarray}
&&{\tt Unprotect[V];}\nonumber\\
&&{\tt U=\{.1,.2,.3,.4,.5,.6,.7,.8\} ;}\nonumber\\
&&{\tt V=\{1.1,1.2,1.3,1.4,1.5,1.6,1.7,1.8\};}\nonumber\\
&&{\tt U.V}\nonumber\\
&&{\tt u=\{1,3,5\};}\nonumber\\
&&{\tt v=\{6,4,2\};}\nonumber\\
&&{\tt U[[u]]}\nonumber\\
&&{\tt V[[v]]}\nonumber\\
&&{\tt U[[u]].V[[v]]}\nonumber\\
&&5.64\nonumber\\
&&\{.1,.3,.5\}\nonumber\\
&&\{1.6,1.4,1.2\}\nonumber\\
&&1.18
\end{eqnarray}
As before, $\tt V$ must be unprotected.  See line 1.
The rest of the first part this run (lines 2 through 4) defines two vectors $\tt U$ and $\tt V$ and then computes their dot product.
Note that if we multiply the entries in $\tt U$ and $\tt V$ pairwise and add, we get the result 
\begin{equation}
.1\times1.1+.2\times1.2 +\cdots+.8\times1.8=5.64,\nonumber\\
\end{equation}
 which agrees with the {\em Mathematica} result for ${\tt U}\cdot {\tt V}$. See line 10. 

The second part of this {\em Mathematica} run, lines 5 through 9, illustrates a powerful feature of the {\em Mathematica} language.  Suppose, as illustrated, we define two arrays $\tt u$ and $\tt v$ of integers, and use these arrays as {\em arguments} for the vectors by writing ${\tt U[[u]]}$ and ${\tt V[[v]]}$.  Then {\em Mathematica} uses the integers in the two arrays $\tt u$ and $\tt v$ as labels to select the corresponding entries in $\tt U$ and $\tt V$, and from these entries it makes new corresponding vectors.  In this example, the 1$^{st}$, 3$^{rd}$, and 5$^{th}$ entries in $\tt U$ are $.1$, $.3$, and $.5$.  And the 6$^{th}$, 4$^{th}$, and 2$^{nd}$ entries in $\tt V$ are $1.6$, $1.4$, and $1.2$.  Consequently, we find that 
\begin{equation}
{\tt U[[u]]}=\{.1,.3,.5\},\nonumber\\
\end{equation}
\begin{equation}
{\tt V[[v]]}=\{1.6,1.4,1.2\},\nonumber\\
\end{equation}
in agreement with lines 11 and 12 of the {\em Mathematica} results.
Correspondingly, we expect that ${\tt U[[u]]}\cdot {\tt V[[v]]}$ will have the value
\begin{equation}
{\tt U[[u]]}\cdot {\tt V[[v]]}=.1\times1.6+.3\times1.4+.5\times1.2=1.18,\nonumber\\
\end{equation}
in agreement with the last line of the {\em Mathematica} output.

\item Now suppose, as an example, that we set $k=8$ and use ${\tt B[[k]]}$ and ${\tt Brev[[k]]}$ in place of the arrays $\tt u$ and $\tt v$.  The {\em Mathematica} fragment below shows what happens when this is done.
\begin{eqnarray}
&&{\tt  k=8;}\nonumber\\
&&{\tt B[[k]]}\nonumber\\
&&{\tt Brev[[k]]}\nonumber\\
&&{\tt U[[B[[k]]]]}\nonumber\\
&&{\tt V[[Brev[[k]]]]}\nonumber\\
&&{\tt U[[B[[k]]]]\cdot B[[Brev[[k]]]]}\nonumber\\
&&\{1,3,8\}\nonumber\\
&&\{8,3,1\}\nonumber\\
&&\{.1,.3,.8\}\nonumber\\
&&\{1.8,1.3,1.1\}\nonumber\\
&&1.45
\end{eqnarray}
From (7.58) we see that ${\tt B[[8]]}=\{1,3,8\}$ and ${\tt Brev[[8]]}=\{8,3,1\}$ in agreement with lines 7 and 8 of the {\em Mathematica} output above.  Also, the 1$^{st}$, 3$^{rd}$, and 8$^{th}$ entries in $\tt U$ are .1, .3, and .8.  And the 
8$^{th}$, 3$^{rd}$, and 1$^{st}$ entries in $\tt V$ are 1.8, 1.3, and 1.1.
Therefore we expect the results
\begin{equation}
{\tt U[[B[[k]]]]}=
\{.1,.3,.8\},\nonumber\\
\end{equation}
\begin{equation}
{\tt V[[Brev[[k]]]]}=
\{1.8,1.3,1.1\},\nonumber\\
\end{equation}
\begin{equation}
{\tt U[[B[[k]]]]}\cdot {\tt V[[Brev[[k]]]]}=
.1\times1.8+.3\times1.3+.8\times1.1=1.45,\nonumber\\
\end{equation}
in agreement with the last three lines of (7.60).

\item Finally, suppose we carry out the operation ${\tt U[[B[[k]]]]}\cdot {\tt V[[Brev[[k]]]]}$ for all $k\in[1,L]$ and put the results together in a Table with entries labeled by $k$.  According to (7.45), the result will be the pyramid for the product of the two polynomials whose individual pyramids are $\tt U$ and $\tt V$.  The {\em Mathematica} fragment below shows how this can be done to define a {\em product} function, called \p{PROD}, that acts on general pyramids $\tt U$ and $\tt V$, using the command Table with an implied Do loop over $k$.
\begin{equation}
{\tt PROD[U\_, V\_] := Table[U[[B[[k]]]]\cdot V[[Brev[[k]]]], \{k, 1, L, 1\}]};
\nonumber\\
\end{equation}

\end{itemize}

\subsubsection{Implementation of Powers}
With operation of multiplication in hand, it is easy to implement the operation of raising a pyramid to a power.  The code shown below in (7.61) demonstrates how this can be done.
\begin{eqnarray}
&&{\tt POWER[U\_, 0] := C1;}\nonumber\\
&&{\tt POWER[U\_, 1] := U;}\nonumber\\
&&{\tt POWER[U\_, 2] := PROD[U, U];}\nonumber\\
&&{\tt POWER[U\_, 3] := PROD[U, POWER[U, 2]];}\nonumber\\
&&...
\end{eqnarray}
Here ${\tt C1}$ is the pyramid for the Taylor series having {\em one} as its {\em constant} term and all other terms zero,
\begin{equation}
{\tt C1=\{1,0,0,0,\cdots\}}.
\end{equation}
It can be set up by the {\em Mathematica} code
\begin{equation}
{\tt C1=Table[KroneckerDelta[k,1],\{k,1,L,1\}];}
\end{equation}
which employs the Table command, the Kronecker delta function, and an implied Do loop over $k$.
This code should be executed before executing (7.61), but after the value of $L$ has been established.

\subsubsection{Replacement Rule}
\begin{mydef}
The transformation $A(\bmv z) \leadsto {\tt A} $ means replacement of every real variable $z_a$ in the {\it arithmetic expression} $A(\bmv z)$ with an associated pyramid, and of every operation on real variables in $A(\bmv z)$ with the associated operation on pyramids. 
\end{mydef}
Automatic Differentiation is based on the following corollary: if $A(\bmv z) \leadsto {\tt A} $, then  ${\tt A}$ is the pyramid of  $A(\bmv z)$.  

For simplicity, we will begin our discussion of the replacement rule with examples involving only a single variable $z$.  In this case monomial labeling, the relation between labels and exponents, is given directly by the simple rules
\begin{equation}
r(j)=1+j{\;}{\rm{and}}{\;}j(r)=r-1.
\end{equation}
See table 7.

\begin{table}[h,t]
\caption{A labeling scheme for monomials in one variable.}
\begin{center}
\begin{tabular}{cc}
$r$ & $j$  \\ \hline
1 & 0 \\
2 & 1 \\
3 & 2 \\
4 & 3 \\
$\cdot$ & $\cdot$ \\
$\cdot$ & $\cdot$ \\
\end{tabular}
\end{center}
\end{table}
As a first example, consider the expression
\begin{equation}
A=2+3(z*z). 
\end{equation}
We have agreed to consider the case $m=1$.  Suppose we also set $p=2$, in which case $L=3$.  In ascending glex order, see Table 7, the pyramid for $A$ is then
\begin{equation}
2+3z^2\leadsto{\tt A}=(2,0,3).
\end{equation}
Now imagine that $A$ was not such a simple polynomial, but some complicated expression. Then the pyramid ${\tt A}$ could be generated by computing derivatives of $A$ at $z=0$ and dividing them by the appropriate factorials.  Automatic differentiation offers another way to find \p{A}. Assume that all operations in the arithmetic expression $A$ have been encoded according to {\it Definition 1}. For our example, these are $+$ and ${\tt PROD}$. Let $\tt C1$ and $\tt Z$ be the pyramids associated with $1$ and $z$,
\begin{equation}
1\leadsto{\tt C1}=(1,0,0),
\end{equation}
\begin{equation}
z\leadsto\tt Z=(0,1,0).
\end{equation}
The quantity $2+3z^2$ results from performing various arithmetic operations on $1$ and $z$.  {\it Definition 1} says that the pyramid of  $2+3z^2$ is identical to the pyramid obtained by performing the same operations on the pyramids ${\tt C1}$ and $\tt Z$. That is, suppose we replace  $1$ and $z$ with correctly associated pyramids $\tt C1$ and $\tt Z$, and also replace $*$ with ${\tt PROD}$. Then, upon evaluating  ${\tt PROD}$, multiplying by the appropriate scalar coefficients, and summing, the result will be the same pyramid $\tt A$,
\begin{equation}
2{\;}{\tt C1}+3{\;}{\tt PROD[Z,Z]}= {\tt A}.
\end{equation}
In this way, by knowing only the basic pyramids ${\tt C1}$ and ${\tt Z}$  (prepared beforehand), one can compute the pyramid of an arbitrary $A(z)$. Finally, in contrast to numerical differentiation, all numerical operations involved are accurate to machine precision.  {\em Mathematica} code that implements (7.69) will be presented shortly in (7.70).

Frequently, if $A(z)$ is some complicated expression, the replacement rule will result in a long chain of nested pyramid operations.  At every step in the chain, the pyramid resulting from the previous step will be combined with some other pyramid to produce a new pyramid. Each such operation has two arguments, and {\it Definition 1} applies to each step in the chain. Upon evaluating all pyramid operations, the final result will be the pyramid of $A(z)$.

By using the replacement operator the above procedure can be represented as:
$$1\leadsto {\tt C1}, \ \ 
z\leadsto {\tt Z}, \ \ 
A\leadsto {\tt A}.
$$
The following general recipe then applies: In order to derive the pyramid associated with some arithmetic expression, apply the $\leadsto$ rule to all its variables, or parts, and replace all operations with operations on pyramids. Here ``apply the $\leadsto$ rule'' to something means replace that something with the associated pyramid.  And the term ``parts'' means sub-expressions.  {\it Definition 1} guarantees that the result will be the same pyramid $\tt A$ no matter how we split the arithmetic expression $A$ into sub-expressions. It is only necessary to recognize, in case of using sub-expressions, that one pyramid expression should be viewed as a function of another.

For illustration, suppose we regard the $A$ given by (7.65) to be the composition of two functions, $F(z)=2+3z$ and  $G(z)=z^2$, so that  $A(z)=F(G(z))$. Instead of associating a constant and a single variable with their respective pyramids, let us now associate whole sub-expressions.  In addition, let us label the pyramid expressions on the right of $\leadsto$ with with some names, $\tt F$ and $\tt G$:
$$2+3z\leadsto2{\;}{\tt C1}+3{\;}{\tt Z}={\tt F[Z]}$$
$$z^2 \leadsto{\tt PROD[Z,Z]=G[Z] }$$
$$A(z) \leadsto{\tt F[G[Z]] =A}.$$
We have indicated the explicit dependence on $\tt Z$. It is important to note that  $\tt F[Z]$ is a pyramid {\em expression} prior to executing any the pyramid operations, i.e it is not yet a pyramid, but is simply the result of formal replacements that follow the association rule. 

{\em Mathematica} code for the simple example (7.69) is shown below,
\begin{eqnarray}
&&{\tt C1=\{1,0,0 \};  }\nonumber\\
&&{\tt  Z=\{0,1,0\}; }\nonumber\\
&&{\tt2\;C1+3\;PROD[Z,Z] }\nonumber\\
&& \{2,0,3\} 
\end{eqnarray}
Note that the result (7.70) agrees with (7.66).  This example does not use any nested expressions.  We will now illustrate how the same results can be obtained using nested expressions. 

We begin by displaying a simple {\em Mathematica} program/execution, that employs ordinary variables, and uses {\em Mathematica}'s intrinsic abilities to handle nested expressions.  
The program/execution is
\begin{eqnarray}
&&{\tt f[z\_]:=2+3z;}\nonumber\\
&&{\tt g[z\_]:=z^2;}\nonumber\\
&&{\tt f[g[z]]}\nonumber\\
&&2+3z^2
\end{eqnarray}
With {\it Mathematica} the underscore in ${\tt z\_}$ indicates that $\tt z$ is a dummy variable name, and the symbols ${\tt :=}$ indicate that $\tt f$ is defined with a delayed assignment.  That is what is done in line one above.  The same is done in line two for $\tt g$.  Line three requests evaluation of the nested function $ f(g(z))$, and the result of this evaluation is displayed in line four.  Note that the result agrees with (7.65).

With this background, we are ready to examine a program with analogous nested pyramid operations. The same comments apply regarding the use of underscores and delayed assignments.  The program is
\begin{eqnarray}
&&{\tt C1=\{1,0,0 \};}  \nonumber\\
&&{\tt  Z=\{0,1,0\}; }\nonumber\\
&&{\tt F[Z\_]:=2{\;}C1+3{\;}Z;}\nonumber\\
&&{\tt G[Z\_]:=PROD[Z,Z];}\nonumber\\
&&{\tt F[G[Z]]}\nonumber\\
&&{\tt \{2,0,3\}}
\end{eqnarray}
Note that line (7.72) agrees with line (7.70), and is consistent with line (7.66).

We close this subsection with an important consequence of the replacement rule and nested operations, which we call the {\em Taylor} rule.  We begin by considering functions of a single variable.  Suppose the function $G(x)$ has the special form
\begin{equation}
G(x)=z^d+x
\end{equation}
where $z^d$ is some constant.  Let $F$ be some other function.  Consider the composite (nested) function $A$ defined by
\begin{equation}
A(x)=F(G(x))=F(z^d+x).
\end{equation}
Then, assuming the necessary analyticity, by the chain rule $A$ evidently has a Taylor expansion in $x$ about the origin of the form
\begin{eqnarray}
A&=&A(0)+A^\prime(0)x+(1/2)A^{\prime\prime}(0)x^2+\cdots\nonumber\\
&=&F(z^d)+F^\prime(z^d)x+(1/2)F^{\prime\prime}(z^d)x^2+\cdots.
\end{eqnarray}
We conclude that if we know the Taylor expansion of $A$ about the origin, then we also know the Taylor expansion of $F$ about $z^d$, and vice versa.  Suppose, for example, that
\begin{equation}
F(z)=1+2z+3z^2
\end{equation}
and
\begin{equation}
z^d=4.
\end{equation}
Then there is the result
\begin{equation}
A(x)=F(G(x))=F(z^d+x)=1+2(4+x)+3(4+x)^2=57+26x+3x^2.
\end{equation}

We now show that this same result can be obtained using pyramids. 
The {\em Mathematica} fragment below illustrates how this can be done.
\begin{eqnarray}
&&{\tt C1 = \{1, 0, 0\};}\nonumber\\
&&{\tt X = \{0, 1, 0\};}\nonumber\\
&&{\tt zd = 4;}\nonumber\\
&&{\tt F[Z\_] := 1\;C1 + 2\;Z + 3\;PROD[Z, Z];}\nonumber\\
&&{\tt G[X\_] := zd{\;}C1 + X;}\nonumber\\
&&{\tt F[G[X]]}\nonumber\\
&&\{57, 26, 3\}
\end{eqnarray}
Note that (7.79) agrees with (7.78).

Let us also illustrate the Taylor rule in the two-variable case.  Let $F(z_1,z_2)$ be some function of two variables.  Introduce the functions $G(x_1)$ and $H(x_1)$ having the special forms
\begin{equation}
G(x_1)=z_1^d+x_1,
\end{equation}
\begin{equation}
H(x_2)=z_2^d+x_2,
\end{equation}
where $z_1^d$ and $z_2^d$ are some constants.
Consider the function $A$ defined by
\begin{equation}
A(x_1,x_2)=F(G(x_1),H(x_2))=F(z_1^d+x_1,z_2^d+x_2).
\end{equation}
Then, again assuming the necessary analyticity, by the chain rule $A$ evidently has a Taylor expansion in $x_1$ and $x_2$ about the origin $(0,0)$ of the form
\begin{eqnarray}
A&=&A(0,0)+[\partial_1A(0,0)]x_1+[\partial_2A(0,0)]x_2\nonumber\\
&&+(1/2)[(\partial_1)^2A(0,0)]x_1^2+[\partial_1\partial_2A(0,0)]x_1x_2+(1/2)[(\partial_2)^2A(0,0)]x_2^2+\cdots\nonumber\\
&=&F(z_1^d,z_2^d)+[\partial_1F(z_1^d,z_2^d)]x_1+[\partial_2F(z_1^d,z_2^d)]x_2\nonumber\\
&&+(1/2)[(\partial_1)^2F(z_1^d,z_2^d)]x_1^2+[\partial_1\partial_2AF(z_1^d,z_2^d)]x_1x_2+(1/2)[(\partial_2)^2A(F(z_1^d,z_2^d))]x_2^2+\cdots\nonumber\\
\end{eqnarray}
where 
\begin{equation}
\partial_1=\partial/\partial x_1, {\;}{\;}\partial_2=\partial/\partial x_2
\end{equation}
when acting on $A$, and
\begin{equation}
\partial_1=\partial/\partial z_1, {\;}{\;}\partial_2=\partial/\partial z_2
\end{equation}
when acting on $F$.
We conclude that if we know the Taylor expansion of $A$ about the origin $(0,0)$, then we also know the Taylor expansion of $F$ about $(z_1^d,z_2^d)$, and vice versa.

As a concrete example, suppose that
\begin{equation}
F(z_1,z_2)=1 + 2 z_1 + 3 z_2 + 4 z_1^2 + 5 z_1 z_2 + 6 z_2^2
\end{equation}
and 
\begin{equation}
z_1^d=7,{\;}{\;}z_2^d=8.
\end{equation}
Then, hand calculation shows that
$F(G(x_1),H(x_2))$ takes the form
\begin{eqnarray}
F(z_1^d+x_1,z_2^d+x_2)&=&F(G(x_1),H(x_2))\nonumber\\
&=&899 + 98 x_1 + 4 x_1^2 + 134 x_2 + 5 x_1 x_2 + 6 x_2^2.
\end{eqnarray}
Below is a {\em Mathematica} execution that finds the same result,
\begin{eqnarray}
&&{\tt F[z1\_, z2\_] := 1 + 2{\;}z1 + 3{\;}z2 + 4{\;}z1^2 
+ 5{\;}z1{\;} z2 + 6{\;} z2^2}\nonumber\\
&&{\tt G[x1\_] := zd1 + x1;}\nonumber\\
&&{\tt H[x2\_] := zd2 + x2;}\nonumber\\
&&{\tt zd1 = 7;}\nonumber\\
&&{\tt zd2 = 8;}\nonumber\\
&&{\tt A=F[G[x1], H[x2]]}\nonumber\\
&&{\tt Expand[A]}\nonumber\\
&&1 + 2{\;}(7 + x1) + 4{\;}(7 + x1)^2 + 3{\;}(8 + x2) + 5{\;}(7 + x1){\;}(8 + x2) + 
 6{\;}(8 + x2)^2\nonumber\\
&& 899 + 98{\;}x1 + 4{\;}x1^2 + 134{\;}x2 + 5{\;}x1{\;}x2 + {\;} x2^2\nonumber\\
\end{eqnarray}

\newpage

The calculation above dealt with the case of a function of two ordinary variables.  We now illustrate, for the same example, that there is an analogous result for pyramids.  For future reference, Table 8 shows our standard modified glex sequencing applied to the case of two variables.

\begin{table}[h,t]
\caption{A labeling scheme for monomials in two variables.}
\begin{center}
\begin{tabular}{ccc}
$r$ & $j_1$ & $j_2$ \\ \hline
1 & 0 & 0  \\
2 & 1 & 0  \\
3 & 0 & 1 \\
4 & 2 & 0  \\
5 & 1 & 1  \\
6 & 0 & 2  \\
7 & 3 & 0  \\
8 & 2 & 1  \\
9 & 1 & 2  \\
10 & 0 & 3  \\
$\cdot$ & $\cdot$ & $\cdot$  \\
$\cdot$ & $\cdot$ & $\cdot$  \\
\end{tabular}
\end{center}
\end{table}

Following the replacement rule, we should make the substitutions
\begin{equation}
z^d_1 + x_1\leadsto{\tt zd1{\;}C1+X1},
\end{equation}
\begin{equation}
z^d_2 + x_2\leadsto{\tt zd2{\;}C1+X2},
\end{equation}
\begin{eqnarray}
&&1 + 2{\;}z_1 + 3{\;}z_2 + 4{\;}z_1^2 + 5{\;}z_1{\;}z_2 + 6{\;}z_2^2\leadsto\nonumber\\
&&{\tt C1+2{\;}Z1+3{\;}Z2+4{\;}PROD[Z1,Z1]+5{\;}PROD[Z1,Z2]+6{\;}PROD[Z2,Z2]}.
\nonumber\\
\end{eqnarray}
The {\em Mathematica} fragment below, executed for the case $m=2$ and $p=2$, in which case $L=6$, illustrates how the analogous result is obtained using pyramids,
\begin{eqnarray}
&&{\tt C1 = \{1, 0, 0, 0, 0, 0\};}\nonumber\\
&&{\tt X1 = \{0, 1, 0, 0, 0, 0\};}\nonumber\\
&&{\tt X2 = \{0, 0, 1, 0, 0, 0\};}\nonumber\\
&&{\tt F[Z1\_, Z2\_] := 
 C1 + 2{\;}Z1 + 3{\;}Z2 + 4{\;}PROD[Z1, Z1] + 5{\;}PROD[Z1, Z2]}\nonumber\\ 
 &&{\tt + 6{\;}PROD[Z2, Z2];}\nonumber\\
&&{\tt G[X1\_] := z01{\;}C1 + X1;}\nonumber\\
&&{\tt H[X2\_] := z02{\;}C1 + X2;}\nonumber\\
&&{\tt zd1 = 7;}\nonumber\\
&&{\tt zd2 = 8;}\nonumber\\
&&{\tt F[G[X1], H[X2]]}\nonumber\\
&&\{899, 98, 134, 4, 5, 6\}
\end{eqnarray}
Note that, when use is made of Table 8, the last line of (7.93) agrees with (7.88) and the last line of (7.89).

\subsection{Replacement Rule and Numerical Integration}
\subsubsection{Numerical Integration}
Consider the set of differential equations (1.5).  A standard procedure for their numerical integration from an initial time $t^i=t^0$ to some final time $t^f$ is to divide the time axis into a large number of steps $N$, each of small duration $h$, thereby introducing successive times $t^n$ defined by the relation
\begin{equation}
t^n = t^0 + nh {\;}{\;}{\rm{with}}{\;}{\;}n=0,1,\cdots,N.
\end{equation}
By construction, there will also be the relation
\begin{equation}
Nh=t^f-t^i.
\end{equation}
The goal is to compute the vectors $\bm{z}^n$, where 
\begin{equation}
\bm{z}^n = \bm{z} (t^n),
\end{equation}
starting from the vector $\bm{z}^0$.  The vector $\bm{z}^0$ is assumed given as a set of definite numbers, i.e. the
initial conditions at
$t^0$.  

If we assume Poincar\'{e} analyticity in $t$, we may convert the set of differential equations (1.5) into a set of recursion relations for the  
$\bm{z}^n$ in such a way that the $\bm{z}^n$ obtained by solving the recursion relations differ from the true $\bm{z}^n$ by only small truncation errors of order $h^m$.  (Here $m$ is {\em not} the number of variables, but rather some fixed integer describing the accuracy of the integration method.)  One such procedure, fourth-order {\em Runge Kutta} (RK4), is the set of marching/recursion rules
\begin{equation}
\bm{z}^{n+1} = \bm{z}^n + \frac{1}{6}
(\bm{a} + 2\bm{b} + 2\bm{c} +
\bm{d})
\end{equation}
where, at each step,
\begin{equation}
\bm{a} = h \bm{f} (\bm{z}^n, t^n),
\end{equation}
\[
\bm{b} = h \bm{f} (\bm{z}^n +
\frac{1}{2} \bm{a}, t^n + \frac{1}{2} h),
\]
\[
\bm{c} = h \bm{f} (\bm{z}^n +
\frac{1}{2} \bm{b}, t^n + \frac{1}{2} h),
\]
\[
\bm{d} = h \bm{f} (\bm{z}^n +
\bm{c}, t^n + h).
\]
Thanks to the genius of Runge and Kutta, the relations (7.97) and (7.98) have been constructed in such a way that the method is locally (at each step) correct through order $h^4$, and makes local truncation errors of order $h^5$.

In the case of a single variable, and therefore a single differential equation, the relations (7.97) and (7.98) may be encoded in the {\em Mathematica} form shown below.  Here ${\tt Zvar}$ is the dependent variable, $\tt t$ is the time, ${\tt Zt}$ is a temporary variable, ${\tt tt}$ is a temporary time, and $\tt ns$ is the number of integration steps. 
The program employs a Do loop over $\tt i$ so that the operations (7.97) and (7.98) are carried out ${\tt ns}$ times.

\begin{eqnarray}
&&\hspace{-.5cm}{\tt RK4:=(} \nonumber  \\
&&{\tt t0=t};\nonumber\\
&&\hspace{-.5cm}{\tt Do[             } \nonumber \\
&&{\tt Aa=h{\;}F[Zvar,t]; }\nonumber  \\
&&{\tt Zt=Zvar+(1/2)Aa;    }\nonumber  \\
&&{\tt tt=t+h/2;        }\nonumber  \\
&&{\tt Bb=h{\;}F[Zt,tt];    }\nonumber  \\
&&{\tt Zt=Zvar+(1/2)Bb;  } \nonumber  \\
&&{\tt Cc=h{\;}F[Zt,tt];    }\nonumber  \\
&&{\tt Zt=Zvar+Cc;    }\nonumber  \\
&&{\tt tt=t+h;}\nonumber  \\
&&{\tt Dd=h{\;}F[Zt,tt];   }\nonumber  \\
&&{\tt Zvar=Zvar+(1/6)(Aa+2{\;}Bb +2{\;}Cc+Dd);  }\nonumber  \\
&&{\tt t=t0+i{\;}h;},\nonumber  \\
&&{\tt \{i,1,ns,1\}  }\nonumber  \\
&&{\tt ] } \nonumber  \\
&&\hspace{-.3cm}{\tt ) } \nonumber\\  
\end{eqnarray}

\subsubsection{Replacement Rule, Single Equation/Variable Case}

We now make what, for our purposes, is a fundamental observation:  The operations that occur in the Runge Kutta recursion rules (7.97) and (7.98) and realized in the code above can be extended to pyramids by application of the replacement rule.  In particular, the dependent variable $\bm{z}$ can be replaced by a pyramid, and the various operations involved in the recursion rules can be replaced by pyramid operations.  Indeed if we look at the code above, apart from the evaluation of $\tt F$, we see that the quantities ${\tt Zvar}$, ${\tt Zt}$, ${\tt Aa}$, ${\tt Bb}$, ${\tt Cc}$, and ${\tt Dd}$ can be viewed, if we wish, as pyramids since the only operations involved are scalar multiplication and addition.  The only requirement for a pyramidal interpretation of the {\tt RK4} {\em Mathematica} code is that the right side of the differential equation, ${\tt F[*,*]}$, be defined for pyramids.  Finally, we remark that the features that make it possible to interpret the {\tt RK4} {\em Mathematica} code either in terms of ordinary variables or pyramidal variables will hold for {\em Mathematica} realizations of many other familiar numerical integration methods including other forms of Runge Kutta, predictor-corrector methods, and extrapolation methods.

To make these ideas concrete, and to understand their implications, let us begin with a simple example.  Suppose, in the single variable case, that the right side of the differential equation has the simple form
\begin{equation}
f(z,t)=-2tz^2.
\end{equation}
The differential equation with this right side can be integrated analytically to yield the solution
\begin{equation}
z(t)=z^0/[1+z^0(t-t^0)^2].
\end{equation}
In particular, for the case $t^0=0$, $z^0=1$, and $t=1$, there is the result
\begin{equation}
z(1)=z^0/[1+z^0]=1/2.
\end{equation}

Let us also integrate the differential equation with the right side (7.100) 
numerically.  Shown below is the result of running the associated {\em Mathematica} Runge Kutta code for this case.

\begin{eqnarray}
&&{\tt Clear[\texttt{"}Global`*\texttt{"}];}\nonumber  \\
&&{\tt F[Z\_,t\_]:=-2{\;} t{\;} Z^2;}\nonumber\\
&&{\tt h=.1;}\nonumber\\
&&{\tt ns=10;}\nonumber\\
&&{\tt t=0;}\nonumber\\
&&{\tt Zvar=1.;}\nonumber\\
&&{\tt RK4;}\nonumber\\
&&{\tt t}\nonumber\\
&&{\tt Zvar}\nonumber\\
&&1.\nonumber\\
&&0.500001\nonumber\\
\end{eqnarray}
Note that the last line of (7.103) agrees with (7.102) save for a ``1" in the last entry.  As expected, and as experimentation shows, this small difference, due to accumulated truncation error, becomes even smaller if ${\tt h}$ is decreased (and correspondingly, ${\tt ns}$ is increased).

Suppose we expand the solution (7.102) about the design initial condition  $z^{d0}=1$ by replacing $z^0$ by $z^{d0}+x$ and expanding the result in a Taylor series in $x$ about the point $x$=0.  Below is a {\em Mathematica} run that performs this task.
\begin{eqnarray}
&&{\tt zd0 = 1;}\nonumber\\
&&{\tt Series[(zd0 + x)/(1 + zd0 + x), \{x, 0, 5\}]}\nonumber\\
&&\frac{1}{2}+\frac{x}{4}-\frac{x^2}{8}+\frac{x^3}{16}-\frac{x^4}{32}+\frac{x^5}{64}+O[x]^6\nonumber\\
\end{eqnarray}

We will now see that the same Taylor series can be obtained by the operation of numerical integration applied to pyramids.
The {\em Mathematica} code below shows, for our example differential equation, the application of numerical integration to pyramids.  
\begin{eqnarray}
&&{\tt Clear[\texttt{"}Global`*\texttt{"}];}\nonumber\\
&&{\tt Needs[\texttt{"}Combinatorica`\texttt{"}];}\nonumber\\
&&{\tt m = 1; p = 5;}\nonumber\\
&&{\tt GAMMA = Compositions[0, m];}\nonumber\\
&&{\tt Do[GAMMA = Join[GAMMA, Reverse[Compositions[d, m]]], \{d, 1, p, 1\}];}\nonumber\\
&&{\tt L = Length[GAMMA];}\nonumber\\
&&{\tt JSK[list\_,k\_]\;:=}\nonumber\\
&&{\tt Position[Apply[And,Thread[\texttt{\#1\;<=\;\#2\;\&\;}[\texttt{\#},k]]]\;\texttt{\&\;/@}\;list,True]\texttt{//}Flatten;}\nonumber\\
&&{\tt B = Table[JSK[GAMMA, GAMMA[[r]]], \{r, 1, L, 1\}];}\nonumber\\
&&{\tt Brev = Reverse \texttt{/@}\; B;}\nonumber\\
&&{\tt PROD[U\_, V\_] := Table[U[[B[[k]]]].V[[Brev[[k]]]], \{k, 1, L, 1\}];}\nonumber\\
&&{\tt F[Z\_, t\_] := -2{\;}t{\;}PROD[Z, Z];}\nonumber\\
&&{\tt h = .01;}\nonumber\\
&&{\tt ns = 100;}\nonumber\\
&&{\tt t = 0;}\nonumber\\
&&{\tt zd0 = 1;}\nonumber\\
&&{\tt C1 = \{1, 0, 0, 0, 0, 0\};}\nonumber\\
&&{\tt X = \{0, 1, 0, 0, 0, 0\};}\nonumber\\
&&{\tt Zvar = zd0{\;}C1 + X;}\nonumber\\
&&{\tt RK4;}\nonumber\\
&&{\tt t}\nonumber\\
&&{\tt Zvar}\nonumber\\
&&1.\nonumber\\
&&\{0.5, 0.25, -0.125, 0.0625, -0.03125, 0.015625\}
\end{eqnarray}
The first 11 lines of the code set up what should be by now the familiar procedure for labeling and multiplying pyramids.  In particular, $m=1$ because we are dealing with a single variable, and $p=5$ since we wish to work through fifth order.  The line 
\begin{equation}
{\tt F[Z\_, t\_] := -2{\;}t{\;}PROD[Z, Z]}
\end{equation}
defines ${\tt F[*,*]}$ for the case of pyramids, and is the result of applying the replacement rule to the right side of $f$ as given by (7.100), 
\begin{equation}
-2{\;}t{\;}z^2\leadsto -2{\;}{\tt t}{\;}{\tt PROD[Z, Z]}.
\end{equation}
Lines 13 through 15 are the same as lines 3 through 5 in (7.103) except that, in order to improve numerical accuracy, the step size $\tt h$ has been decreased and correspondingly the number of steps ${\tt ns}$ has been increased.  Lines 16 through 19 now initialize ${\tt Zvar}$ as a pyramid with constant part ${\tt zd0}$ and first-order monomial part 1, 
\begin{equation}
{\tt Zvar = zd0{\;}C1 + X}.
\end{equation}
These lines are the pyramid equivalent of line 6 in (7.103).  Finally lines 20 through 22 are the same as lines 7 through 9 in (7.103).  In particular, the line {\tt RK4} in (7.103) and the line {\tt RK4} in (7.105) refer to exactly the {\em same} code, namely that in (7.99).  

Let us now compare the outputs of (7.103) and (7.105).  Comparing the penultimate lines in each we see that the final time $t=1$ is the same in each case.  Comparing the last lines shows that the output ${\tt Zvar}$ for (7.105) is a pyramid whose first entry agrees with the last line of (7.103).  Finally, all the entries in the pyramid output agree with the Taylor coefficients in the expansion (7.104).  We see, in the case of numerical integration (of a single differential equation), that replacing the dependent variable by a pyramid, with the initial value of the pyramid given by (7.108), produces a Taylor expansion of the final condition in terms of the initial condition. 

What accounts for this near miraculous result?  It's the Taylor rule described at the end of Section 7.1.6.  We have already learned that to expand some function $F(z)$ about some point $z^d$ we must evaluate $F(z^d+x)$.  See (7.74).  We know that the final $Zvar$, call it $Zvar^{\rm{fin}}$, is an analytic function of the initial $Zvar$, call it $Zvar^{\rm{in}}$, so that we may write
\begin{equation}
Zvar^{\rm{fin}}=Zvar^{\rm{fin}}(Zvar^{\rm{in}})=g(Zvar^{\rm{in}})
\end{equation}
where $g$ is the function that results from following the trajectory from $t=t^{\rm{in}}$ to $t=t^{\rm{fin}}$.  Therefore, by the Taylor rule, to expand $Zvar^{\rm{fin}}$ about $Zvar^{\rm{in}}=z^{d0}$, we must evaluate $Zvar^{\rm{fin}}(z^{d0}+x)$.  That, with the aid of pyramids, is
what the code (7.105) accomplishes.

\subsubsection{Multi Equation/Variable Case}

Because of {\em Mathematica's} built-in provisions for handling arrays, the work of the previous section can easily be extended to the case of several differential equations.  Consider, as an example, the two-variable case for which $\bm{f}$ has the form
\begin{eqnarray}
&&f_1(\bm{z},t)=-z_1^2,\nonumber\\
&&f_2(\bm{z},t)=+2z_1z_2.
\end{eqnarray}
The differential equations associated with this $\bm{f}$ can be solved in closed form to yield, with the understanding that $t^0=0$, the solution
\begin{eqnarray}
&&z_1(t)=z_1^0/(1+tz_1^0),\nonumber\\
&&z_2(t)=z_2^0(1+tz_1^0)^2.
\end{eqnarray}
For the final time $t=1$ we find the result
\begin{eqnarray}
&&z_1(1)=z_1^0/(1+z_1^0),\nonumber\\
&&z_2(1)=z_2^0(1+z_1^0)^2.
\end{eqnarray}

Let us expand the solution (7.112) about the design initial conditions 
\begin{eqnarray}
&&z_1^{d0}=1,\nonumber\\
&&z_2^{d0}=2,
\end{eqnarray}
by writing
\begin{eqnarray}
&&z^0_1=z_1^{d0}+x_1=1+x_1,\nonumber\\
&&z^0_2=z_2^{d0}+x_2=2+x_2.
\end{eqnarray}
Doing so gives the results
\begin{eqnarray}
z_1(1)&=&(1+x_1)/(2+x_1)=(2+x_1-1)/(2+x_1)=1-1/(2+x_1)=\nonumber\\
&=&1-(1/2)(1+x_1/2)^{-1}=1-(1/2)[1-x_1/2+(x_1/2)^2-(x_1/2)^3+\cdots\nonumber\\
&=&(1/2)+(1/4)x_1-(1/8)x_1^2+(1/16)x_1^3+\cdots,\nonumber\\
\end{eqnarray}
\begin{eqnarray}
z_2(1)&=&(2+x_2)(2+x_1)^2\nonumber\\
&=&8+8x_1+4x_2+2x_1^2+4x_1x_2+x_1^2x_2.
\end{eqnarray}

We will now explore how this same result can be obtained using the replacement rule applied to the operation of numerical integration.  As before, we will label individual monomials by an integer $r$.  Recall that Table 8 shows our standard modified glex sequencing applied to the case of two variables.

The {\em Mathematica} code below shows, for our two-variable example differential equation, the application of numerical integration to pyramids.  Before describing the code in some detail, we take note of the bottom two lines.  When interpreted with the aid of Table 8, we see that the penultimate line of (7.117) agrees with (7.115), and the last line of (7.117) nearly agrees with (7.116).  The only discrepancy is that for the monomial with label $r=7$ in the last line of (7.117).  In the {\em Mathematica} output it has the value $-1.16563\times10^{-7}$ while, according to (7.116), the true value should be zero.  This small discrepancy arises from the truncation error inherent in the RK4 algorithm, and becomes smaller as the step size {\tt h} is decreased (and {\tt ns} is correspondingly increased), or if some more accurate integration algorithm is used.  We conclude that, with the use of pyramids, it is also possible in the two-variable case to obtain Taylor expansions of the final conditions in terms of the initial  conditions.  Indeed, what is involved is again the Taylor rule applied, in this instance, to the case of two variables.

\begin{eqnarray}
&&{\tt Clear[\texttt{"}Global`*\texttt{"}];}\nonumber\\
&&{\tt Needs[\texttt{"}Combinatorica`\texttt{"}];}\nonumber\\
&&{\tt m = 2; p = 3;}\nonumber\\
&&{\tt GAMMA = Compositions[0, m];}\nonumber\\
&&{\tt Do[GAMMA = Join[GAMMA, Reverse[Compositions[d, m]]], \{d, 1, p, 1\}];}\nonumber\\
&&{\tt L = Length[GAMMA];}\nonumber\\
&&{\tt JSK[list\_,k\_]\;:=}\nonumber\\
&&{\tt Position[Apply[And,Thread[\texttt{\#1\;<=\;\#2\;\&\;}[\texttt{\#},k]]]\;\texttt{\&\;/@}\;list,True]\texttt{//}Flatten;}\nonumber\\
&&{\tt B = Table[JSK[GAMMA, GAMMA[[r]]], \{r, 1, L, 1\}];}\nonumber\\
&&{\tt Brev = Reverse \texttt{/@}\; B;}\nonumber\\
&&{\tt PROD[U\_, V\_] := Table[U[[B[[k]]]].V[[Brev[[k]]]], \{k, 1, L, 1\}];}\nonumber\\
&&{\tt F[Z\_, t\_] := \{-PROD[Z[[1]], Z[[1]]], 2.{\;}PROD[Z[[1]], Z[[2]]]\};}\nonumber\\
&&{\tt h = .01;}\nonumber\\
&&{\tt ns = 100;}\nonumber\\
&&{\tt t = 0;}\nonumber\\
&&{\tt zd0 = \{1.,2.\};}\nonumber\\
&&{\tt C1 = Table[KroneckerDelta[k,1],\{k,1,L,1\}];}\nonumber\\
&&{\tt X[1] = Table[KroneckerDelta[k,2],\{k,1,L,1\}];}\nonumber\\
&&{\tt X[2] = Table[KroneckerDelta[k,3],\{k,1,L,1\}];}\nonumber\\
&&{\tt Zvar = \{zd0[[1]]{\;}C1 + X[1],zd0[[2]]{\;}C1 + X[2]\};}\nonumber\\
&&{\tt RK4;}\nonumber\\
&&{\tt t}\nonumber\\
&&{\tt Zvar}\nonumber\\
&&1.\nonumber\\
&&\{\{0.5, 0.25,0., -0.125, 0., 0., 0.0625, 0.,0.,0,\},\nonumber\\
&&{\;}{\;}\{8.,8.,4.,2.,4.,0.,-1.16563\times10^{-7},1.,0.,0.\}\}
\end{eqnarray}

Let us compare the structures of the routines for the single variable case and multi (two) variable case as illustrated in (7.105) and (7.117).  The first difference occurs at line 3 where the number of variables $m$ and the maximum degree $p$ are specified.  In (7.117) $m$ is set to 2 because we wish to treat the case of two variables, and $p$ is set to 3 simply to limit the lengths of the output arrays.  The next difference occurs in line 12 where the right side $\tt F$ of the differential equation is specified.  The major feature of the definition of $\tt F$ in (7.117) is that it is specified as two pyramids because the right side of the definition has the structure ${\tt \{*,*\}}$ where each item $ *$ is an instruction for computing a pyramid.  In particular, the two pyramids are those for the two components of  $\bmv f$ as given by (7.110) and use of the replacement rule,
\begin{equation}
-z_1^2\leadsto {\tt -PROD[Z[[1]], Z[[1]]]},
\end{equation}
\begin{equation}
2z_1z_2\leadsto 2.{\tt {\;}PROD[Z[[1]], Z[[2]]]}.
\end{equation}
The next differences occur in lines 16 through 20 of (7.117).  In line 16, since specification of the initial conditions now requires two numbers, see (7.113), 
${\tt zd0}$ is specified as a two-component array.  In lines 17 and 18 of (7.105) the pyramids ${\tt C1}$ and $\tt X$ are set up explicitly for the case $p=5$.  By contrast, in lines 17 through 19 of (7.117), the pyramids ${\tt C1}$, ${\tt X[1]}$, and ${\tt X[2]}$ are set up for general $p$ with the aid of the Table command and the Kronecker delta function.  Recall (7.62) and observe from Tables 4, 7, and 8 that, no matter what the values of $m$ and $p$, the constant monomial has the label $r=1$ and the monomial $x_1$ has the label $r=2$.  Moreover, as long as $m\ge2$ and no matter what the value of $p$, the $x_2$ monomial has the label $r=3$.  Finally, compare line 19 in (7.105) with line 20 in (7.117), both of which 
define the initial ${\tt Zvar}$.  We see that the difference is that in (7.105) ${\tt Zvar}$ is defined as a single pyramid while in (7.117) it is defined as a pair of pyramids of the form $\{*,*\}$.  Most remarkably, all other corresponding lines in (7.105) and (7.117) are the same.  In particular, the {\em same} RK4 code, namely that given by (7.99), is used in the scalar case (7.103), the single pyramid case (7.105), and the two-pyramid case (7.117).  This multi-use is possible because of the convenient way in which {\em Mathematica} handles arrays.  

We conclude that the pattern for the multivariable case is now clear. Only the following items need to be specified in an $m$ dependent way:
\begin{itemize}
\item The value of $m$.
\item The entries in $\tt F$ with entries entered as an array $\{*,*,\cdots\}$ of $m$ pyramids.
\item The design initial condition array ${\tt zd0}$.
\item The pyramids for ${\tt C1}$ and ${\tt X[1]}$ through ${\tt X[m]}$.
\item The entries for the initial ${\tt Zvar}$ specified as an array 

${\tt \{zd0[[1]]} \;{\tt C1 + X[1]}{\tt ,zd0[[2]]} \;{\tt C1 + X[2]}{\tt ,\cdots,zd0[[m]]} \;{\tt C1 + X[m]\}}$ of $m$ pyramids.
\end{itemize}
 
\subsection{Duffing Equation Application}

Let us now apply the methods just developed to the case of the Duffing equation with parameter dependence as described by the relations (6.12) through (6.17).  {\em Mathematica} code for this purpose is shown below.  By looking at the final lines that result from executing this code, we see that the final output is an array of the form ${\tt \{\{*\},\{*\},\{*\}\}}$.  That is, the final output is an array of three pyramids.  This is what we expect, because now we are dealing with three variables.  See line 3 of the code, which sets m=3.  Also, for convenience of viewing, results are calculated and displayed only through third order as a consequence of setting ${\tt p=3}$. 

\begin{eqnarray*}
&&{\tt Clear[\texttt{"}Global`*\texttt{"}];}\nonumber\\
&&{\tt Needs[\texttt{"}Combinatorica`\texttt{"}];}\nonumber\\
&&{\tt m = 3; p = 3;}\nonumber\\
&&{\tt GAMMA = Compositions[0, m];}\nonumber\\
&&{\tt Do[GAMMA = Join[GAMMA, Reverse[Compositions[d, m]]], \{d, 1, p, 1\}];}\nonumber\\
&&{\tt L = Length[GAMMA];}\nonumber\\
&&{\tt JSK[list\_,k\_]\;:=}\nonumber\\
&&{\tt Position[Apply[And,Thread[\texttt{\#1\;<=\;\#2\;\&\;}[\texttt{\#},k]]]\;\texttt{\&\;/@}\;list,True]\texttt{//}Flatten;}\nonumber\\
&&{\tt B = Table[JSK[GAMMA, GAMMA[[r]]], \{r, 1, L, 1\}];}\nonumber\\
&&{\tt Brev = Reverse \texttt{/@}\; B;}\nonumber\\
&&{\tt PROD[U\_, V\_] := Table[U[[B[[k]]]].V[[Brev[[k]]]], \{k, 1, L, 1\}];}\nonumber\\
&&{\tt POWER[U\_, 2] := PROD[U, U];}\nonumber\\
&&{\tt POWER[U\_, 3] := PROD[U, POWER[U, 2]];}\nonumber\\
&&{\tt C0 = Table[0, \{k, 1, L, 1\}];}\nonumber\\
&&{\tt F[Z\_, t\_] := \{Z[[2]],}\nonumber\\
&&{\tt -2.{\;}beta{\;}PROD[Z[[3]], Z[[2]]] - PROD[POWER[Z[[3]], 2], Z[[1]]] -}\nonumber\\
&&{\tt POWER[Z[[1]], 3] - eps{\;}Sin[t]{\;}POWER[Z[[3]], 3],}\nonumber\\
&&{\tt C0\};}\nonumber\\
\end{eqnarray*}
\begin{eqnarray}
&&{\tt ns = 100;}\nonumber\\
&&{\tt t = 0;}\nonumber\\
&&{\tt h = (2 Pi)/ns;}\nonumber\\
&&{\tt beta = .1;eps = 1.5;}\nonumber\\
&&{\tt zd0 = \{.3,.4,.5\};}\nonumber\\
&&{\tt C1 = Table[KroneckerDelta[k,1],\{k,1,L,1\}];}\nonumber\\
&&{\tt X[1] = Table[KroneckerDelta[k,2],\{k,1,L,1\}];}\nonumber\\
&&{\tt X[2] = Table[KroneckerDelta[k,3],\{k,1,L,1\}];}\nonumber\\
&&{\tt X[3] = Table[KroneckerDelta[k, 4], \{k, 1, L, 1\}];}\nonumber\\
&&{\tt Zvar = \{zd0[[1]]{\;}C1 + X[1], zd0[[2]]{\;}C1 + X[2], zd0[[3]]{\;}C1 + X[3]\};}\nonumber\\
&&{\tt RK4;}\nonumber\\
&&{\tt t}\nonumber\\
&&{\tt Zvar}\nonumber\\
&&2\pi\nonumber\\
&&\{\{-0.0493158, 0.973942, -0.110494, 5.51271, 3.54684, 3.46678,
\nonumber\\
&&{\;}{\;}{\;}11.2762, 2.36463, 1.0985, 23.3332, -1.03541, -3.23761, -12.8064,
\nonumber\\
&&{\;}{\;}{\;}4.03421, -23.4342, -17.8967, 1.96148, 5.07403, -36.9009, 25.1379\},\nonumber\\
&&{\;}{\;}\{0.439713, 1.05904, 0.427613, 3.3177, 0.0872459, 0.635397, -3.02822,
\nonumber\\
&&{\;}{\;}{\;}1.77416, -4.10115, 3.16981, -2.43002, -5.33643, -7.77038, -6.08476,\nonumber\\
&&{\;}{\;}{\;}-0.541465, -21.1672, -1.4091, -9.54326, 14.6334, -39.2312\},\nonumber\\
&&{\;}{\;}\{0.5, 0, 0, 1, 0, 0, 0, 0, 0, 0, 0, 0, 0, 0, 0, 0, 0, 0, 0, 0\}\}\nonumber\\
\end{eqnarray}
  
The first unusual fragments in the code are lines 12 and 13, which define functions that implement the calculation of second and third powers of pyramids.  Recall Section 7.1.5.  The first new fragment is line 14, which defines the pyramid ${\tt C0}$ with the aid of the Table command and an implied Do loop.  As a result of executing this code, ${\tt C0}$ is an array of $L$ zeroes.  The next three lines, lines 15 through 18, define $\tt F$, which specifies the right sides of equations (6.12) through (6.14).  See (6.15) through (6.18).  The right side of $\tt F$ is of the form ${\tt \{*,*,*\}}$, an array of three pyramids.  By looking at (6.15) and recalling the replacement rule, we see that the first pyramid should be ${\tt Z[[2]]}$,
\begin{equation}
z_2\leadsto {\tt Z[[2]]}.
\end{equation}
The second pyramid on the right side of $\tt F$ is more complicated.  It arises by applying the replacement rule to the right side of (6.16) to obtain the associated pyramid,
\begin{eqnarray}
&&- \ 2\beta z_3z_2 -z^2_3z_1-z^3_1-\epsilon z^3_3 \sin t\leadsto\nonumber\\
&&{\tt -2.{\;}beta{\;}PROD[Z[[3]], Z[[2]]] - PROD[POWER[Z[[3]], 2], Z[[1]]] -}\nonumber\\
&&{\tt POWER[Z[[1]], 3] - eps{\;}Sin[t]{\;}POWER[Z[[3]], 3]}.
\end{eqnarray}
The third pyramid on the right side of $F$ is simplicity itself.  From (6.17) we see that this pyramid should be the result of applying the replacement rule to the number $0$.  Hence, this pyramid is $\tt C0$,
\begin{equation}
0\leadsto {\tt C0=\{0,0,\cdots,0\}}.
\end{equation}

The remaining lines of the code require little comment.  Line 20 sets the initial time to $0$, and line 21 defines $h$ in such a way that the final value of $t$ will be $2\pi$.  Line 22 establishes the parameter values $\beta=.1$ and $\epsilon=1.5$, which are those for Figure 4.
Line 23 specifies that the design initial condition is
\begin{equation}
z_1(0)=z_1^{d0}=.3,{\;}z_2(0)=z_2^{d0}=.4,{\;}z_3(0)=z_3^{d0}=.5=\sigma,
\end{equation}
and consequently
\begin{equation}
\omega=1/\sigma=2.
\end{equation}
See (6.3).  Also, it follows from (6.2) and (6.5) that
\begin{equation}
q(0)=\omega Q(0)=\omega z_1(0)=(2)(.3)=.6,
\end{equation}
\begin{equation}
q^\prime(0)=\omega^2\dot{Q}(0)=\omega^2z_2(0)=(2^2)(.4)=1.6.
\end{equation}
Next, lines 24 through 28 specify that the expansion is to be carried out about the initial conditions (7.124).  Finally, line 29 invokes the {\tt RK4} code given by (7.99).  That is, as before, {\em no} modifications are required in the integration code.

A few more comments about the output are appropriate.  Line 32 shows that the final time $t$ is indeed $2\pi$, as desired.  The remaining output lines display the three pyramids that specify the final value of {\tt Zvar}.  From the first entry in each pyramid we see that 
\begin{equation}
z_1(2\pi)=-0.0493158,
\end{equation}
\begin{equation}
z_2(2\pi)=0.439713,
\end{equation}
\begin{equation}
z_3(2\pi)=.5,
\end{equation}
when there are no deviations in the initial conditions.  The remaining entries in the pyramids are the coefficients in the Taylor series that describe the changes in the final conditions that occur when changes are made in the initial conditions (including the parameter $\sigma$).
We are, of course, particularly interested in the first two pyramids.  The third pyramid has entries only in the first place and the fourth place,
and these entries are the same as those in the third pyramid pyramid for ${\tt Zvar}$ at the start of the integration, namely those in ${\tt zd0[3]{\;}C1+X[3]}$.  The fact that the third pyramid in ${\tt Zvar}$ remains constant is the expected consequence of (6.17). 

We should also describe how the ${\cal{M}}_8$ employed in Section 6.2 was actually computed.  It could have been computed by setting $p=8$ in (7.120) and specifying a large number of steps $ns$ to insure good accuracy.  Of course, when $p=8$, the pyramids are large.  Therefore, one does not usually print them out, but rather writes them to files or sends them directly to other programs for further use.

However, rather than using RK4 in (7.120), we replaced it with an 
adaptive 4-$5^{\rm{th}}$ order Runge-Kutta-Fehlberg routine that dynamically adjusts the time step $h$ during the course of integration to achieve a specified local accuracy, and we required that the error at each step be no larger than $10^{-12}$.  Like the RK4 routine, the Runge-Kutta-Fehlberg routine, when implemented in {\em Mathematica}, has the property that it can integrate any number of equations both in scalar variable and pyramid form without any changes in the code.\footnote{A {\em Mathematica} version of this code is available from the first author upon request.}  

\subsection{Relation to the Complete Variational Equations}

At this point it may not be obvious to the reader that the use of pyramids in integration routines to obtain Taylor expansions is the same as integrating the complete variational equations.  We now show that 
the integration of pyramid equations is equivalent to the forward integration of the complete variational equations.  For simplicity, we will examine the single variable case with no parameter dependence.  The reader  who has mastered this case should be able to generalize the results obtained to the general case.

In the single variable case with no parameter dependence (2.1) becomes
\begin{equation}
\dot{z}=f(z,t).
\end{equation}
Let $z^d(t)$ be some design solution and introduce a deviation variable $\zeta$ by writing
\begin{equation}
z=z^d+\zeta.
\end{equation}
Then the equation of motion (7.131) takes the form
\begin{equation}
\dot{z}^d+\dot{\zeta}=f(z^d+\zeta,t).
\end{equation}
Also, the relations (2.4) and (2.5) take the form
\begin{equation}
f(z^d+\zeta,t)=f(z^d,t)+g(z^d,t,\zeta)
\end{equation}
where $g$ has an expansion of the form
\begin{equation}
g(z^d,t,\zeta)=\sum_{j=1}^\infty g^j(t)\zeta^j.
\end{equation}
Finally, (2.6) and (2.7) become
\begin{equation}
\dot{z}^d=f(z^d,t),
\end{equation}
\begin{equation}
\dot{\zeta}=g(z^d,t,\zeta)=\sum_{j=1}^\infty g^j(t)\zeta^j,
\end{equation}
and (2.8) becomes
\begin{equation}
\zeta=\sum_{j=1}^\infty h^j(t)(\zeta_i)^j.
\end{equation}
Insertion of (7.138) into both sides of (7.137) and equating like powers of $\zeta_i$ now yields the set of differential equations
\begin{equation}
\dot{h}^{j^{\prime \prime}} (t)  = \sum_{j=1}^\infty g^j(t) U^{j^{\prime \prime}}_j(h^s){\;}{\rm{with}}{\;}j,j^{\prime\prime}\ge1
\end{equation}
where the (universal) functions $U^{j^{\prime \prime}}_j(h^s)$ are given by the relations
\begin{equation}
\left(\sum_{j^\prime=1}^\infty h^{j^\prime}(\zeta_i)^{j^\prime}\right)^j=
\sum_{j^{\prime\prime}=1}^\infty U^{j^{\prime \prime}}_j (h^s)(\zeta_i)^{j^{\prime \prime}}.
\end{equation}
The equations (7.136) and (7.139) are to be integrated from $t=t^{\rm{in}}=t^0$ to $t=t^{\rm{fin}}$ with the initial conditions
\begin{equation}
z^d(t^0)=z^{d0},
\end{equation}
\begin{equation}
h^1(t^0)=1,
\end{equation}
\begin{equation}
h^{j^{\prime\prime}}(t^0)=0{\;}{\rm{for}}{\;}j^{\prime\prime}>1.
\end{equation}

Let us now consider the numerical integration of pyramids.  Upon some reflection, we see that the numerical integration of pyramids is equivalent to finding the numerical solution to a differential equation with pyramid arguments.  For example, in the single-variable case, let ${\tt {Zvar}}(t)$ be the pyramid appearing in the integration process.  Then, its integration is equivalent to solving numerically the 
pyramid differential equation
\begin{equation}
(d/dt){\tt {Zvar}}(t)={\tt {F}}({\tt {Zvar}},t).
\end{equation}

We now work out the consequences of this observation.  By the inverse of the replacement rule, we may associate a Taylor series with the pyramid ${\tt {Zvar}}(t)$ by writing
\begin{equation}
{\tt {Zvar}}(t)\leadsto c_0(t)+\sum_{j\ge 1}c_j(t)x^j.
\end{equation}
By (1.45) it is intended that the entries in the pyramid  
${\tt {Zvar}}(t)$ be used to construct a corresponding Taylor series with variable $x$.  
In view of (7.108), there are the initial conditions 
\begin{equation}
c_0(t_0)=z^d(t_0),
\end{equation}
\begin{equation}
c_1(t_0)=1,
\end{equation}
\begin{equation}
c_j(t_0)=0{\;}{\rm{for}}{\;}j>1.
\end{equation}

We next seek the differential equations that determine the time evolution of the $c_j(t)$.
Under the inverse replacement rule, there is also the correspondence 
\begin{equation}
(d/dt){\tt {Zvar}}(t)\leadsto {\dot{c}}_0(t)+\sum_{j\ge 1}{\dot{c}}_j(t)x^j.
\end{equation}
We have found a representation for the left side of (7.144).  We need to do the same for the right side.
That is, we need the Taylor series associated with the pyramid 
${\tt {F}}({\tt {Zvar}},t)$.  By the inverse replacement rule, it will be given by the relation
\begin{equation}
{\tt {F}}({\tt {Zvar}},t)\leadsto 
f(\sum_{j\ge0}c_j(t)x^j,t).
\end{equation}
Here  it is understood that the right side of (7.150) is to be expanded in a Taylor series about $x=0$.  From (7.134), (7.135), and (7.140) we have the relations
\begin{eqnarray} 
f(\sum_{j\ge0}c_j(t)x^j,t)&=&f(c_0(t))+g(c_0(t),t,\sum_{j\ge1}c_j(t)x^j)\nonumber\\
&=&f(c_0(t))+\sum_{k\ge1}g^k(t)(\sum_{j\ge1}c_j(t)x^j))^k\nonumber\\
&=&f(c_0(t))+\sum_{k\ge1}g^k(t)\sum_{j\ge1}U_k^j(c_\ell)x^j.\nonumber\\
\end{eqnarray}
Therefore, there is the inverse replacement rule
\begin{equation}
{\tt {F}}({\tt {Zvar}},t)\leadsto 
f(c_0(t))+\sum_{k\ge1}g^k(t)\sum_{j\ge1}U_k^j(c_\ell)x^j.
\end{equation}

Upon comparing like powers of $x$ in (7.149) and 
(7.152), we see that the pyramid differential equation 
(7.144) is equivalent to the set of differential equations 
\begin{equation}
{\dot{c}}_0(t)=f(c_0(t)),
\end{equation}
\begin{equation}
{\dot{c}}_j(t)=\sum_{k\ge1}g^k(t)U_k^j(c_\ell).
\end{equation}

Finally, compare the initial conditions (7.141) through (7.143) with the initial conditions (7.146) through (7.148), and compare the differential equations (7.136) and (7.139) with the differential equations (7.153) and (7.154).  We conclude that that there must be the relations
\begin{equation}
c_0(t)=z^d(t),
\end{equation}
\begin{equation}
c_j(t)=h^j(t){\;}{\rm{for}}{\;}j\ge1.
\end{equation}
We have verified, in the single variable case, that the use of pyramids in integration routines is equivalent to the solution of the complete variational equations using forward integration.  As stated earlier, verification of the analogous $m$-variable result is left to the reader.  We also observe the wonderful  convenience that, when pyramid operations are implemented and employed, it is not necessary to explicitly work out the forcing terms $g_a^r(t)$ and the functions $U^{r^{\prime\prime}}_r(h^s_n)$, nor is it necessary to set up the equations (3.6).  All these complications are handled implicitly and automatically by the pyramid routines.

\section{Concluding Summary} 
\setcounter{equation}{0}
Poincar\'{e} analyticity implies that transfer maps arising from ordinary differential equations can be expanded as Taylor series in the initial conditions and also in whatever parameters may be present.  Section 2 showed that the determination of these expansions is equivalent to solving the complete variational equations, and Sections 3 and 4 showed that the complete variational equations can be solved either by forward or backward integration.  Sections 5 and 6 applied this procedure for the Duffing stroboscopic map and found, remarkably,  that an $8^{\rm{th}}$ order polynomial approximation to this map produced an infinite period doubling cascade and apparent strange attractor that closely resembled those of the exact map.  A final section described computer methods for automatically setting up and numerically integrating the complete variational equations.


\begin{thebibliography}{99}
\bibitem{} J. Barrow-Green, {\em Poincar\'{e} and the Three Body Problem}, American Mathematical Society (1997).
\bibitem{} F. Browder, Edit., {\em The Mathematical Heritage of Henri Poincar\'{e}, Proceedings of Symposia in Pure Mathematics of the American Mathematical Society} {\bf{39}}, Parts 1 and 2,  American Mathematical Society (1983).
\bibitem{} H. Poincar\'{e}, {\em New Methods of Celestial Mechanics,} Parts 1, 2, and 3.  (Originally published as  Les M\'{e}thodes nouvelles de la M\'{e}chanique c\'{e}leste.)  American Institute of Physics {\em History of Modern Physics and Astronomy}, Volume 13, D. L. Goroff, Edit., American Institute of Physics (1993).
\bibitem{} Francis J. Murray and Kenneth S. Miller, {\em Existence Theorems for Ordinary Differential Equations}, New York University Press and Interscience Publishing Co. (1954).
\bibitem{} The method of backward integration was discovered by F. Neri circa 1986.
\bibitem{} A. Dragt, {\em Lie Methods for Nonlinear Dynamics with Applications to Accelerator Physics},  (2011), available at http://www.physics.umd.edu/dsat/
\bibitem{neidinger} R.~Neidinger, ``Computing Multivariable Taylor 
Series to Arbitrary Order'', {\em Proc. of Intern. Conf. on Applied programming 
languages}, San Antonio,  pp.~134-144  (1995).

\bibitem{mathematica} Wolfram Research, Inc., {\em Mathematica}, Version 7.0, Champaign, IL (2008).

\bibitem{kalman} D. Kalman and R. Lindell, 
``A recursive approach to multivariate automatic differentiation", 
{\em Optimization Methods and Software}, Volume 6, Issue 3,  pp.~161-192 (1995).
\bibitem{berz} M. Berz, ``Differential algebraic description of 
beam dynamics to very high orders'', {\em Particle Accelerators} 24, p. 109 (1989).
\end{thebibliography}
\end{document}